# Surface-enhanced Raman scattering and density functional theory study of selected-lanthanide-citrate complexes (lanthanide: Tb, Dy, Ho, Er, Tm, Yb and Lu)

Hao Jin[1,*] and Yuko S. Yamamoto[1,*]

[1] Graduate School of Advanced Science and Technology, Japan Advanced Institute of Science and Technology (JAIST), Nomi, Ishikawa 923-1292, Japan

*Corresponding authors: chnliefeng@gmail.com; yamayulab@gmail.com

**Abstract**

In this study, surface-enhanced Raman scattering (SERS) and density functional theory (DFT) calculations were combined to investigate the SERS spectra of Ln-citrate complexes (Ln: Tb, Dy, Ho, Er, Tm, Yb, and Lu) under 488 and 532 nm excitation. Peak assignment was supported by simulated SERS spectra calculated with an optimized DFT method using large-core effective core potentials. The main bands near 935, 1060, 1315, and 1485 cm$^{-1}$ were assigned to $\nu$(C-COO$^-$) + $\delta$(CH$_2$), $\gamma$(CH$_2$), $\gamma$(CH$_2$) + $\nu$(C-O$\cdots$Ln), $\nu_{sym}$(COO$^-$) + $\delta$(CH$_2$), and $\nu_{asym}$(COO$^-$) + $\gamma$(CH$_2$), respectively. Relative peak intensities were evaluated by normalizing the bands near 935, 1060, and 1485 cm$^{-1}$ to that near 1315 cm$^{-1}$. The ratios $I_{935}/I_{1315}$ and $I_{1485}/I_{1315}$ generally increased from Dy-citrate to Lu-citrate, whereas the $I_{1060}/I_{1315}$ ratio decreased. These trends were observed under both excitation wavelengths. The decrease in relative SERS peak intensity of the 1060 cm$^{-1}$ band is attributed to stronger Ln-O interaction and reduced polarizability change, whereas the increases of the 935 and 1485 cm$^{-1}$ bands are likely related to changes in local electronic distribution and effective symmetry sensitivity.

**Introduction**

Lanthanides (Ln), which include 15 elements from La (atomic number = 57) to Lu (atomic number = 71), are important rare earth elements[1]. As the atomic number increases, electrons are gradually filled into the 4f orbitals, and the electronic configuration of $Ln^{3+}$ ions can be written as $[Xe]4f^n$ (n = 0–14).[1] Owing to this unique electronic configuration, lanthanides show similar chemical properties while also possessing rich energy levels and high-spin states[1]. These features give rise to unusual optical and magnetic properties and support a wide range of applications, including luminescent materials[2-4], biomedical probes[3, 4], magnetic resonance imaging (MRI)[4, 5], quantum technologies[6], and spin-related studies.[7–9] Another important feature of lanthanides is the lanthanide contraction[1, 11]. Due to the weak shielding effect of 4f electron, the effective nuclear attraction on the outer electrons increases across the series, which leads to a gradual decrease in the ionic radius of $Ln^{3+}$ ions[1,10,11]. Since $Ln^{3+}$ ions and their molecular complexes are widely used in many fields, reliable methods are needed to detect their characteristic signals and structural features. In addition, conventional techniques, such as nuclear magnetic resonance (NMR)[12], fluorescence spectroscopy[2-4], and electron spin resonance (ESR) spectroscopy[13], still face difficulties in the rapid and sensitive analysis of Ln molecular complexes, especially at low concentrations. Therefore, a sensitive and non-contact method is needed for the detection of Ln molecular systems. Raman spectroscopy is a non-destructive and non-contact technique that provides molecular fingerprint information and vibrational structural information[14, 15]. When molecules are adsorbed on noble metal surfaces, the Raman signal can be greatly enhanced by electromagnetic and chemical effects[16-18], including charge-transfer (CT) effects[16-18]. This phenomenon is known as surface-enhanced Raman scattering (SERS)[16–18]. Because SERS can detect very low concentrations and even single molecules[18], it is a promising method for the analysis of Ln-molecular complexes.

However, the application of SERS to Ln-molecular complexes remains challenging. Because $Ln^{3+}$ ions have similar chemical properties, their small molecular complexes often have similar structures and similar vibrational features, which makes spectral discrimination difficult[19]. In addition, the 4f orbitals are shielded by the 5s and 5p orbitals and do not directly participate in bonding[1]. Some $Ln^{3+}$ ions, such as $Pr^{3+}$, $Nd^{3+}$, and $Eu^{3+}$, also have dense energy levels[20], which may cause fluorescence interference[21,22] or resonance Raman effects[23], further complicating spectral analysis. As a result, previous SERS studies of lanthanides have mainly focused on individual ions or limited systems, and systematic studies on Ln-small-molecule complexes are still limited.[15, 24,25] Such studies are nevertheless needed, because they are important for understanding the fundamental spectral behavior of lanthanide coordination systems. In our previous work[26], we systematically investigated the SERS spectra of Ln-citrate complexes from La to Gd (La, Ce, Pr, Nd, Sm, Eu, and Gd), excluding Pm. In this part of the lanthanide series, the atomic number increases from 57 to 64, and the 4f electron count of $Ln^{3+}$ ions increases from 0 to 7, corresponding to a progression from an empty 4f shell to a half-filled 4f shell with all electrons unpaired. Pm was not included because it is radioactive and unstable in nature. Additionally, due to the complex 4f electronic configuration of $Ln^{3+}$ ions and pronounced relativistic effects[27], we also developed a simplified DFT-based method for SERS simulation for the peak assignment[26].

In this study, we extended our previous SERS investigation of Ln-citrate complexes to the previously unexplored Tb-Lu region of the lanthanide series[26]. This part includes Tb, Dy, Ho, Er, Tm, Yb, and Lu, with atomic numbers from 65 to 71. In this region, the 4f electron count of $Ln^{3+}$ ions increases from 8 to 14, meaning that the 4f shell becomes progressively filled while the number of unpaired electrons gradually decreases to zero. These differences in electronic configuration led us to expect new spectral features in the SERS spectra of Tb-Lu citrate complexes. Moreover, the SERS spectra of these Tb-Lu small-molecule complexes have not yet been systematically studied. To examine this region systematically, we combined experimental SERS measurements with DFT-based vibrational mode assignment on a consistent measurement platform. The purpose of this study is to clarify how variation in the 4f electronic configuration affects the SERS spectral features of Ln-citrate complexes. The SERS spectra of their citrate complexes were measured using citrate-capped silver nanoparticle (citrate@AgNP) colloids after addition of the corresponding $Ln^{3+}$ ions show in the proposed papers[19,26]. With the aid of DFT-simulated SERS spectra for peak assignment, we analyzed the spectral differences among these complexes and discussed the possible origins of these differences. These results provide fundamental insight into how the 4f electronic configuration of $Ln^{3+}$ ions influences SERS spectral behavior in Ln-citrate complexes. This work also expands the spectral study of the lanthanide series from La–Gd to Tb–Lu, and supports future studies on the SERS measurement, spectral simulation, and coordination chemistry of Ln-molecular complexes.

**Experiments**

**Chemicals.** Eight lanthanide oxides were used to prepare the $Ln^{3+}$ solutions: $Gd_2O_3$, $Tb_4O_7$, $Dy_2O_3$, $Ho_2O_3$, $Er_2O_3$, $Tm_2O_3$, $Yb_2O_3$, and $Lu_2O_3$. All lanthanide oxides were purchased from Fujifilm Wako Pure Chemical Corporation (Japan) and had a purity higher than 99%. Ultrapure water produced by a Direct-Q® UV 3 system (Millipore, USA) was used as the solvent. Stock solutions of $Ln(NO_3)_3$ (Ln = Gd, Tb, Dy, Ho, Er, Tm, Yb, and Lu) with a concentration of 0.2 M were prepared by dissolving the corresponding lanthanide oxides in 1 M $HNO_3$ (Fujifilm Wako Pure Chemical Corporation, Japan) under heating. Because $Tb_4O_7$ was difficult to dissolve in 1 M $HNO_3$, 1-1.5 mL of 30% $H_2O_2$ (Fujifilm Wako Pure Chemical Corporation, Japan) was added during the dissolution process to assist dissolution. The stock solutions were then diluted with ultrapure water to obtain $2 \times 10^{-3}$ M $Ln(NO_3)_3$ solutions. Citrate-capped silver nanoparticles (citrate@AgNPs), which are colloidal AgNPs dispersed in water and stabilized by surface-bound citrate molecules, were prepared by a modified Lee–Meisel method, as described in our previous study[19, 26].

**Characterization.** Samples for SERS measurements were prepared by adding 50 μL of the $2 \times 10^{-3}$ M $Ln(NO_3)_3$ solution to 1 mL of citrate@AgNPs. The final concentration of each Ln-citrate@AgNPs sample was $1 \times 10^{-4}$ M. The samples were stored at room temperature for 12 h before spectroscopic measurements.

For ultraviolet-visible (UV-Vis) measurements, each SERS sample was diluted 10-fold with ultrapure water. The spectra were recorded using a UV-Vis-NIR spectrometer (V-770, JASCO, Japan) with a 1 cm path-length polystyrene cuvette. Ultrapure water was used as the blank.

For SERS measurements, Gd-citrate SERS sample was used to find the SERS measurement parameter that can reproduce the Gd-citrate SERS spectra in our previous study[26]. Especially, all sample solutions were loaded into clean soda glass capillaries (1.1 × 75 mm, DWK Life Sciences, USA) and allowed to stand for approximately 0.5–1 h before measurement. The SERS spectra were recorded using a Raman microscope (T64000, Horiba Scientific, Japan) equipped with a 90× objective lens. The excitation wavelengths were 488 and 532 nm, and the laser powers were 50 and 50 mW, respectively. The corresponding power densities at the sampling points were 101 and 85 mW $\mu m^{-2}$, respectively. To avoid sample damage caused by high laser power and to maintain repeatability, the exposure time was set to 30 s with 2 accumulations. Raman shift calibration was performed using indene. Each Ln-citrate complex was measured with 8 SERS spectra, and such SERS spectra were averaged for spectral analysis.

Spectral analysis was carried out using OriginPro 2022 (OriginLab Corporation, USA). Baseline correction of the raw spectra was performed using the interpolation mode. Noise reduction was then applied twice using the Savitzky-Golay method with an 11-point window and a polynomial order of 1.

**Computational methods** The simulated SERS spectra of Ln-citrate complexes on a silver surface were calculated using a simplified model developed in our previous work[19, 26]. The Ln-citrate complexes were modeled as LnCit⁻ based on the trisodium citrate structure obtained from the Cambridge Crystallographic Data Centre (CCDC, deposition number: 1478188). To represent the silver surface, an $Ag_{11}$ cluster with a size comparable to that of the LnCit⁻ complex was cut from bulk silver along the (111) plane[28].

Geometry optimization and vibrational frequency calculations were carried out by density functional theory (DFT) using Gaussian 16 (Revision C.01, Gaussian Inc., USA)[29]. Analytical Hessians were used in the vibrational calculations. The optimized structures and vibrational modes were analyzed using GaussView 6 (Gaussian Inc., USA)[30]. The B3LYP functional is widely used, but it is often not suitable for transition metals[31,32] or lanthanides[33,34]. In contrast, the PBE0 functional[35] has been reported to give better results for Ln molecular complexes[36] and small molecules[37]. Therefore, the PBE0 functional with DFT-D3(BJ) dispersion correction[38] was used in this work. To reproduce the experimental environment, water was included as the solvent using the SMD model[39].

For the basis sets, aug-cc-pVDZ[40] was used for C, H, O, and Na. The def2-SVPD

basis set[41] was used for Ag, and the def2-TZVPD basis set[42] was used for Ln. Because some Ln-citrate complexes showed convergence difficulties, large-core relativistic effective core potentials (ECPs)[43] were used consistently for all $Ln^{3+}$ ions. In these large-core ECPs, only 10 outer electrons are treated explicitly. According to the number of replaced core electrons, MWB53, MWB54, MWB55, MWB56, MWB57, MWB58, MWB59, and MWB60 were used for $Gd^{3+}$, $Tb^{3+}$, $Dy^{3+}$, $Ho^{3+}$, $Er^{3+}$, $Tm^{3+}$, $Yb^{3+}$, and $Lu^{3+}$, respectively, in place of the corresponding small-core ECPs in the def2-TZVPD basis set.

The Raman spectra of the Ln-citrate-$Ag_{11}$ models, which were used as the simulated SERS spectra of Ln-citrate complexes, were generated by converting the Raman scattering activity of each vibrational mode into Raman intensity using the following equation[44]:

$$I_i = \frac{C(v_0 - v_i)^4 S_i}{v_i B_i}; B_i = 1 - \exp\left(-\frac{hv_i C}{k_B T}\right) \qquad (1)$$

where i is the vibrational mode, C is an arbitrary normalization factor, $v$ is the vibrational frequency, $v_0$ is the frequency of the incident light, and S is the Raman activity calculated by Gaussian 16. h, c, $k_B$, and T are the Planck constant, the speed of light, the Boltzmann constant, and the temperature, respectively. In the present work, $v_0$ and T were set to 532 nm and 298.15 K, respectively, to match the experimental conditions.

The calculated Raman shifts were corrected using literature scaling factors of 0.956 and 0.971 for PBE0-D3(BJ)/aug-cc-pVDZ and PBE0-D3(BJ)/def2-TZVPD, respectively[45,46]. First, the Raman shifts in the 1000-1700 $cm^{-1}$ region were scaled by 0.956. Then, the peaks located in the 1000-1250 $cm^{-1}$ region after the first scaling were further scaled by 0.971 to account for the effect of $Ln^{3+}$ ions. The full width at half maximum (FWHM) of all simulated SERS spectra was set to 25 $cm^{-1}$ to approximate the experimental spectra. To support vibrational mode assignment, we also considered five different adsorption configurations of the Ln-citrate-$Ag_{11}$ models and compared their simulated SERS spectra. This additional analysis was used to further examine the assignment of the main vibrational modes. All DFT results were further analyzed using Multiwfn[47,48] (version 3.8-dev, Beijing Kein Research Center for Natural Sciences, China).

**Results and discussion**

**SERS measurements of Ln-citrate complexes (Ln: Tb, Dy, Ho, Er, Tm, Yb and Lu).**

The experimental results are discussed in the context of the differences in the electronic configurations of the $Ln^{3+}$ ions, the SERS enhancement mechanism, and the UV-Vis extinction spectra, as illustrated in Figure 1. Figure 1a shows the 4f electronic configurations of the selected $Ln^{3+}$ ions from $Tb^{3+}$ to $Lu^{3+}$. In this series, the number of 4f electrons increases from 8 to 14. These 4f electrons are shielded by the outer orbitals and therefore do not directly participate in chemical bonding[1]. Table 1 summarizes the electronic configuration, ionic radius, spin, and the reported absorption and emission wavelengths of these $Ln^{3+}$ ions[1,21,49]. From $Tb^{3+}$ to $Lu^{3+}$, the ionic radius gradually decreases with increasing atomic number. This trend is known as the lanthanide contraction[1]. Because 4f electrons provide only weak shielding of the nuclear charge, the effective nuclear attraction increases across the series, which leads to smaller ionic radii[1]. In the $Tb^{3+}$-$Lu^{3+}$ region, the newly added 4f electrons are mainly introduced as paired electrons.

Figure 1b illustrates the SERS measurement scheme used in this study. When individual $Ln^{3+}$ ions were added to citrate-capped AgNP colloids, they coordinated with the surface-bound citrate molecules to form Ln-citrate complexes on the AgNP surface. These systems are denoted as Ln-citrate@AgNP, where AgNP acts as the core and the Ln-citrate complex forms the outer coordination layer. At the same time, the addition of $Ln^{3+}$ ions induced aggregation of the AgNPs. When the interparticle distance became sufficiently small, localized surface plasmon resonance (LSPR) generated strong local electromagnetic fields between neighboring particles[16-18]. These regions are referred to as hotspots[16-18]. When Ln-citrate complexes were located in these hotspots, their Raman signals were strongly enhanced.

Figure 1c shows the UV-Vis extinction spectra of the SERS samples after 10-fold dilution with ultrapure water. The strong extinction band near 400 nm is mainly attributed to the LSPR of AgNPs, which arises from the collective oscillation of free electrons under light irradiation[16]. The gradual decrease in this extinction band after the addition of $Ln^{3+}$ ions indicates the aggregation of AgNPs in the SERS samples.

We combined experimental SERS measurements with DFT-based peak assignment to analyze the spectral features of the Ln-citrate complexes. Figures 2a and 2b show the normalized and averaged experimental SERS spectra obtained under 488 and 532 nm excitation, respectively, in the 500-2000 $cm^{-1}$ region. Because $Ln^{3+}$ ions have very similar chemical properties, the structural differences among the Ln-citrate complexes formed on the AgNP surface are expected to be small. This is also reflected in their SERS spectral features, especially in the very small differences in the Raman shifts of the main characteristic bands. In addition, based on our previous study, the vibrations observed in the 500-2000 $cm^{-1}$ region mainly originate from the coordination region between citrate and $Ln^{3+}$ ions. Although some Ln-citrate complexes may show pre-resonance Raman effects at 488 or 532 nm according to their absorption features (Table 1), no obvious spectral change attributable to resonance enhancement was observed. This is likely because resonance effects in the coordination region were weak. In addition, the absence of fluorescence from $Tb^{3+}$, $Dy^{3+}$, $Ho^{3+}$, and $Er^{3+}$ can be explained by fluorescence quenching on the AgNP surface[50,51].

Peak assignment was carried out with the aid of DFT simulations (Figure 3 and Table 2). In the 500-900cm$^{-1}$ region, the weak bands near 620, 690 and 700 cm$^{-1}$ were assigned to different $\delta(COO^-)$ modes. The bands near 740 and 820 cm$^{-1}$ were assigned to $\delta(COO^-)$ + $\gamma(COO^-)$ and $\nu(CCCC\text{-}O)$, respectively. Under both 488 and 532 nm excitation, the relative intensity differences in this region were small.

In the 900-1700 cm$^{-1}$ region, the main bands at about 935, 1060, 1315, and 1485 cm$^{-1}$ were assigned to $\nu(C\text{-}COO^-)$ + $\delta(CH_2)$, $\gamma(CH_2)$ + $\nu(C\text{-}O\cdots Ln)$, $\nu_{sym}(COO^-)$ + $\delta(CH_2)$, and $\nu_{asym}(COO^-)$ + $\gamma(CH_2)$, respectively. In these complexes, $Ln^{3+}$ coordinates with the citrate ligand through one -OH group and two of the three -COOH groups[12]. Therefore, bands involving C-O$\cdots$Ln and coordinated $COO^-$ vibrations can be regarded as modes related to the Ln-citrate coordination region, even when the direct vibrational contribution from Ln coordination is not dominant[19, 26]. The peak positions showed slight variation with both $Ln^{3+}$ species and excitation wavelength. This behavior may reflect secondary effects from electromagnetic enhancement, such as plasmon-dependent variation, and chemical enhancement, such as charge-transfer-related bond orientation effects[16,17,19].

For the SERS spectrum of Tb-citrate, the characteristic bands were located at Raman shifts similar to those of the other Ln-citrate complexes, with peaks near 690, 934, 1060, and 1487 $cm^{-1}$. However, some clear differences were also observed. For example, the Tb-citrate band appeared near 1057 $cm^{-1}$, whereas the corresponding band of the other Ln-citrate complexes was located in the 1060-1063 $cm^{-1}$ range. In addition, the overall spectral shape of Tb-citrate was clearly different, especially in the intensity of the band near 690 $cm^{-1}$ and in the spectral profile above 1300 $cm^{-1}$. According to the extinction spectra in Figure 1c, the aggregation of citrate@AgNPs was strongest after the addition of $Tb^{3+}$ ions, whereas the aggregation induced by the other $Ln^{3+}$ ions was generally similar. Therefore, the different spectral shape of Tb-citrate mainly arose from a difference in the state of the SERS sample.

Figure 3 compares the experimental SERS spectra under 532 nm excitation with the simulated spectra and the optimized structures of the Ln-citrate-$Ag_{11}$ models. The simulated spectra reproduced the main experimental bands above 1000 $cm^{-1}$ reasonably well, although some differences remained in other regions. These differences can be explained by several factors. First, the experimental SERS spectra are affected by multiple enhancement mechanisms and by the excitation wavelength, whereas the simulated spectra were calculated under a static external field and mainly reflect the ground-state charge-transfer contribution[16,17,52]. Second, the simulations used one adsorption model at a time and cannot fully represent the more complex adsorption situations in the experimental system (Figure S2-S8). Third, the DFT calculations still include intrinsic limitations, such as the effects of the solvent model and basis set choice, which may introduce errors in Raman shifts, although these can be partly corrected by scaling factors[19,53,54]. In addition, the effect of ECP size on the simulation results was confirmed to be acceptable[29]. Therefore, while the absolute intensities of the simulated SERS spectra cannot be directly compared with the experimental ones, the scaled Raman shifts still provide a reliable basis for peak assignment.

We also analyzed the effect of the excitation wavelength on the SERS spectrum of each individual Ln-citrate complex. To describe the Raman scattering intensity, we adopted the same theoretical framework as that used in our previous study[26]. The Raman scattering intensity $I_{mn}$, corresponding to the transition from state m to state n, is described by the following equations[55,56]:

$$I_{mn} = \frac{128\pi^5}{9C^4}(v_i \pm v_{mn})^4 I_i \sum_{p\sigma} \left|(\alpha_{\rho\sigma})_{mn}\right|^2 \tag{2}$$

$$(\alpha_{\rho\sigma})_{mn} = \frac{1}{h}\sum_e \left[\frac{\langle m|\mu_\sigma|e\rangle\langle e|\mu_\rho|n\rangle}{v_{em}-v_i+\mathrm{i}\Gamma_e} + \frac{\langle m|\mu_\rho|e\rangle\langle e|\mu_\sigma|n\rangle}{v_{en}+v_i+\mathrm{i}\Gamma_e}\right] \tag{3}$$

where, $I_i$ is the intensity of the incident light with frequency $v_i$. $v_{mn}$ represents the Raman shift. $\alpha_{\rho\sigma}$ denotes the component $\rho\sigma$ of the Raman scattering tensor. The summation e includes all quantum mechanical eigenstates of the molecule. $v_{em}$ and $v_{en}$ are the transition frequencies from state m to e and from state e to n, respectively. The terms $\langle m|\mu_\sigma|e\rangle$, $\langle e|\mu_\rho|n\rangle$, $\langle m|\mu_\rho|e\rangle$ and $\langle e|\mu_\sigma|n\rangle$ represent the components of the transition electric dipole moment. $\mu_\sigma$ and $\mu_\rho$ are the electric dipole moment operators in the $\sigma$ (Raman excitation) and $\rho$ (Raman scattering) direction, respectively. $\Gamma_e$ denotes the damping constant of the state. Equation (2) indicates that the Raman scattering intensity is proportional to the fourth power of the scattered light frequency $v_i \pm v_{mn}$. As $v_i \gg v_{mn}$, the Raman scattering intensity is considered proportional to the fourth power of the incident light frequency $v_i$. Therefore, we normalized and averaged the experimental spectra by considering only the major vibrational bands, so as to minimize the influence of laser-wavelength-dependent variations in absolute intensity. On this basis, our analysis focuses exclusively on the relative intensity relationships among the peaks near 935, 1060, 1315, and 1485 cm[-1]. The relative intensity of the peaks at approximately 1315 and 1485 cm[-1] differed significantly at these two excitation wavelengths. This is consistent with our previous study[19], which indicated that as the excitation wavelength moves further away from the LSPR peak (around 400 nm), the peak around 1315 cm[-1] becomes stronger relative to the peak near 1485 cm[-1].

**The relative peak intensity variation of SERS spectra**

To further examine the characteristic peak variations in the SERS spectra of the Ln-citrate complexes, we analyzed the changes in the main bands and discussed their possible origins. As shown in Figure 2, the experimental SERS spectra of the Ln-citrate complexes showed four main characteristic bands near 935, 1060, 1315, and 1485 cm$^{-1}$. Based on the DFT-assisted peak assignment, these bands were assigned to $\nu$(C-COO$^-$) + $\delta$(CH$_2$), $\gamma$(CH$_2$) + $\nu$(C-O$\cdots$Ln), $\nu_{sym}$(COO$^-$) + $\delta$(CH$_2$), and $\nu_{asym}$(COO$^-$) + $\gamma$(CH$_2$), respectively.

To further compare the spectral differences among the Ln-citrate complexes, we analyzed the relative intensities of these characteristic bands under different excitation wavelengths. The band near 1315 cm$^{-1}$ was used as the reference, and the relative intensities of the bands near 935, 1060, and 1485 cm$^{-1}$ were expressed as the SERS intensity ratios $I_{935}/I_{1315}$, $I_{1060}/I_{1315}$, and $I_{1485}/I_{1315}$, respectively. Numerically, these values are equivalent to the relative intensities obtained after normalizing the 1315 cm$^{-1}$ band to 1. The results are summarized in Figure 4.

It should be noted that the spectral shape of Tb-citrate was clearly different from that of the other Ln-citrate complexes. Although the characteristic bands of Tb-citrate appeared at similar Raman shifts, its overall spectral profile was different, especially in the regions near 690 cm$^{-1}$ and above 1300 cm$^{-1}$. This difference is likely related to a different sample state, which is also supported by the extinction spectrum in Figure 1c.

Because the preparation of the Tb-citrate SERS sample could not be further optimized under the present experimental conditions, the Tb-citrate data are shown in Figure 4 but were not included in the comparison and discussion of the spectral trends.

Figures 4a and 4b show the intensity ratio $I_{935}/I_{1315}$ under 488 and 532 nm excitation, respectively. For the Ln-citrate complexes from Dy-citrate to Lu-citrate, this ratio gradually increased. In contrast to our previous La-citrate to Gd-citrate study[26], in which the corresponding band showed little relative change, the present Dy-Lu series exhibited a clear increasing trend.

Figures 4c and 4d show the intensity ratio $I_{1060}/I_{1315}$ under 488 and 532 nm excitation, respectively. From Dy-citrate to Lu-citrate, this ratio gradually decreased. The ratios of Tm-citrate, Yb-citrate, and Lu-citrate were relatively similar. This trend was observed under both excitation wavelengths and was consistent with that observed in our previous study[26].

Figures 4e and 4f show the intensity ratio $I_{1485}/I_{1315}$ under 488 and 532 nm excitation, respectively. From Dy-citrate to Lu-citrate, this ratio gradually increased, and the increase was more obvious under 532 nm excitation. This trend was opposite to that observed previously for the La-citrate to Gd-citrate series[26], where the relative intensity of the band near 1485 cm$^{-1}$ gradually decreased.

Overall, the same variation tendencies were observed under both 488 and 532 nm excitation, although the numerical ranges of the intensity ratios were different, with the exclusion of Tb-citrate. Thus, Figure 4 shows that the relative intensities of the main SERS bands in Ln-citrate complexes vary systematically across the Dy-Lu series. In addition, as shown in Figures S9-S17, the experimental SERS spectra were collected at different measurement points within the same sample solution, and some variation in absolute intensity was unavoidable because of local differences in AgNP aggregation and hotspot distribution. However, as demonstrated in our previous study[26], the relative peak intensity ratios remained within reproducible ranges for each $Ln^{3+}$ ion and could still reflect consistent Ln-dependent spectral trends. Therefore, in this study, these ratios were used as practical parameters for analyzing relative intensity changes rather than as exact quantitative fingerprints for universal ion identification.

**The relative peak intensity variation of SERS spectra**

To further analyze the origin of the Ln-dependent SERS peak's relative intensity trends observed in Figure 4, we examined the possible mechanism of these spectral changes. Figure 5a shows the DFT-optimized structures of representative Tb-citrate@AgNP, Dy-citrate@AgNP, and Lu-citrate@AgNP models, together with the electronic configurations of the corresponding $Ln^{3+}$ ions. The other Ln-citrate systems are omitted for simplicity. As the atomic number increases from Tb to Lu, the number of 4f electrons increases, whereas the number of unpaired 4f electrons decreases. Because these $Ln^{3+}$ ions have very similar chemical properties, and because the Raman shifts of their main SERS bands are also very close, only small structural differences are expected among the Ln-citrate complexes. This expectation is supported by the DFT results. As shown in Figure 5b, the four main characteristic bands near 935, 1060, 1315, and 1485 $cm^{-1}$ were assigned to $\nu(C\text{-}COO^-) + \delta(CH_2)$, $\gamma(CH_2) + \nu(C\text{-}O\cdots Ln)$, $\nu_{sym}(COO^-) + \delta(CH_2)$, and $\nu_{asym}(COO^-) + \gamma(CH_2)$, respectively. In the Ln-citrate complexes, $Ln^{3+}$ coordinates with one hydroxyl group and two of the three carboxyl groups of citrate. Therefore, the main vibrational features include one $C\text{-}O\cdots Ln$-related mode and two $COO^-\cdots Ln$-related modes. These characteristic bands can thus be regarded as mainly originating from the coordination region, and their relative intensity changes are likely affected by the nature of the $Ln^{3+}$ ion. A possible explanation is that, as the atomic number and 4f electron count increase, the effective nuclear charge of $Ln^{3+}$ also increases, which strengthens the attraction between $Ln^{3+}$ and the negative charge in the coordination region. This effect is consistent with lanthanide contraction. As a result, the electron distribution in the coordination region becomes less easily distorted during vibration, which may reduce the polarizability response and hence the Raman intensity of coordination-related modes.

Within this framework, the band near 1315 cm$^{-1}$, which is dominated by the symmetric $\nu_{sym}$(COO$^-$)vibration, is also expected to decrease in Raman intensity as the effective nuclear charge increases. The band near 1060 cm$^{-1}$, which contains the $\nu$(C$-$O$\cdots$Ln) contribution and is more directly related to the C-O$\cdots$Ln coordination unit, appears to be even more strongly affected. Therefore, its intensity decreases more markedly than that of the 1315 cm$^{-1}$ band, leading to the observed decrease in the $I_{1060}/I_{1315}$ ratio.

In contrast, from Dy-citrate to Lu-citrate ,the bands near 935 and 1485 cm$^{-1}$ are associated with $\nu$(C-COO$^-$) + $\delta$(CH$_2$) and $\nu_{asym}$(COO$^-$) + $\gamma$(CH$_2$),, respectively. These modes are mixed or asymmetric vibrations involving COO$^-$ motion. Compared with the symmetric COO$^-$ mode near 1315 cm$^{-1}$, they may be less strongly suppressed by the decrease in polarizability response in the coordination region. As a result, although their absolute Raman intensities may also decrease, their decrease is likely smaller than that of the 1315 cm$^{-1}$ band. This would lead to the apparent increase in the $I_{935}/I_{1315}$ and $I_{1485}/I_{1315}$ ratios.

Therefore, the opposite trends of $I_{935}/I_{1315}$ and $I_{1485}/I_{1315}$ relative to $I_{1060}/I_{1315}$ suggest that the effect of Ln$^{3+}$ ions on the Ln-citrate SERS spectra is not uniform for all vibrational modes. Instead, the stronger Ln-O interaction from Dy to Lu appears to suppress the C-O···Ln-related and symmetric $COO^-$-dominated modes more strongly than the mixed or asymmetric $COO^-$-related modes. Although this interpretation is indirect, it provides a reasonable framework for understanding the systematic intensity changes observed across the Dy-Lu series.

In addition, although Tb-citrate might also be expected to follow the general spectral trends described above, the present analysis was not applied to Tb-citrate because of differences in sample preparation and the resulting sample state under the current experimental conditions. As discussed above, the spectral shape of Tb-citrate was clearly different from those of the other Ln-citrate complexes, which makes direct comparison less reliable.

It should also be noted that the selected $Ln^{3+}$ ions have clear differences in atomic mass. However, this effect was not obvious in the 500-2000 $cm^{-1}$ region, because the main vibrational contributions in this range originate mainly from the citrate ligand rather than from direct metal-centered motion.

**Conclusion**

In this study, we successfully measured the SERS spectra of selected Ln-citrate complexes (Ln = Tb, Dy, Ho, Er, Tm, Yb, and Lu) at a concentration of $1 \times 10^{-4}$ M under 488 and 532 nm excitation. The experimental spectra showed characteristic bands near 935, 1060, 1315, and 1485 cm$^{-1}$, with no detectable resonance Raman effect or fluorescence signal. The simulated SERS spectra were successfully calculated using an optimized DFT method based on large-core ECP replacement, which enabled reliable peak assignment and vibrational analysis. SERS intensity ratios were obtained by normalizing the peak intensities near 935, 1060, and 1485 cm$^{-1}$ to that of the band near 1315 cm$^{-1}$.

The results showed that the relative intensities of the bands near 935 and 1485 cm$^{-1}$ generally increased from Dy-citrate to Lu-citrate, whereas the $I_{1060}/I_{1315}$ ratio decreased. These trends suggest that the vibrational mode of SERS response of Ln-citrate complexes is sensitive to systematic changes in 4f electronic configuration. The decrease of the 1060 cm$^{-1}$ band is likely related to stronger Ln-O interaction and reduced polarizability change in the coordination region, whereas the increases of the 935 and 1485 cm$^{-1}$ bands may reflect changes in the local electronic distribution and effective symmetry that is less sensitive than band near 1315 cm$^{-1}$.

These topics will be investigated in our future work. The findings of the present study provide effective guidance and reference for the vibrational spectroscopy studies of other lanthanides and even actinides. These insights lay the groundwork for future development of SERS-based analytical methods for electronic structure analysis and classification of f-block elements.

**Acknowledgments**

The authors thank Dr. Tamitake Itoh (National Institute of Advanced Industrial Science and Technology (AIST), Japan) for his valuable comments on the manuscript. The authors acknowledge Dr. Tian Lu (Beijing Kein Research Center for Natural Sciences, China) for his fruitful suggestions for the DFT calculations of rare earth elements. The authors acknowledge funding from JSPS KAKENHI Grant-in-Aid for Scientific Research (C), number 21K04935, 25K08506 and 25K08520.

**Supporting information**

Experimental SERS spectra of Gd-citrate. Effect of adsorption modes on the simulated SERS spectra. Experimental SERS spectra. SERS peak assignment. Modeling data.

**References**

(1) Simon Cotton. Lanthanide and Actinide Chemistry;  John Wiley & Sons, Ltd, 2006;

pp 9– 102.

(2)Kotyk, C. M.; Weber, J. E.; Hyre, A. S.; McNeely, J.; Monteiro, J. H. S. K.; Domin, M.; Balaich, G. J.; Rheingold, A. L.; de Bettencourt-Dias, A.; Doerrer, L. H. Luminescence of Lanthanide Complexes with Perfluorinated Alkoxide Ligands. Inorg. Chem. 2020, 59 (14), 9807– 9823

(3)Eliseeva, S. V.; Bünzli, J.-C. G. Lanthanide Luminescence for Functional Materials and Bio-Sciences. Chem. Soc. Rev. 2010, 39 (1), 189– 227

(4)Wang, G.; Peng, Q.; Li, Y. Lanthanide-Doped Nanocrystals: Synthesis, Optical-Magnetic Properties, and Applications. Acc. Chem. Res. 2011, 44 (5), 322– 332

(5)Lee, H. Y.; Jee, H. W.; Seo, S. M.; Kwak, B. K.; Khang, G.; Cho, S. H. Diethylenetriaminepentaacetic Acid–Gadolinium (DTPA-Gd)-Conjugated Polysuccinimide Derivatives as Magnetic Resonance Imaging Contrast Agents. Bioconjugate Chem. 2006, 17 (3), 700– 706

(6)Serrano, D.; Kuppusamy, S. K.; Heinrich, B.; Fuhr, O.; Hunger, D.; Ruben, M.; Goldner, P. Ultra-Narrow Optical Linewidths in Rare-Earth Molecular Crystals. Nature 2022, 603 (7900), 241– 246

(7)Sushkov, A. O.; Chisholm, N.; Igor Lovchinsky; Kubo, M.; Lo, P.-C.; Bennett, S.; Hunger, D.; Akimov, A. V.; Walsworth, R. L.; Park, H.; Lukin, M. D. All-Optical Sensing of a Single-Molecule Electron Spin. Nano Letters 2014, 14 (11), 6443–6448

(8)Li, C.-Y.; Langit Cahya Adi; Paillot, K.; Breslavetz, I.; Long, L.-S.; Zheng, L.-S.; Geert L. J. A. Rikken; Train, C.; Kong, X.-J.; Matteo Atzori. Enhancement of Magneto-Chiral Dichroism Intensity by Chemical Design: The Key Role of Magnetic-Dipole Allowed Transitions. Journal of the American Chemical Society 2024, 146 (24), 16389–16393

(9)Lunkley, J. L.; Dai Shirotani; Kazuaki Yamanari; Sumio Kaizaki; Muller, G. Extraordinary Circularly Polarized Luminescence Activity Exhibited by Cesium Tetrakis(3-Heptafluoro-Butylryl-(+)-Camphorato) Eu(III) Complexes in EtOH and CHCl$_3$ Solutions. Journal of the American Chemical Society 2008, 130 (42), 13814–13815

(10)Geoffroy Ferru; Reinhart, B.; Bera, M. K.; Olvera, M.; Qiao, B.; Ellis, R. J. The Lanthanide Contraction beyond Coordination Chemistry. 2016, 22 (20), 6899–6904.

(11)Seitz, M.; Oliver, A. G.; Raymond, K. N. The Lanthanide Contraction Revisited. Journal of the American Chemical Society 2007, 129 (36), 11153–11160

(12)Ivanova, V. Yu.; Shurygin, I. D.; Chevela, V. V.; Ajsuvakova, O. P.; Semenov, V. E.; Bezryadin, S. G. New Aspects of Complex Formation in the Gadolinium(III)–Citric Acid System in Aqueous Solution. Comments on Inorganic Chemistry 2022, 42 (2), 109– 144

(13)Baggio, R.; Calvo, R.; Garland, M. T.; Peña, O.; Perec, M.; Rizzi, A. C. Gadolinium

and neodymium citrates: Evidence for weak ferromagnetic exchange between gadolinium (III) cations. Inorg. Chem. 2005, 44 (24), 8979– 8987

(14)Cao, Y. C. Nanoparticles with Raman Spectroscopic Fingerprints for DNA and RNA Detection. Science 2002, 297 (5586), 1536–1540.

(15)Rudolph, W. W.; Irmer, G. Raman Spectroscopic Characterization of Light Rare Earth Ions: $La^{3+}$, $Ce^{3+}$, $Pr^{3+}$, $Nd^{3+}$ and $Sm^{3+}$– Hydration and Ion Pair Formation. Dalton Transactions 2017, 46 (13), 4235–4244.

(16)Itoh, T.; Procházka, M.; Dong, Z.-C.; Ji, W.; Yamamoto, Y. S.; Zhang, Y.; Ozaki, Y. Toward a New Era of SERS and TERS at the Nanometer Scale: From Fundamentals to Innovative Applications. Chem. Rev. 2023, 123 (4), 1552– 1634

(17)Yamamoto, Y. S.; Itoh, T. Why and How Do the Shapes of Surface-Enhanced Raman Scattering Spectra Change? Recent Progress from Mechanistic Studies. Journal of Raman Spectroscopy 2016, 47 (1), 78–88.

(18)Nie, S.; Emory, S. R. Probing Single Molecules and Single Nanoparticles by Surface-Enhanced Raman Scattering. Science 1997, 275 (5303), 1102– 1106

(19)Jin, H.; Itoh, T.; Yamamoto, Y. S. Classification of $La^{3+}$ and $Gd^{3+}$ Rare-Earth Ions Using Surface-Enhanced Raman Scattering. The Journal of Physical Chemistry C 2024, 128 (13), 5611–5620.

(20)Peijzel, P. S.; Meijerink, A.; Wegh, R. T.; Reid, M. F.; Burdick, G. W. A Complete Energy Level Diagram for All Trivalent Lanthanide Ions. Journal of Solid State Chemistry 2005, 178 (2), 448–453.

(21)Hagemann, H.; Ayoubipour, S.; Delgado, T.; Schnyder, C.; Gnos, E. Probing Luminescence of Rare Earth Ions in Natural Pink Fluorites Using Raman Microscopes. Journal of Raman Spectroscopy 2022, 53 (8), 1464–1470.

(22)Lenz, C.; Nasdala, L.; Talla, D.; Hauzenberger, C.; Seitz, R.; Kolitsch, U. Laser-Induced REE$^{3+}$ Photoluminescence of Selected Accessory Minerals — an "Advantageous Artefact" in Raman Spectroscopy. Chemical Geology 2015, 415, 1–16.

(23)Albrecht, A. On the Theory of Raman Intensities. Journal of Chemical Physics 1961, 34 (5), 1476–1484.

(24)Yang, S.; Yao, J.; Quan, Y.; Hu, M.; Su, R.; Gao, M.; Han, D.; Yang, J. Monitoring the Charge-Transfer Process in a Nd-Doped Semiconductor Based on Photoluminescence and SERS Technology. Light: Sci. Appl. 2020, 9 (1), 117

(25)López-Neira, J. P.; Galicia-Hernández, J. M.; Reyes-Coronado, A.; Pérez, E.; Castillo-Rivera, F. Surface Enhanced Raman Scattering of Amino Acids Assisted by Gold Nanoparticles and Gd$^{3+}$ Ions. J. Phys. Chem. A 2015, 119 (18), 4127– 4135

(26)Hao Jin, Tamitake Itoh, Yuko S Yamamoto; Surface-enhanced Raman scattering and density functional theory study of selected-lanthanide-citrate complexes (lanthanide: La, Ce, Pr, Nd, Sm, Eu, and Gd), submitted to The Journal of Physical Chemistry C

(27)Seth, M.; Dolg, M.; Fulde, P.; Schwerdtfeger, P. Lanthanide and Actinide Contractions: Relativistic and Shell Structure Effects. Journal of the American Chemical Society 1995, 117 (24), 6597–6598

(28)Grimme, S.; Antony, J.; Ehrlich, S.; Krieg, H. A Consistent and Accurate Ab Initio Parametrization of Density Functional Dispersion Correction (DFT-D) for the 94 Elements H-Pu. J. Chem. Phys. 2010, 132 (15), 15410

(29)Gaussian 16, Revision C.01, Frisch, M. J.; Trucks, G. W.; Schlegel, H. B.; Scuseria, G. E.; Robb, M. A.; Cheeseman, J. R.; Scalmani, G.; Barone, V.; Petersson, G. A.; Nakatsuji, H. et al . Gaussian, Inc., Wallingford CT, 2016.

(30)GaussView, Version 6.1, Roy Dennington, Todd A. Keith, and John M. Millam, Semichem Inc., Shawnee Mission, KS, 2016.

(31)Paier, J.; Marsman, M.; Kresse, G. Why Does the B3LYP Hybrid Functional Fail for Metals?. J. Chem. Phys. 2007, 127 (2), 024103

(32)Minenkov, Y.; Singstad, Å.; Occhipinti, G.; Jensen, V. R. The Accuracy of DFT-Optimized Geometries of Functional Transition Metal Compounds: A Validation Study of Catalysts for Olefin Metathesis and Other Reactions in the Homogeneous Phase. Dalton

Transactions 2012, 41 (18), 5526.

(33)Martins, C.; Jorge, F. E.; Franco, M. L.; Ferreira, I. B. All-Electron Gaussian Basis Sets of Double Zeta Quality for the Actinides. Journal of chemical physics 2016, 145 (24).

(34)Roca-Sabio, A.; Regueiro-Figueroa, M.; Esteban-Gómez, D.; de Blas, A.; Rodríguez-Blas, T.; Platas-Iglesias, C. Density Functional Dependence of Molecular Geometries in Lanthanide(III) Complexes Relevant to Bioanalytical and Biomedical Applications. Computational and Theoretical Chemistry 2012, 999, 93–104.

(35)Adamo, C.; Barone, V. Toward Reliable Density Functional Methods without Adjustable Parameters: The PBE0Model. J. Chem. Phys. 1999, 110 (13), 6158– 6170

(36)Chen, X.; Chen, T.-T.; Li, W.-L.; Lu, J.-B.; Zhao, L.-J.; Jian, T.; Hu, H.-S.; Wang, L.-S.; Li, J. Lanthanides with Unusually Low Oxidation States in the PrB3- and PrB4- Boride Clusters. Inorg. Chem. 2019, 58 (1), 411– 418

(37)Brémond, É.; Savarese, M.; Su, N. Q.; Pérez-Jiménez, Á. J.; Xu, X.; Sancho-García, J. C.; Adamo, C. Benchmarking Density Functionals on Structural Parameters of Small-/Medium-Sized Organic Molecules. J. Chem. Theory Comput. 2016, 12 (2), 459– 465

(38)Grimme, S.; Ehrlich, S.; Goerigk, L. Effect of the damping function in dispersion corrected density functional theory. J. Comput. Chem. 2011, 32, 1456– 1465,

(39)Marenich, A. V.; Cramer, C. J.; Truhlar, D. G. Universal Solvation Model Based on Solute Electron Density and on a Continuum Model of the Solvent Defined by the Bulk Dielectric Constant and Atomic Surface Tensions. The Journal of Physical Chemistry B 2009, 113 (18), 6378–6396.

(40)Dunning, T. H. Gaussian Basis Sets for Use in Correlated Molecular Calculations. I. The Atoms Boron through Neon and Hydrogen. J. Chem. Phys. 1989, 90 (2), 1007– 1023,

(41)Andrae, D.; Häußermann, U.; Dolg, M.; Stoll, H.; Preuß, H. Energy-Adjustedab Initio Pseudopotentials for the Second and Third Row Transition Elements. Theoretica Chimica Acta 1990, 77 (2), 123– 141

(42)Rappoport, D. Property-Optimized Gaussian Basis Sets for Lanthanides. J. Chem. Phys. 2021, 155 (12), 124102

(43)Dolg, M.; Stoll, H.; Savin, A.; Preuss, H. Energy-Adjusted Pseudopotentials for the Rare Earth Elements. Theor. Chim. Acta 1989, 75 (3), 173– 194

(44)Liu, Z.; Lu, T.; Chen, Q. Vibrational Spectra and Molecular Vibrational Behaviors of All-Carboatomic Rings, Cyclo[18]Carbon and Its Analogues. Chemistry – An Asian Journal 2020, 16 (1), 56–63.

(45)Kashinski, D. O.; Chase, G. M.; Nelson, R. G.; Di Nallo, O. E.; Scales, A. N.; VanderLey, D. L.; Byrd, E. F. C. Harmonic Vibrational Frequencies: Approximate Global Scaling Factors for TPSS, M06, and M11 Functional Families Using Several Common

Basis Sets. Journal of Physical Chemistry. A 2017, 121 (11), 2265– 2273

(46)Straßner, A.; Wiehn, C.; Klein, M. P.; Fries, D. V.; Dillinger, S.; Mohrbach, J.; Prosenc, M. H.; Armentrout, P. B.; Gereon Niedner-Schatteburg. Cryo Spectroscopy of N2 on Cationic Iron Clusters. The Journal of Chemical Physics 2021, 155 (24).

(47)Lu, T. A Comprehensive Electron Wavefunction Analysis Toolbox for Chemists, Multiwfn. The Journal of Chemical Physics 2024, 161 (8).

(48)Lu, T.; Chen, F. Multiwfn: A Multifunctional Wavefunction Analyzer. J. Comput. Chem. 2011, 33 (5), 580–592.

(49) Sharma, S. K.; Behm, T.; Köhler, T.; Beyer, J.; Gloaguen, R.; Heitmann, J. Library of UV-Visible Absorption Spectra of Rare Earth Orthophosphates, LnPO4 (Ln = La-Lu, except Pm). Crystals 2020, 10 (7), 593.

(50) Nerambourg, N.; Werts, N.; Charlot, M.; Blanchard-Desce, M. Quenching of Molecular Fluorescence on the Surface of Monolayer-Protected Gold Nanoparticles Investigated Using Place Exchange Equilibria. Langmuir 2007, 23 (10), 5563–5570.

(51)Hildebrandt, P.; Stockburger, M. Surface-Enhanced Resonance Raman Spectroscopy of Rhodamine 6G Adsorbed on Colloidal Silver. The Journal of Physical Chemistry 1984, 88 (24), 5935–5944.

(52)Polavarapu, P. L. Ab Initio Vibrational Raman and Raman Optical Activity Spectra.

The Journal of Physical Chemistry 1990, 94 (21), 8106–8112.

(53)Ahuja, T.; Chaudhari, K.; Paramasivam, G.; Ragupathy, G.; Mohanty, J. S.; Pradeep, T. Toward Vibrational Tomography of Citrate on Dynamically Changing Individual Silver Nanoparticles. J. Phys. Chem. C 2021, 125 (6), 3553– 3566

(54)Leonard, J.; Haddad, A.; Green, O.; Birke, R. L.; Kubic, T.; Kocak, A.; Lombardi, J. R. SERS , Raman , and DFT Analyses of Fentanyl and Carfentanil: Toward Detection of Trace Samples. Journal of Raman Spectroscopy 2017, 48 (10), 1323–1329.

(55)Udagawa, Y.; Mikami, N.; Kaya, K.; Ito, M. Absolute Intensity Ratios of Raman Lines of Benzene and Ethylene Derivatives with 5145 Å and 3371 Å Excitation. Journal of Raman Spectroscopy 1973, 1 (4), 341–346.

(56)Albrecht, A. C.; Hutley, M. C. On the Dependence of Vibrational Raman Intensity on the Wavelength of Incident Light. Journal of Chemical Physics 1971, 55 (9), 4438–4443.

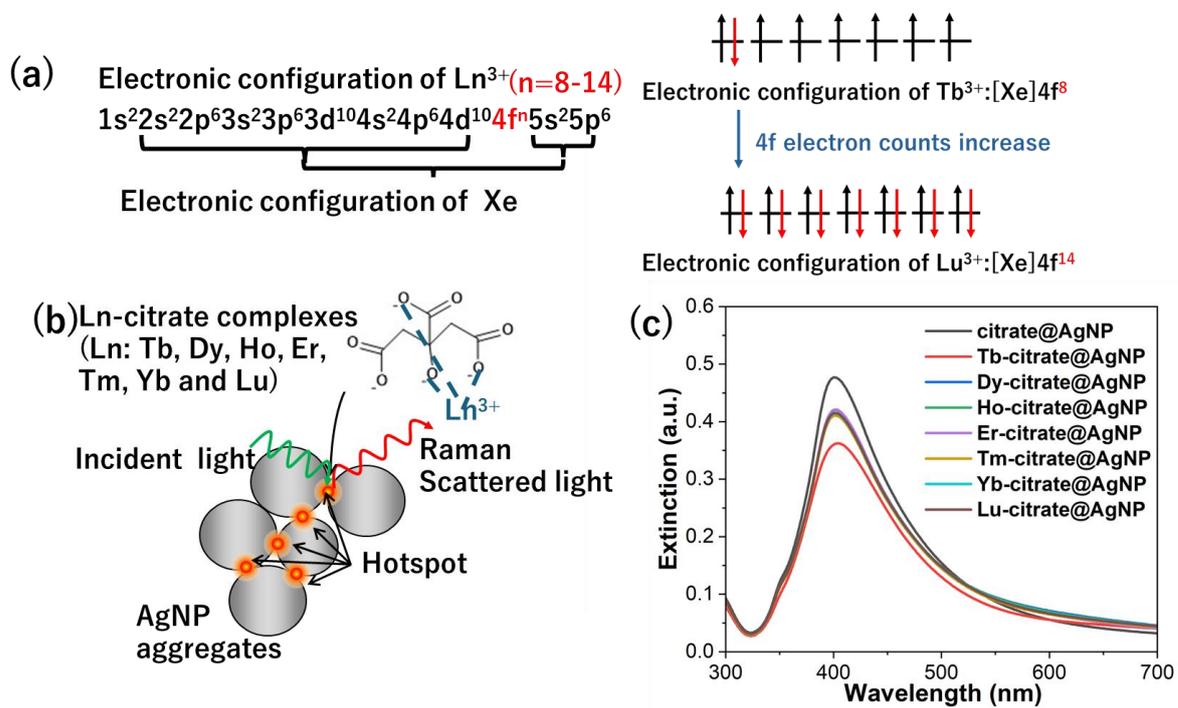

Figure 1. (a) Schematic of the electronic configuration of $Ln^{3+}$ ion(Ln: Tb, Dy, Ho, Er, Tm, Yb and Lu). (b) Schematic of SERS measurement. (c) UV-Vis extinction spectra of original citrate@AgNP and Ln-citrate@AgNP samples.

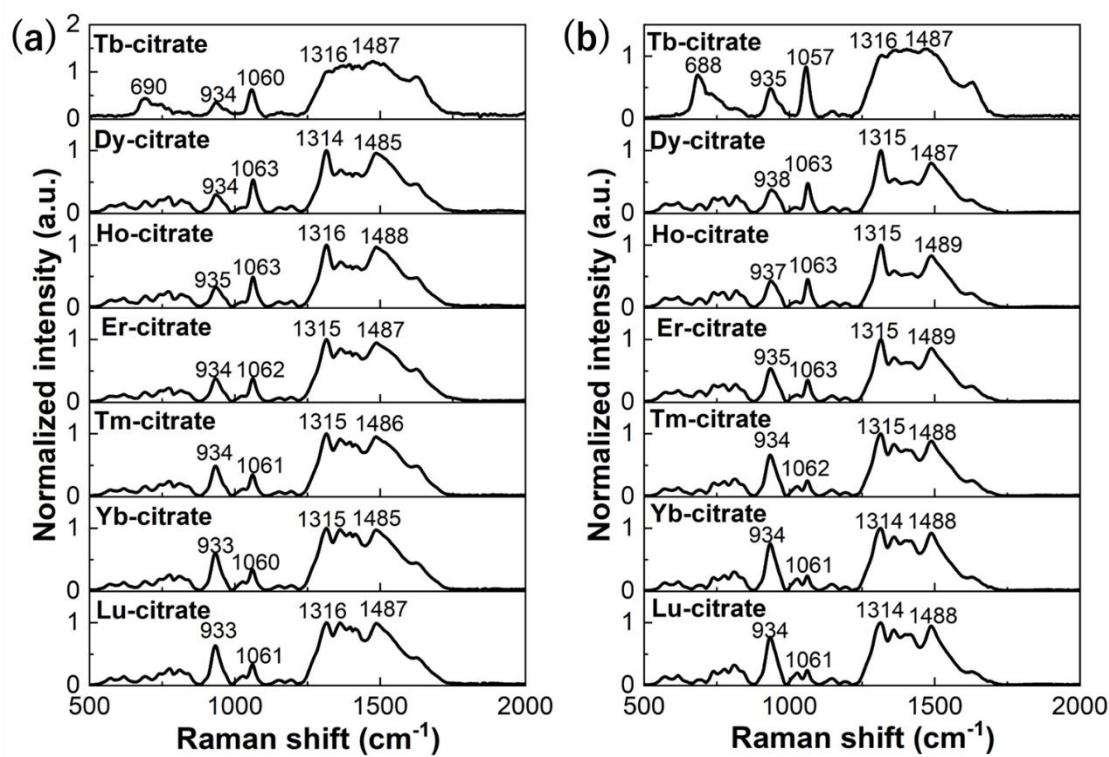

Figure 2. Normalized and averaged SERS spectra of Ln-citrate complexes under excitation at (a) 488 nm, (b) 532 nm. The averaged number (N) of experiment SERS spectra for Tb-citrate through Lu-citrate were N = 8 under 488 nm and 532 nm excitation respectively. All experimental SERS spectra shown in Figure S3-S9.

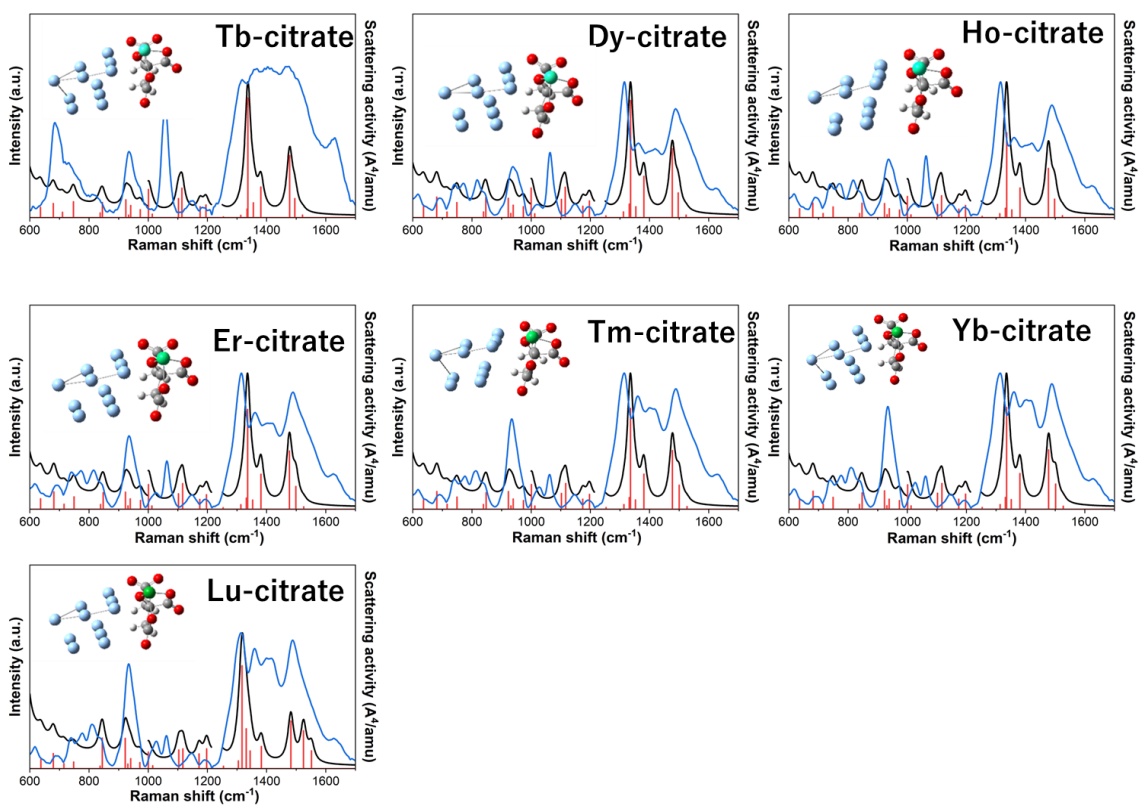

Figure 3. Experimental SERS spectra (red curves) at $1 \times 10^{-4}$ M, calculated vibrational modes (black curves), and simulated SERS spectra (blue curves) for (a) Tb-citrate, (b) Dy-citrate, (c) Ho-citrate, (d) Er-citrate, (e) Tm-citrate, (f) Yb-citrate, and (g) Lu-citrate. As shown in the structures, all simulations were based on a single-molecule SERS model. Some regions between 1000 cm$^{-1}$ and 1213-1250 cm$^{-1}$ were missing due to double scaling when the spectra were exported to Multiwfn. However, this has no effect on the results because there are no vibrational modes in these missing regions.

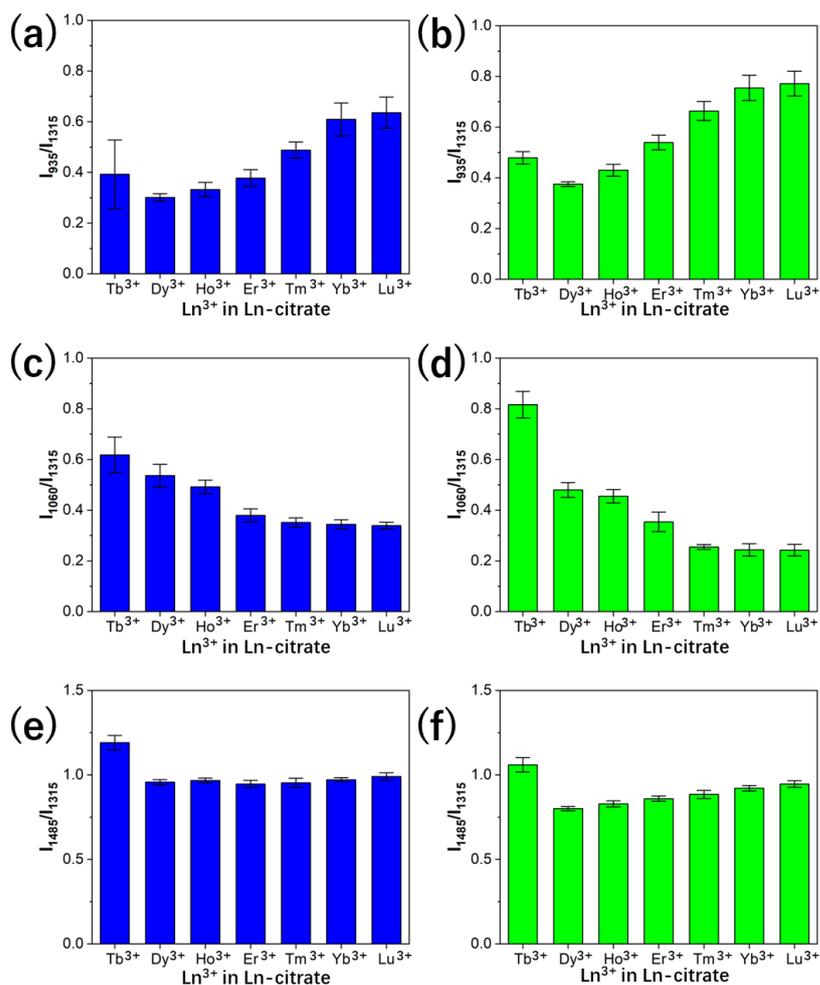

Figure 4. (a-b): SERS intensity ratio $I_{935}/I_{1315}$ under excitation at (a) 488 nm, (b) 532 nm.

(c-d): SERS intensity ratio $I_{1060}/I_{1315}$ under excitation at (a) 488 nm, (b) 532 nm.  (e-f):

SERS intensity ratio $I_{1485}/I_{1315}$ under excitation at (a) 488 nm, (b) 532 nm. The SERS

intensity ratios were obtained by normalizing the peak intensity around 935, 1060 or 1485

cm$^{-1}$ to that near 1315 cm$^{-1}$. The measurement times (N) for Tb-citrate through Lu-citrate were N = 8 under 488 nm and 532 nm excitation, respectively. Error bars indicate ± SD.

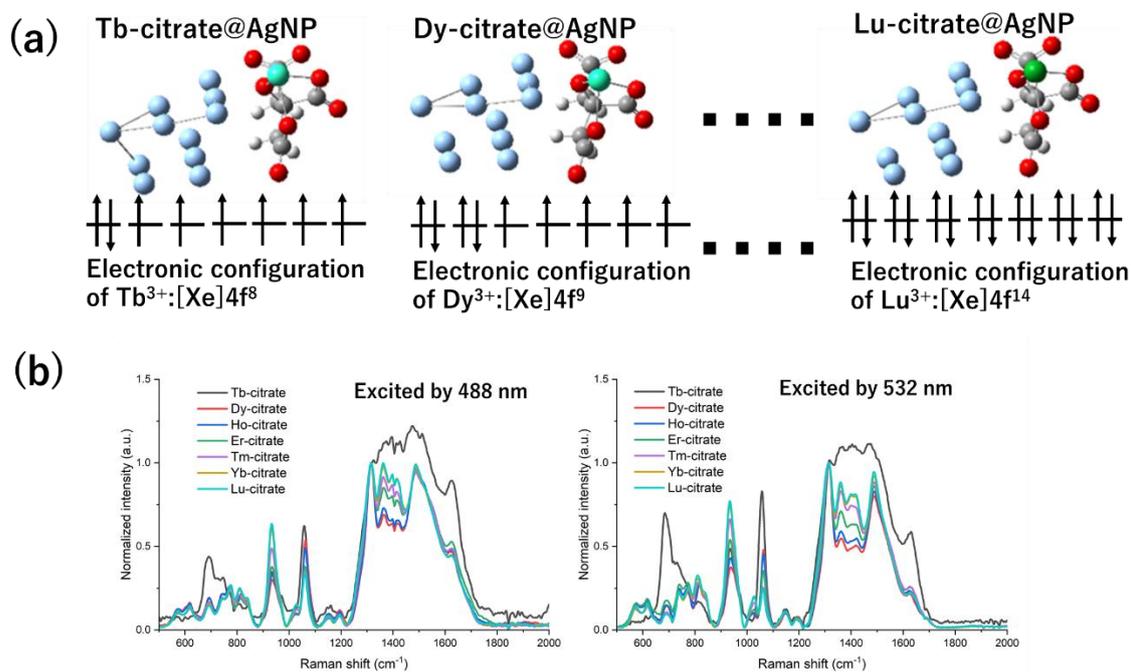

Figure 5. (a) DFT-optimized example structures of Tb-citrate@AgNP, Dy-citrate@AgNP, and Lu-citrate@AgNP, together with the electronic configurations of the corresponding Ln³⁺ ions. (b) Experimental SERS spectra of Ln-citrate complexes (Ln = Tb-Lu) under 488 and 532 nm excitation.

Table 1. Basic information on $Ln^{3+}$[1,21,52]

| Type of $Ln^{3+}$ | Electronic configuration | Ionic radius (pm) | Total spin S | Absorption wavelength (nm) | Emission wavelength (nm) |
|---|---|---|---|---|---|
| $Tb^{3+}$ | $[Xe]4f^8$ | 92.3 | 3 | 532, 490 | 345, 450, 492, 549, 580, 654 |
| $Dy^{3+}$ | $[Xe]4f^9$ | 91.2 | 5/2 | 367, 452 | 478, 574, 670, 758, 833, 913, 984, 1142, 1256, 1349 |
| $Ho^{3+}$ | $[Xe]4f^{10}$ | 90.1 | 2 | 289, 485, 656 | 553, 645, 660, 750, 890, 970, 1000 |
| $Er^{3+}$ | $[Xe]4f^{11}$ | 89 | 3/2 | 380, 665, 800 | 380, 410, 523, 540–554, 657, 725, 840, 978, 1529 |
| $Tm^{3+}$ | $[Xe]4f^{12}$ | 88 | 1 | 361, 467, 792 | 453, 680, 780, 1200, 1611, 1666, 1820, 1856 |
| $Yb^{3+}$ | $[Xe]4f^{13}$ | 86.8 | 1/2 | 980 | 1035 |
| $Lu^{3+}$ | $[Xe]4f^{14}$ | 86.1 | 0 | None | None |

Table 2. Peak assignments of Ln-citrate complexes

| Type of Ln-citrate | Raman shift (cm$^{-1}$) | | | | | | |
|---|---|---|---|---|---|---|---|
| | 690*, 700 | 740 | 820 | 935 | 1060 | 1315 | 1485 |
| Tb-citrate | $\delta(COO^-)$ | $\delta(COO^-)$, $\gamma(COO^-)$ | $\nu(CCCC\text{-}O)$ | $\nu(C\text{–}COO^-)$, $\delta(CH_2)$ | $\nu(C\text{-}O\cdots Tb)$, $\gamma(CH_2)$ | $\nu_{sym}(COO^-)$, $\delta(CH_2)$ | $\nu_{asym}(COO^-)$, $\gamma(CH_2)$ |
| Dy-citrate | $\delta(COO^-)$ | $\delta(COO^-)$, $\gamma(COO^-)$ | $\nu(CCCC\text{-}O)$ | $\nu(C\text{–}COO^-)$, $\delta(CH_2)$ | $\nu(C\text{-}O\cdots Dy)$, $\gamma(CH_2)$ | $\nu_{sym}(COO^-)$, $\delta(CH_2)$ | $\nu_{asym}(COO^-)$, $\gamma(CH_2)$ |
| Ho-citrate | $\delta(COO^-)$ | $\delta(COO^-)$, $\gamma(COO^-)$ | $\nu(CCCC\text{-}O)$ | $\nu(C\text{–}COO^-)$, $\delta(CH_2)$ | $\nu(C\text{-}O\cdots Ho)$, $\gamma(CH_2)$ | $\nu_{sym}(COO^-)$, $\delta(CH_2)$ | $\nu_{asym}(COO^-)$, $\gamma(CH_2)$ |
| Er-citrate | $\delta(COO^-)$ | $\delta(COO^-)$, $\gamma(COO^-)$ | $\nu(CCCC\text{-}O)$ | $\nu(C\text{–}COO^-)$, $\delta(CH_2)$ | $\nu(C\text{-}O\cdots Er)$, $\gamma(CH_2)$ | $\nu_{sym}(COO^-)$, $\delta(CH_2)$ | $\nu_{asym}(COO^-)$, $\gamma(CH_2)$ |
| Tm-citrate | $\delta(COO^-)$ | $\delta(COO^-)$, $\gamma(COO^-)$ | $\nu(CCCC\text{-}O)$ | $\nu(C\text{–}COO^-)$, $\delta(CH_2)$ | $\nu(C\text{-}O\cdots Tm)$, $\gamma(CH_2)$ | $\nu_{sym}(COO^-)$, $\delta(CH_2)$ | $\nu_{asym}(COO^-)$, $\gamma(CH_2)$ |
| Yb-citrate | $\delta(COO^-)$ | $\delta(COO^-)$, $\gamma(COO^-)$ | $\nu(CCCC\text{-}O)$ | $\nu(C\text{–}COO^-)$, $\delta(CH_2)$ | $\nu(C\text{-}O\cdots Yb)$, $\gamma(CH_2)$ | $\nu_{sym}(COO^-)$, $\delta(CH_2)$ | $\nu_{asym}(COO^-)$, $\gamma(CH_2)$ |
| Lu-citrate | $\delta(COO^-)$ | $\delta(COO^-)$, $\gamma(COO^-)$ | $\nu(CCCC\text{-}O)$ | $\nu(C\text{–}COO^-)$, $\delta(CH_2)$ | $\nu(C\text{-}O\cdots Lu)$, $\gamma(CH_2)$ | $\nu_{sym}(COO^-)$, $\delta(CH_2)$ | $\nu_{asym}(COO^-)$, $\gamma(CH_2)$ |

Note: $\nu$ indicates stretching, $\nu_{sym}$ is symmetric stretching, $\nu_{asym}$ is asymmetric stretching, $\delta$ is in-plane bending and rocking, and $\gamma$ is out-of-plane wagging and twisting. The calculated SERS frequencies were scaled by using scaling factors. * Tb-citrate's peak.

# Supporting information

**Surface-enhanced Raman scattering and density functional theory study of selected-lanthanide-citrate complexes (lanthanide: Tb, Dy, Ho, Er, Tm, Yb and Lu)**

Hao Jin[1],* and Yuko S. Yamamoto[1],*

[1] Graduate School of Advanced Science and Technology, Japan Advanced Institute of Science and Technology (JAIST), Nomi, Ishikawa 923-1292, Japan

*Corresponding author: chnliefeng@gmail.com; yamayulab@gmail.com

# Contents

## 1.   Experimental SERS spectra of Gd-citrate

The Gd-citrate SERS sample was used to test and optimize the experimental parameters. On the AgNP surface, citrate molecules coordinate with $Ln^{3+}$ ions to form Ln-citrate complexes. Because this adsorption process can vary depending on the synthesis batch of citrate-capped AgNPs, it was necessary to first identify experimental conditions that could reproduce the Gd-citrate SERS spectrum observed in our previous study[1, 2] before carrying out the formal measurements. Through repeated tests using the Gd-citrate sample, we found that the previously reported Gd-citrate SERS spectrum could be reproduced when the mixed SERS solution was loaded into a capillary and then kept at room temperature for 0.5-1 h before measurement. Therefore, these conditions were adopted as the standard experimental parameters in this work. The SERS spectrum of Gd-citrate measured under 532 nm excitation is shown in Figure S1.

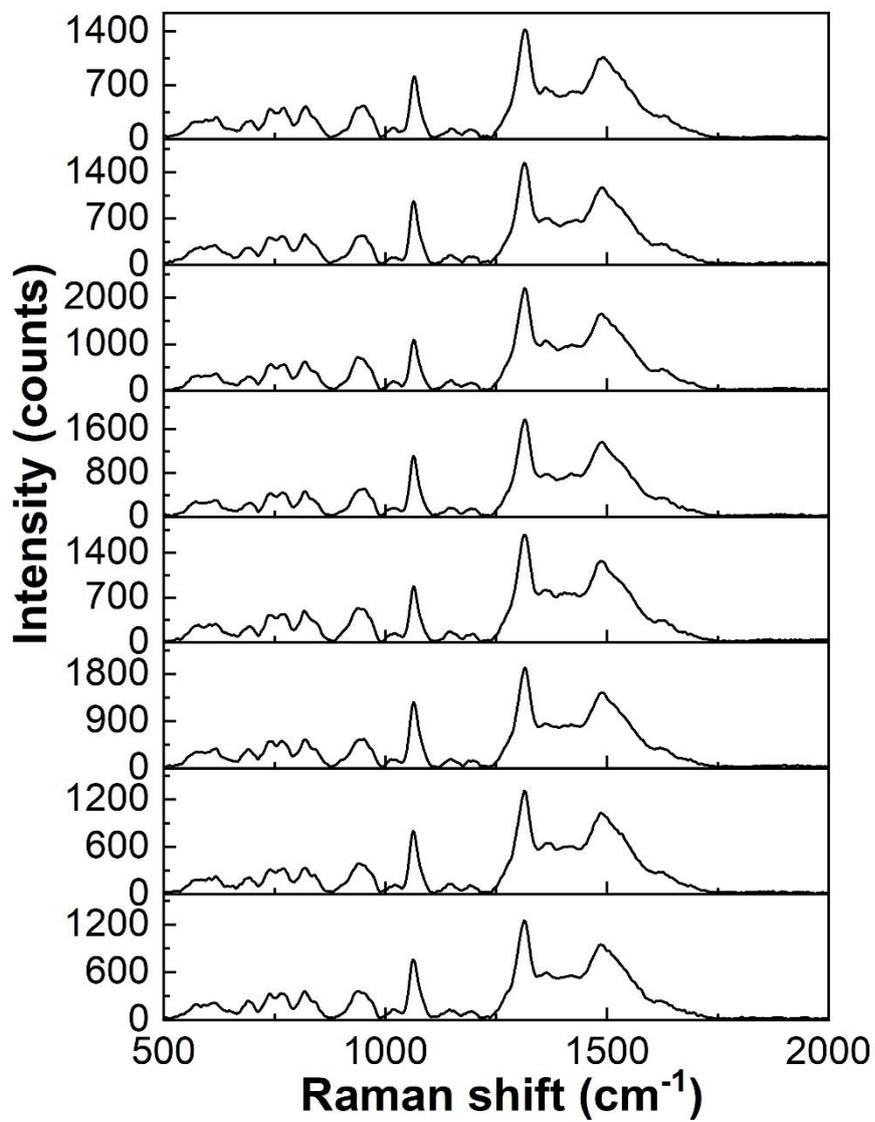

**Figure S1.** SERS spectra of Gd-citrate excited at 532 nm.

## 2. Effect of adsorption modes on the simulated SERS spectra

Figures S2-S8 compare the experimental SERS spectra with the simulated SERS spectra of Ln-citrate-Ag$_{11}$ cluster (Ln: Tb, Dy, Ho, Er, Tm, Yb, and Lu) models in five different adsorption modes. The simulated SERS spectra were used to support peak assignment of the experimental SERS spectra. Although some differences in spectral shape were observed in certain characteristic SERS peak regions, likely because the mechanisms of simulated and experimental SERS are different [3-5], analysis of the vibrational modes (Tables S1-S7) showed that the assigned vibrational modes were consistent among the different adsorption models. The main difference among the simulated SERS spectra was in relative peak intensity rather than vibrational assignment.

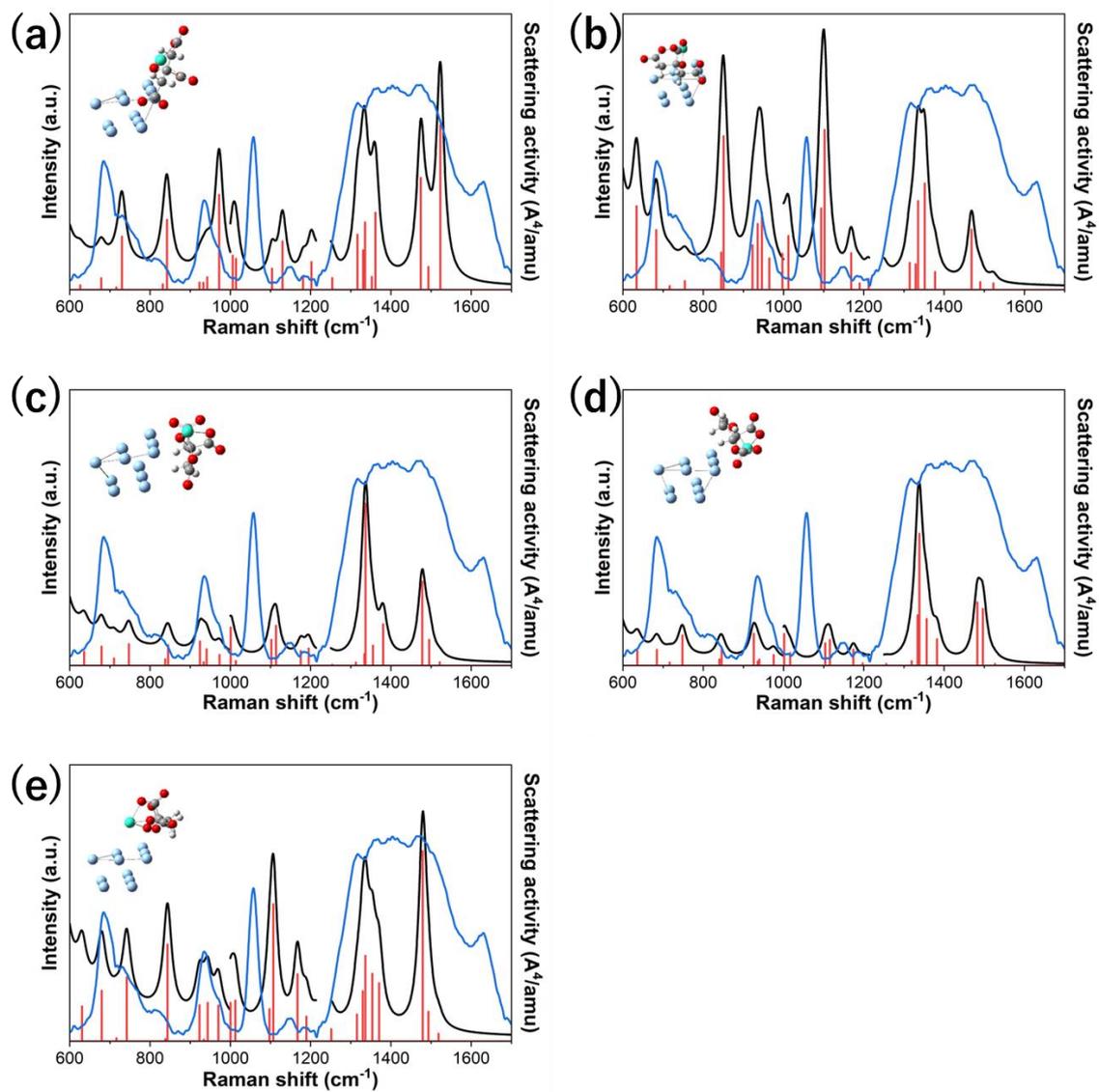

**Figure S2.** (a)-(e) Comparison of the simulated SERS spectra (black), assigned vibrational modes (red), and the experimental SERS spectrum measured under 532 nm excitation (blue) of Tb-citrate for adsorption modes 1-5, respectively.

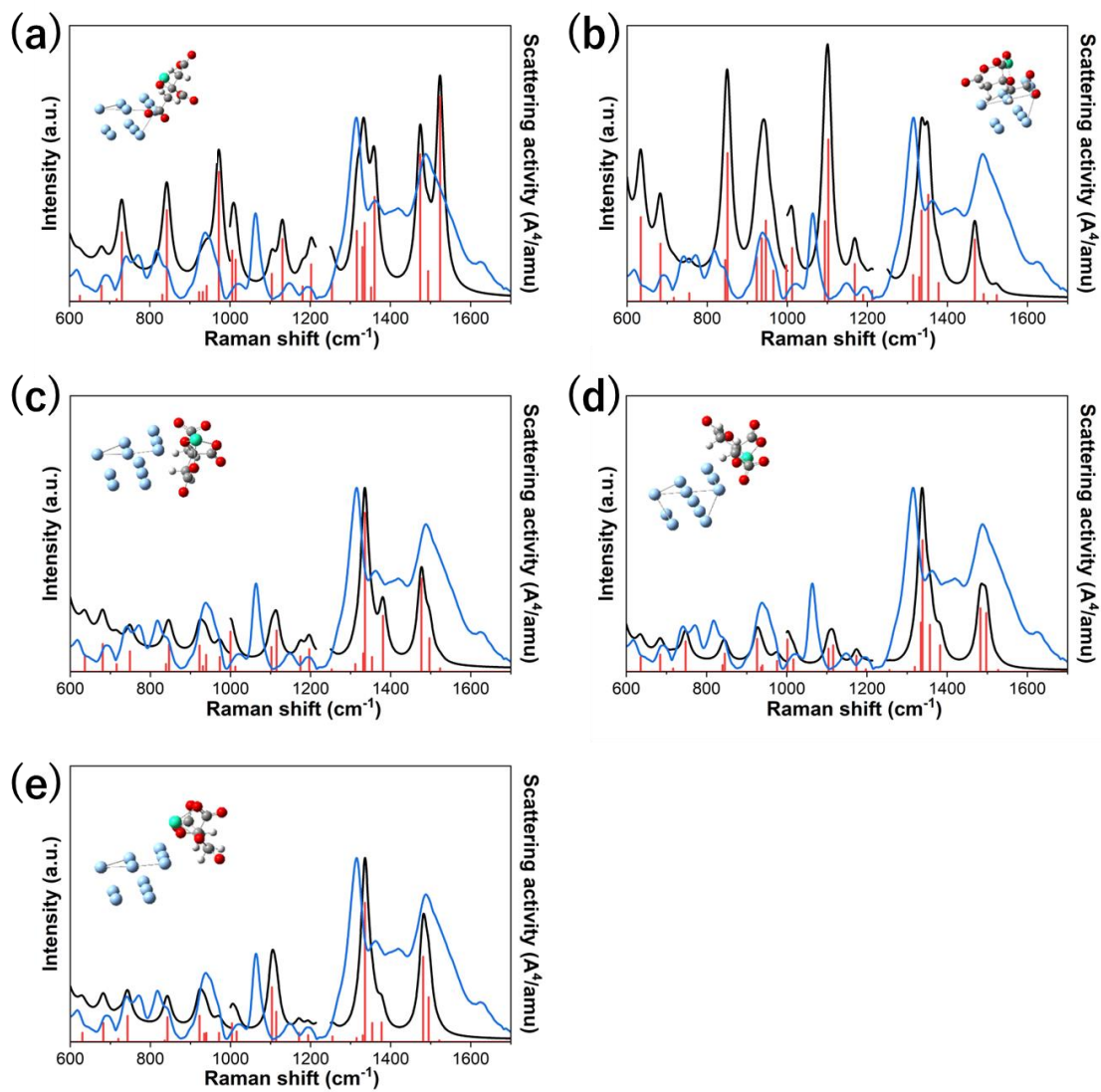

**Figure S3.** (a)-(e) Comparison of the simulated SERS spectra (black), assigned vibrational modes (red), and the experimental SERS spectrum measured under 532 nm excitation (blue) of Dy-citrate for adsorption modes 1-5, respectively.

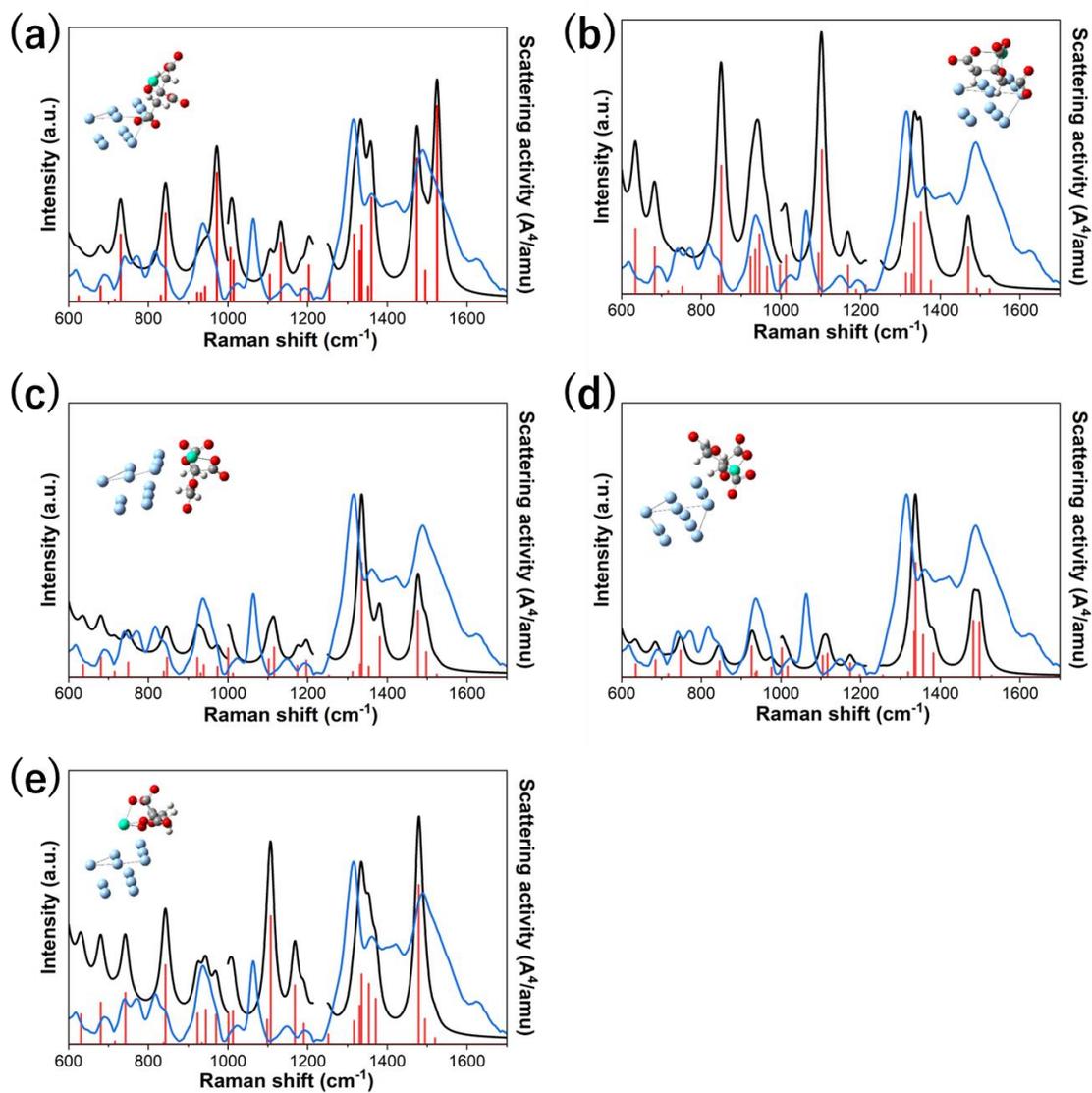

**Figure S4.** (a)-(e) Comparison of the simulated SERS spectra (black), assigned vibrational modes (red), and the experimental SERS spectrum measured under 532 nm excitation (blue) of Ho-citrate for adsorption modes 1-5, respectively.

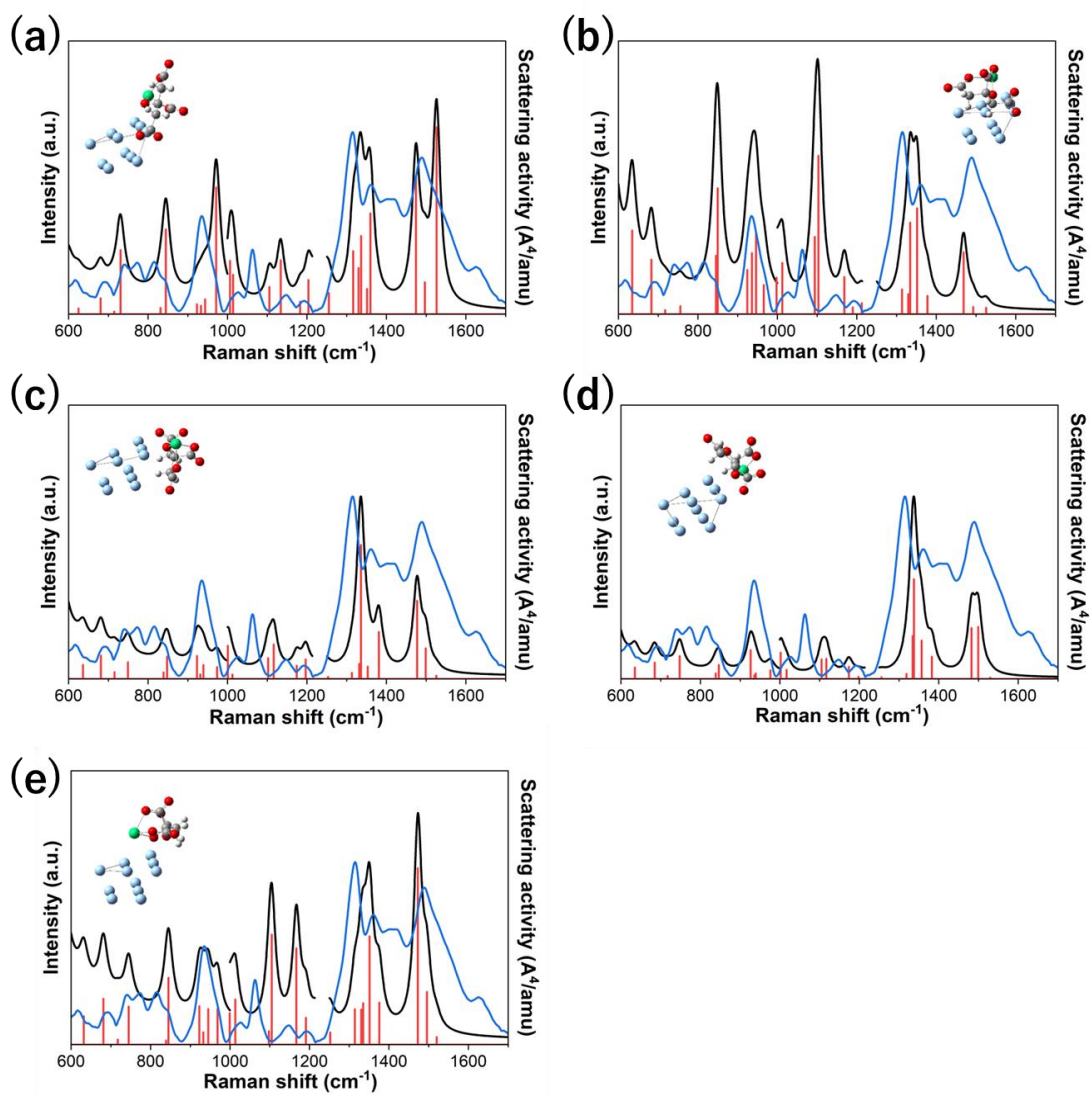

**Figure S5.** (a)-(e) Comparison of the simulated SERS spectra (black), assigned vibrational modes (red), and the experimental SERS spectrum measured under 532 nm excitation (blue) of Er-citrate for adsorption modes 1-5, respectively.

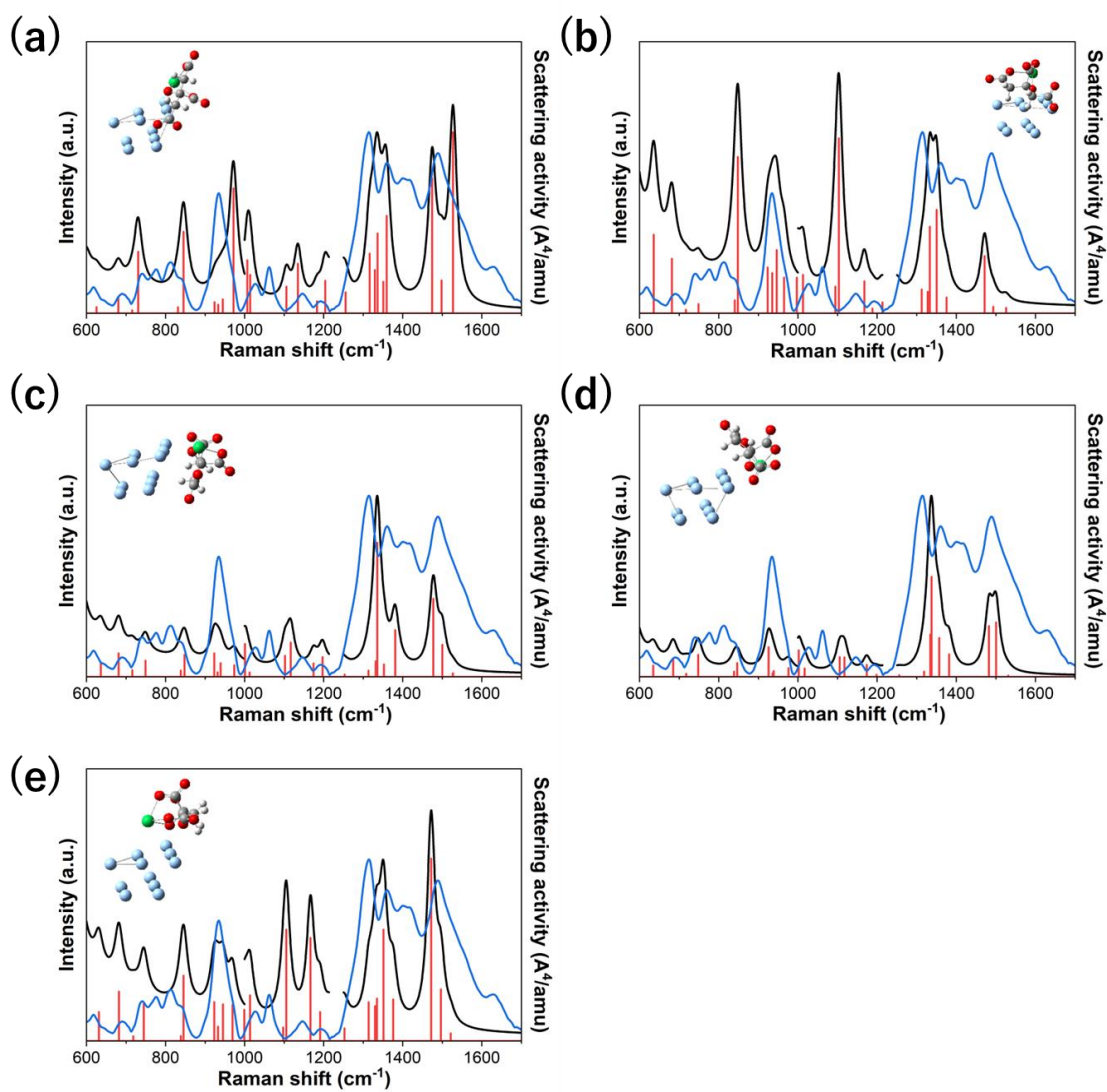

**Figure S6.** (a)-(e) Comparison of the simulated SERS spectra (black), assigned vibrational modes (red), and the experimental SERS spectrum measured under 532 nm excitation (blue) of Tm-citrate for adsorption modes 1-5, respectively.

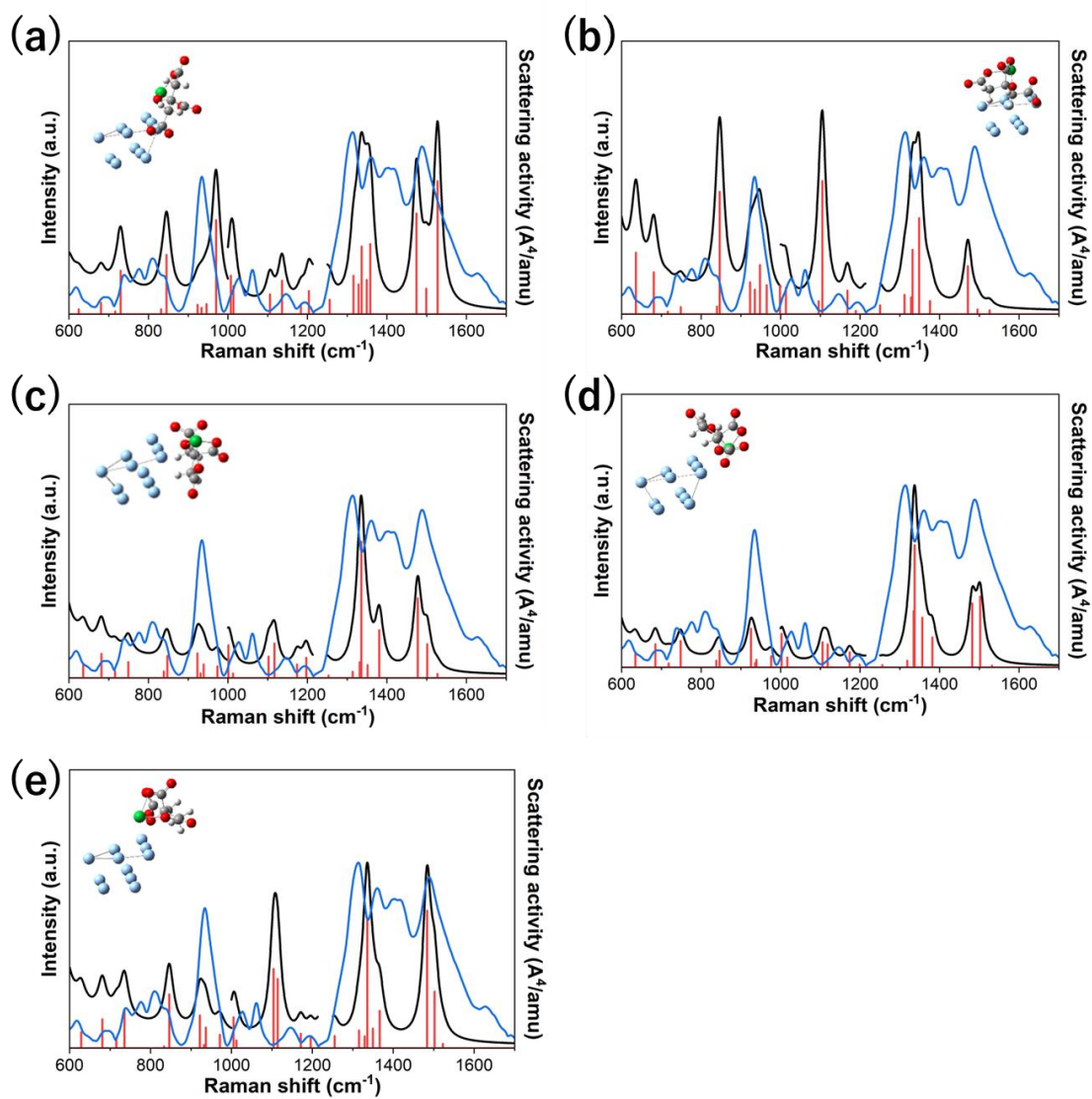

**Figure S7.** (a)-(e) Comparison of the simulated SERS spectra (black), assigned vibrational modes (red), and the experimental SERS spectrum measured under 532 nm excitation (blue) of Yb-citrate for adsorption modes 1-5, respectively.

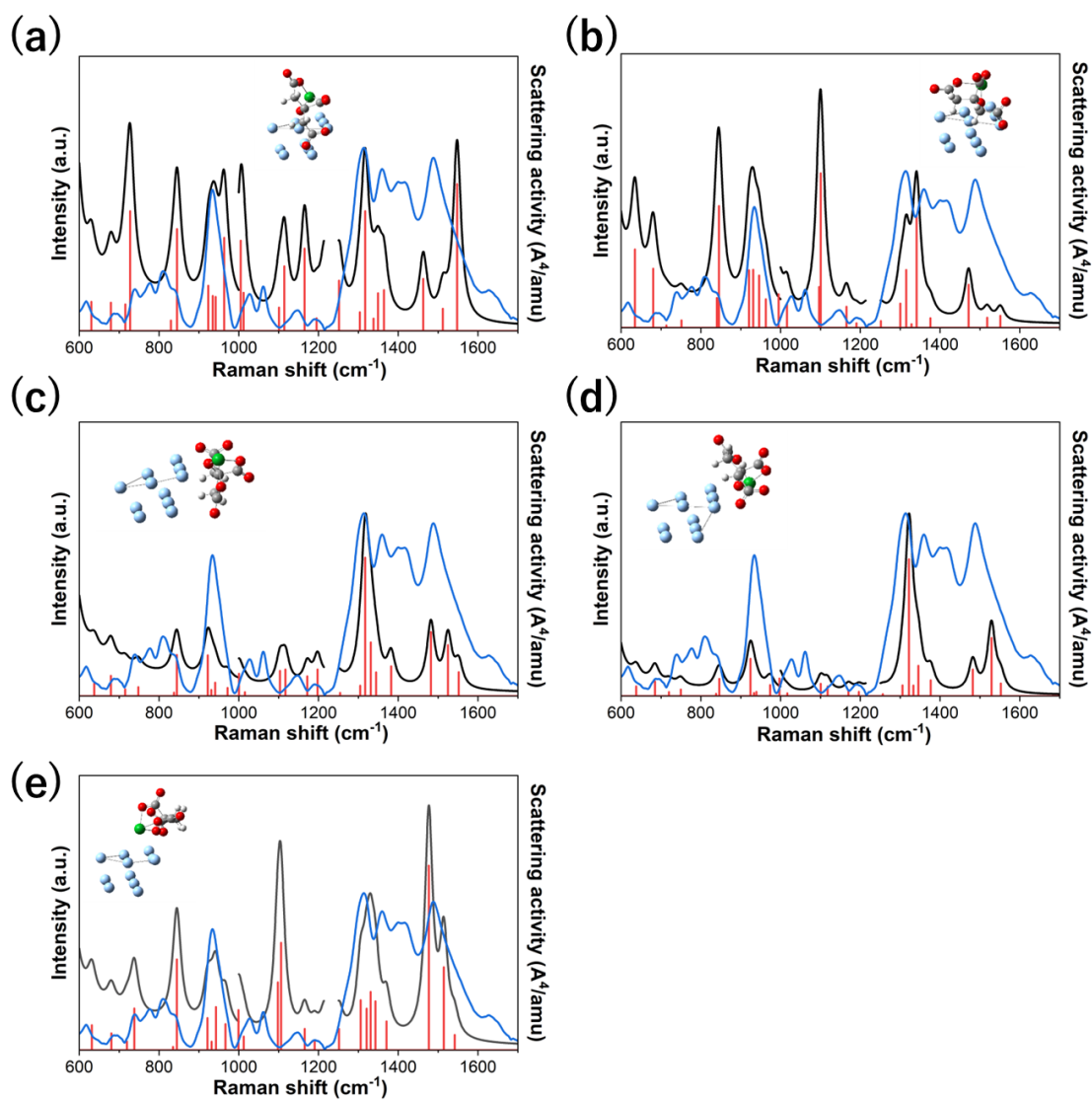

**Figure S8.** (a)-(e) Comparison of the simulated SERS spectra (black), assigned vibrational modes (red), and the experimental SERS spectrum measured under 532 nm excitation (blue) of Lu-citrate for adsorption modes 1-5, respectively.

### 3. Experimental SERS spectra of Ln-citrate

Figures S9-S15 show the experimental SERS spectral data sets of the different Ln-citrate (Ln: Tb, Dy, Ho, Er, Tm, Yb, and Lu) complexes under 488 and 532 nm excitation, together with the corresponding normalized spectral data sets. For normalization, the intensity of the band near 1315 $cm^{-1}$ was set to 1.

Baseline correction of the raw spectra was performed using the interpolation mode. Noise reduction was then applied twice using the Savitzky-Golay method with an 11-point window and a polynomial order of 1.

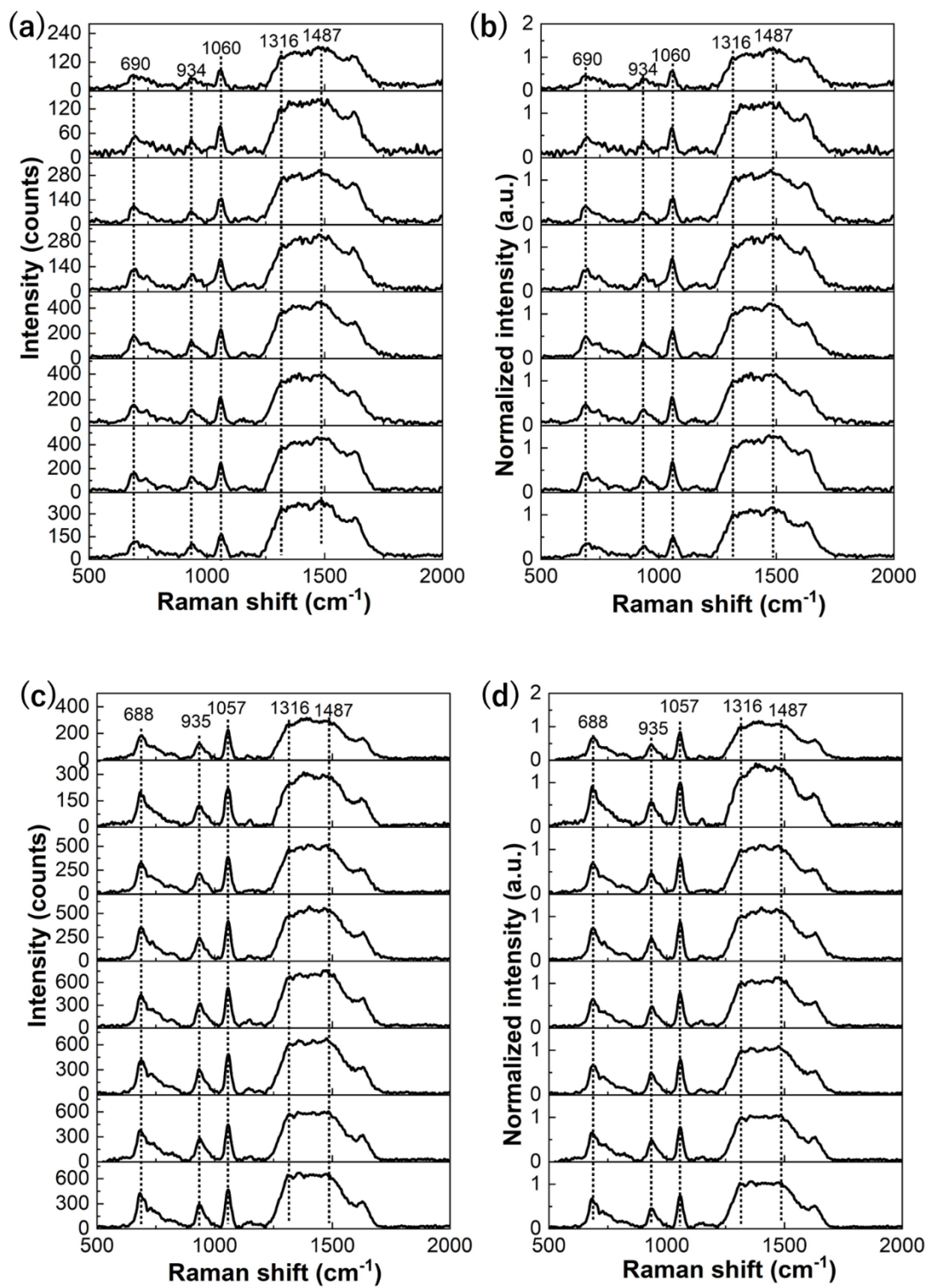

**Figure S9**. (a) and (b): (a)SERS spectra of Tb-citrate complexes under 488 nm

excitation and (b) spectra normalized to the 1316 cm$^{-1}$ peak.(c) and (d): (c)SERS spectra under 532 nm excitation and (d) spectra normalized to the 1316 cm$^{-1}$ peak.

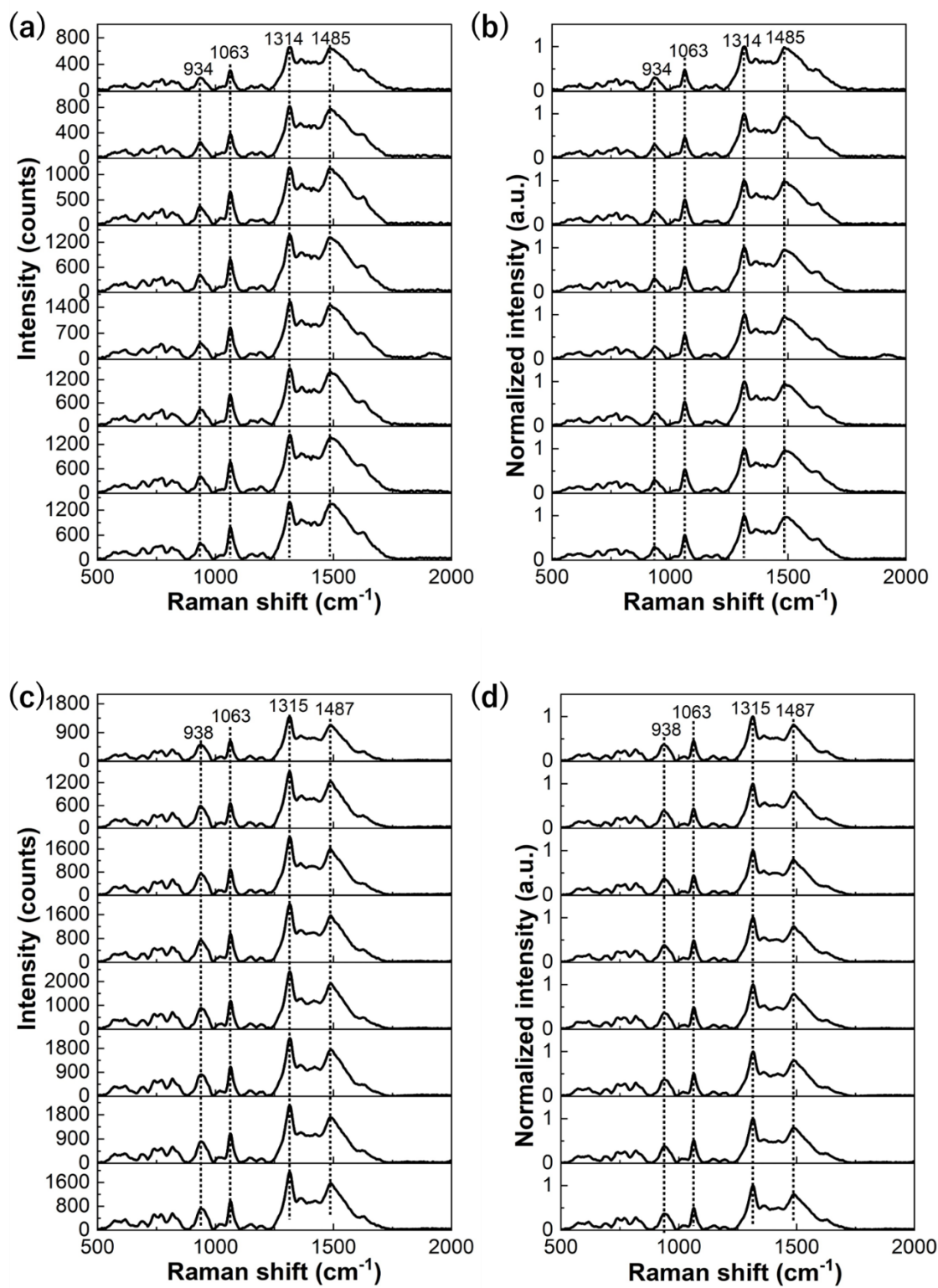

**Figure S10**. (a) and (b): (a)SERS spectra of Dy-citrate complexes under 488 nm

excitation and (b) spectra normalized to the 1314 cm$^{-1}$ peak.(c) and (d): (c)SERS spectra under 532 nm excitation and (d) spectra normalized to the 1315 cm$^{-1}$ peak.

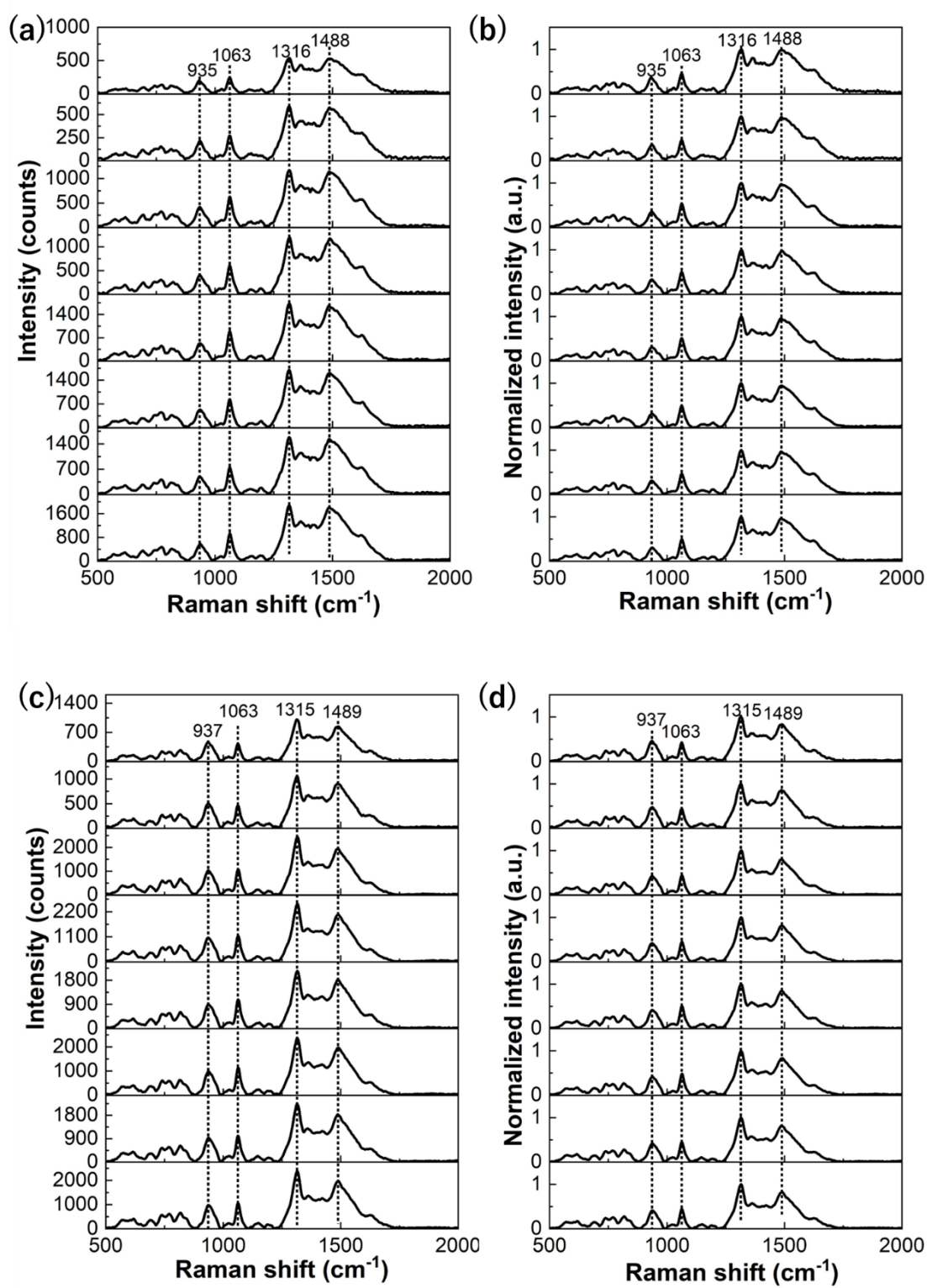

**Figure S11.** (a) and (b): (a)SERS spectra of Ho-citrate complexes under 488 nm

excitation and (b) spectra normalized to the 1316 cm$^{-1}$ peak.(c) and (d): (c)SERS spectra under 532 nm excitation and (d) spectra normalized to the 1315 cm$^{-1}$ peak.

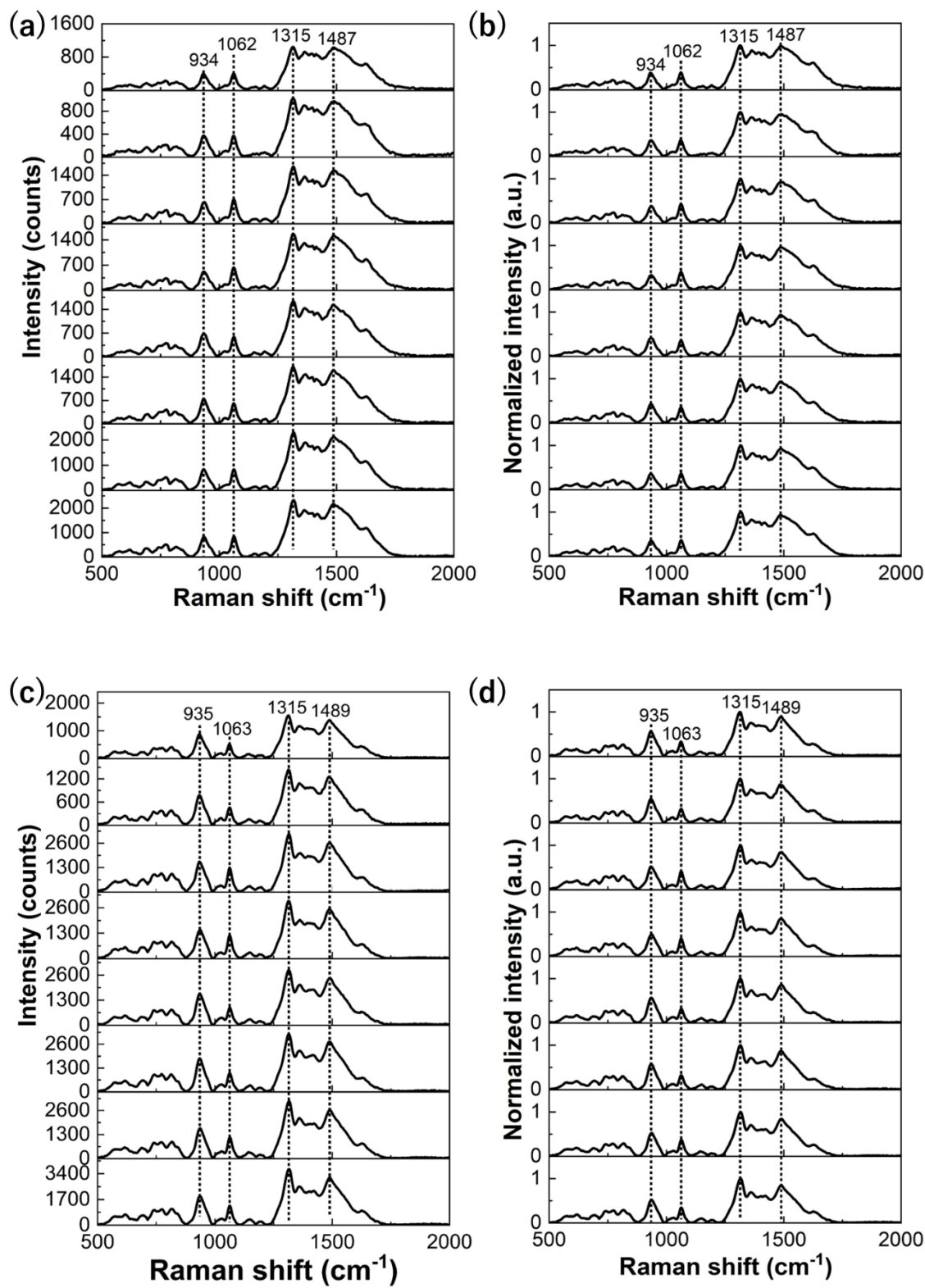

**Figure S12**. (a) and (b): (a)SERS spectra of Er-citrate complexes under 488 nm

excitation and (b) spectra normalized to the 1315 cm⁻¹ peak.(c) and (d): (c)SERS spectra under 532 nm excitation and (d) spectra normalized to the 1315 cm⁻¹ peak.

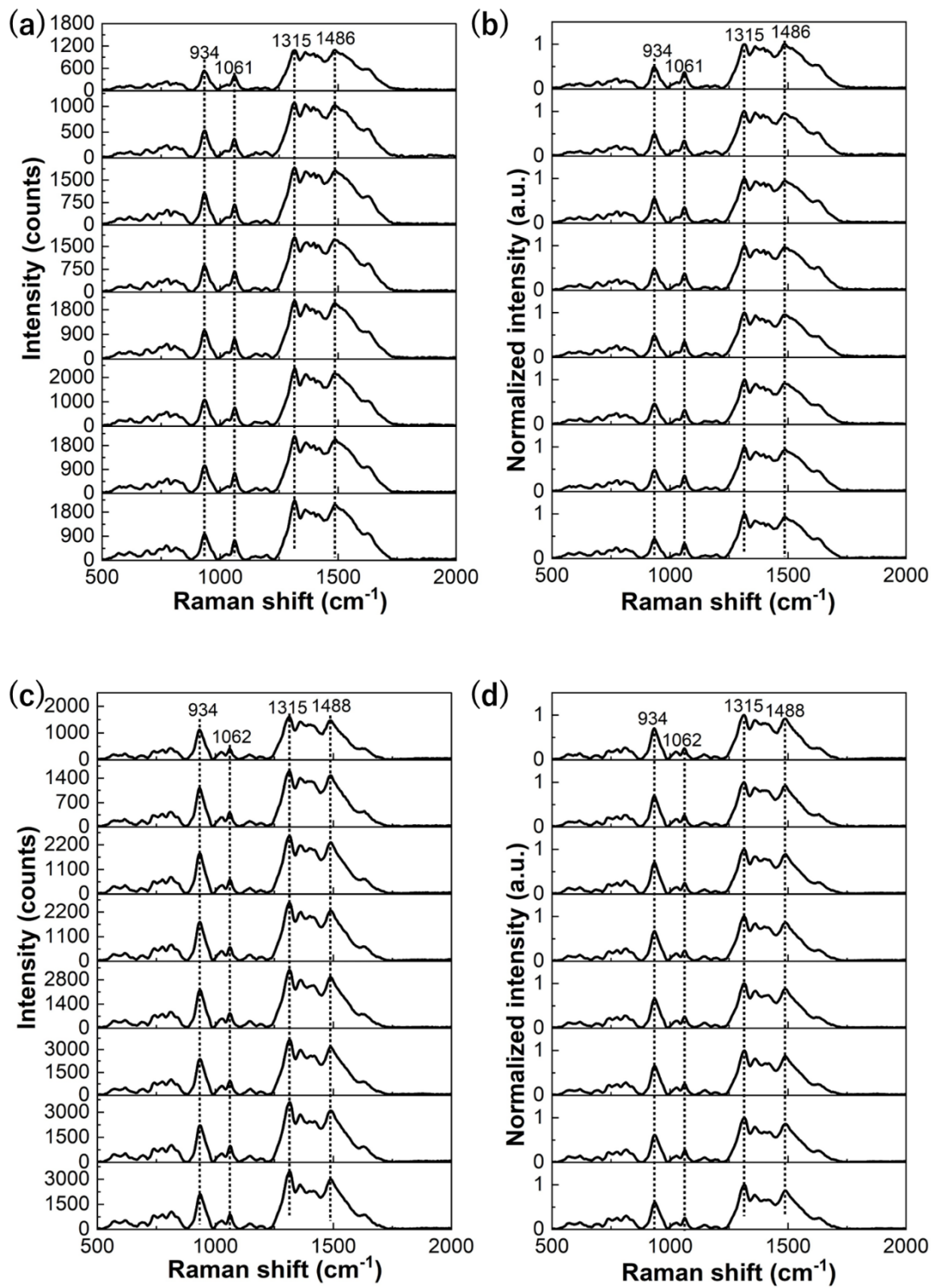

**Figure S13**. (a) and (b): (a)SERS spectra of Tm-citrate complexes under 488 nm excitation and (b) spectra normalized to the 1315 cm$^{-1}$ peak.(c) and (d): (c)SERS spectra under 532 nm excitation and (d) spectra normalized to the 1315 cm$^{-1}$ peak.

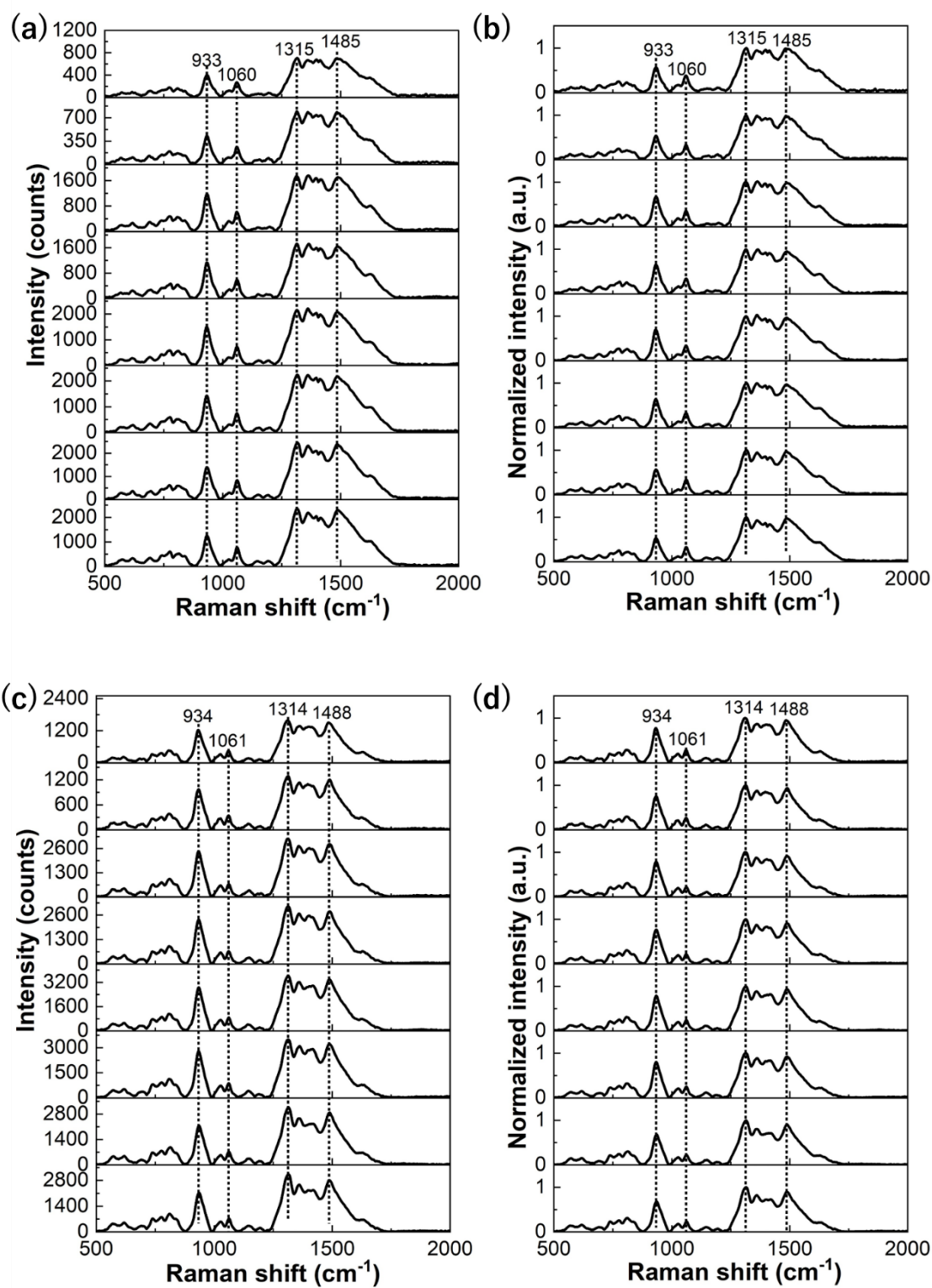

**Figure S14**. (a) and (b): (a)SERS spectra of Yb-citrate complexes under 488 nm

excitation and (b) spectra normalized to the 1315 cm$^{-1}$ peak.(c) and (d): (c)SERS spectra under 532 nm excitation and (d) spectra normalized to the 1314 cm$^{-1}$ peak.

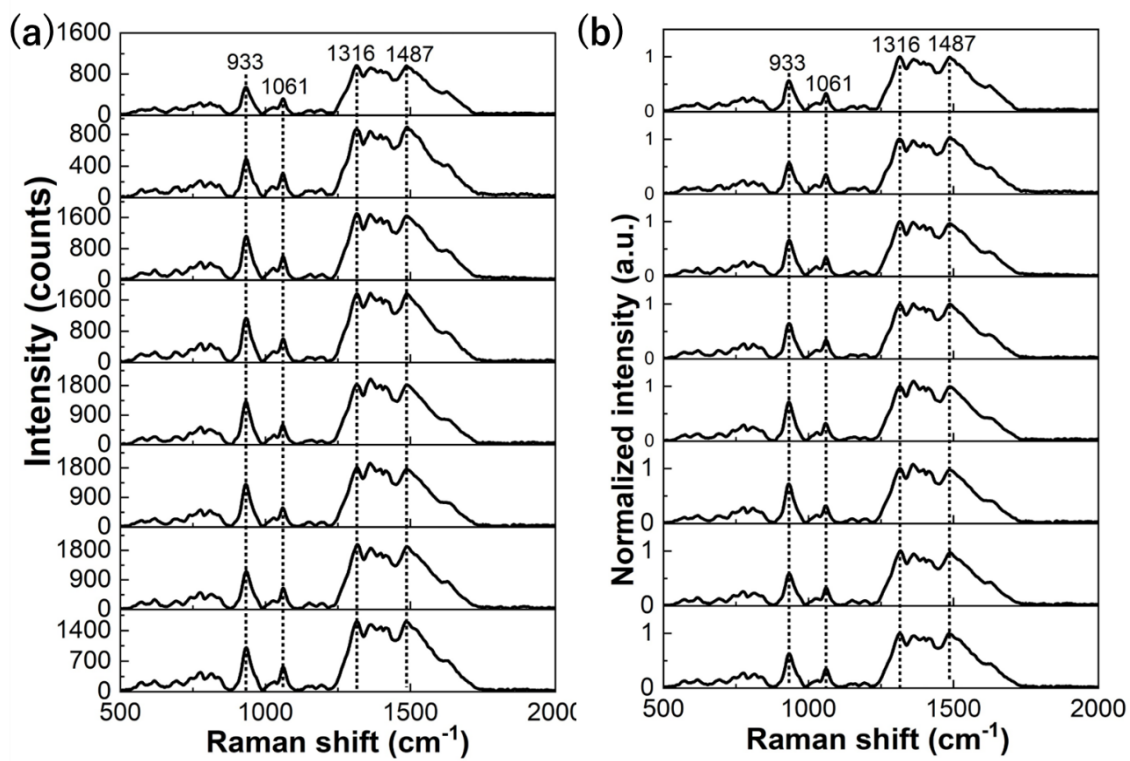

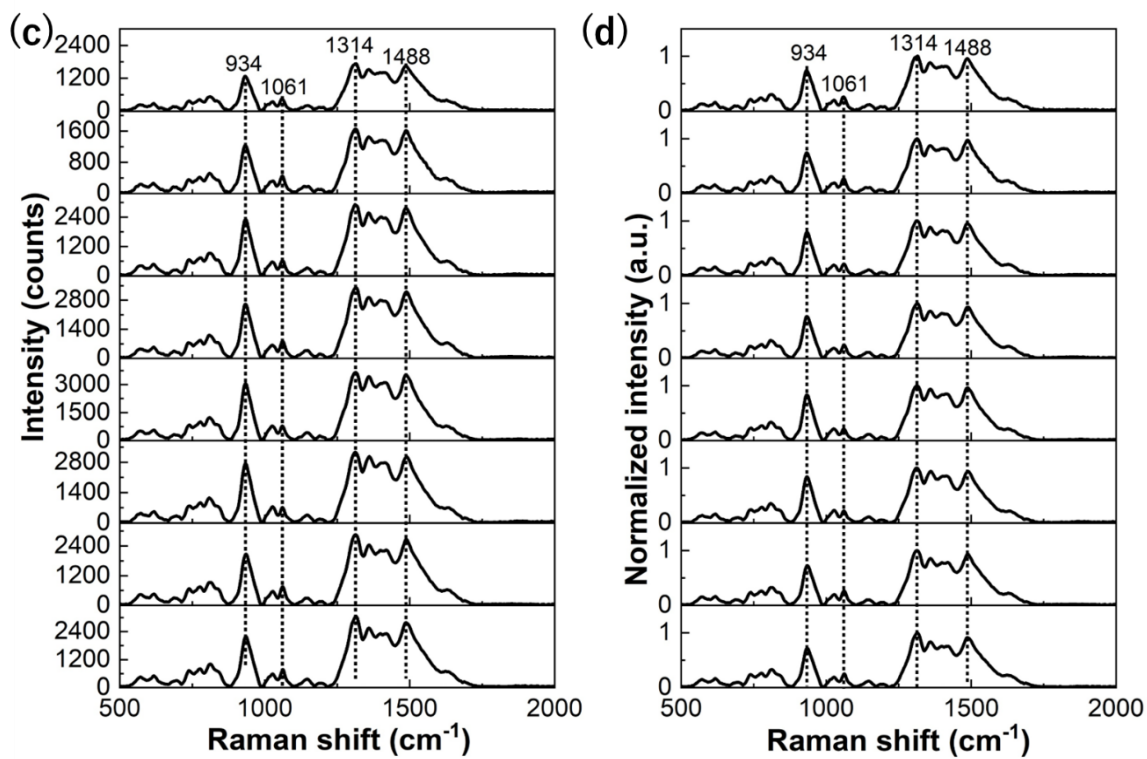

**Figure S15**. (a) and (b): (a)SERS spectra of Lu-citrate complexes under 488 nm excitation and (b) spectra normalized to the 1316 cm$^{-1}$ peak.(c) and (d): (c)SERS spectra under 532 nm excitation and (d) spectra normalized to the 1314 cm$^{-1}$ peak.

4. Peak assignments for SERS spectra of Ln-citrate (Ln: Tb, Dy, Ho, Er, Tm, Yb, and Lu) by DFT calculations

We also analyzed the simulated SERS spectra (mode 3) of selected Ln-citrate (Ln: Tb, Dy, Ho, Er, Tm, Yb, and Lu) to assign the SERS peaks based on each vibrational mode. All tables used the same symbols for peak assignment: $\nu$ indicates stretching, $\nu_{sym}$ is symmetric stretching, $\nu_{asym}$ is asymmetric stretching, $\delta$ is in-plane bending and rocking, and $\gamma$ is out-of-plane wagging and twisting. The calculated SERS frequencies were scaled by using scaling factors.

**Table S1**. The vibrational modes contained in each simulated SERS peak of Tb-citrate

| Simulated SERS peak (cm$^{-1}$) | Vibrational mode (number) | Frequency (cm$^{-1}$) | Assignment |
|---|---|---|---|
| 636 | 32 | 636 | $\delta(COO^-)$ |
| 679 | 33 | 679 | $\delta(COO^-)$ |
| 710 | 34 | 710 | $\delta(COO^-)+$ $\gamma(COO^-)$ |
| 747 | 35 | 747 | $\delta(COO^-)+$ $\gamma(COO^-)$ |
| 843 | 36 | 838 | $\nu(CCCC-O)$ |
| | 37 | 845 | $\nu(CCCC-O)$ |
| 928, 939 | 38 | 924 | $\nu(C-COO^-)$ |
| | 39 | 934 | $\delta(CH_2)$ |
| | 40 | 941 | $\delta(CH_2)$ |
| 973 | 41 | 973 | $\nu(C-COO^-)$ |
| 1001 | 42 | 1001 | $\gamma(CH_2)$ |
| | 43 | 1013 | $\gamma(CH_2)$ |
| 1111 | 44 | 1102 | $\gamma(CH_2)$ |
| | 45 | 1114 | $\nu(C-O\cdots Tb)$ |
| 1175, 1195 | 46 | 1175 | $\gamma(CH_2)$ |

| | 47 | 1195 | $\gamma(CH_2)$ |
|---|---|---|---|
| 1254 | 48 | 1254 | $\gamma(CH_2)$ |
| 1338, 1378 | 49 | 1313 | $\nu_{sym}(COO^-)$, $\delta(CH_2)$ |
| | 50 | 1334 | $\nu_{sym}(COO^-)$, $\delta(CH_2)$ |
| | 51 | 1337 | $\nu_{sym}(COO^-)$, $\delta(CH_2)$ |
| | 52 | 1355 | $\nu_{sym}(COO^-)$, $\delta(CH_2)$ |
| | 53 | 1381 | $\nu_{sym}(COO^-)$, $\delta(CH_2)$ |
| 1477 | 54 | 1478 | $\nu_{asym}(COO^-)$, $\gamma(CH_2)$ |
| | 55 | 1495 | $\nu_{asym}(COO^-)$, $\gamma(CH_2)$ |
| | 56 | 1522 | $\nu_{asym}(COO^-)$, $\gamma(CH_2)$ |

**Table S2**. The vibrational modes contained in each simulated SERS peak of Dy-citrate

| Simulated SERS peak ($cm^{-1}$) | Vibrational mode (number) | Frequency ($cm^{-1}$) | Assignment |
|---|---|---|---|
| 636 | 32 | 636 | $\delta(COO^-)$ |
| 681 | 33 | 681 | $\delta(COO^-)$ |
| 715 | 34 | 715 | $\delta(COO^-)+$ $\gamma(COO^-)$ |
| 749 | 35 | 749 | $\delta(COO^-)+$ $\gamma(COO^-)$ |
| 845 | 36 | 839 | $\nu(CCCC\text{-}O)$ |
| | 37 | 846 | $\nu(CCCC\text{-}O)$ |
| 923, 939 | 38 | 923 | $\nu(C\text{-}COO^-)$ |
| | 39 | 931 | $\delta(CH_2)$ |
| | 40 | 939 | $\delta(CH_2)$ |
| 973 | 41 | 973 | $\nu(C\text{-}COO^-)$ |
| 1001 | 42 | 1000 | $\gamma(CH_2)$ |
| | 43 | 1012 | $\gamma(CH_2)$ |
| 1113 | 44 | 1102 | $\gamma(CH_2)$ |
| | 45 | 1115 | $\nu(C\text{-}O\cdots Dy)$ |
| 1174, 1197 | 46 | 1174 | $\gamma(CH_2)$ |
| | 47 | 1197 | $\gamma(CH_2)$ |
| 1253 | 48 | 1253 | $\gamma(CH_2)$ |
| 1336, 1379 | 49 | 1312 | $\nu_{sym}(COO^-)$, $\delta(CH_2)$ |
| | 50 | 1331 | $\nu_{sym}(COO^-)$, $\delta(CH_2)$ |
| | 51 | 1336 | $\nu_{sym}(COO^-)$, $\delta(CH_2)$ |
| | 52 | 1354 | $\nu_{sym}(COO^-)$, $\delta(CH_2)$ |
| | 53 | 1381 | $\nu_{sym}(COO^-)$, $\delta(CH_2)$ |
| 1478 | 54 | 1477 | $\nu_{asym}(COO^-)$, $\gamma(CH_2)$ |

| | 55 | 1497 | $\nu_{asym}(COO^-)$, $\gamma(CH_2)$ |
|---|---|---|---|
| | 56 | 1523 | $\nu_{asym}(COO^-)$, $\gamma(CH_2)$ |

**Table S3**. The vibrational modes contained in each simulated SERS peak of Ho-citrate

| Simulated SERS peak (cm$^{-1}$) | Vibrational mode (number) | Frequency (cm$^{-1}$) | Assignment |
|---|---|---|---|
| 636 | 32 | 636 | $\delta(COO^-)$ |
| 681 | 33 | 681 | $\delta(COO^-)$ |
| 715 | 34 | 715 | $\delta(COO^-)+\gamma(COO^-)$ |
| 749 | 35 | 749 | $\delta(COO^-)+\gamma(COO^-)$ |
| 845 | 36 | 839 | $\nu(CCCC-O)$ |
| | 37 | 847 | $\nu(CCCC-O)$ |
| 923, 939 | 38 | 923 | $\nu(C-COO^-)$ |
| | 39 | 931 | $\delta(CH_2)$ |
| | 40 | 939 | $\delta(CH_2)$ |
| 974 | 41 | 974 | $\nu(C-COO^-)$ |
| 1001 | 42 | 1000 | $\gamma(CH_2)$ |
| | 43 | 1012 | $\gamma(CH_2)$ |
| 1113 | 44 | 1102 | $\gamma(CH_2)$ |
| | 45 | 1116 | $\nu(C-O\cdots Ho)$ |
| 1174, 1197 | 46 | 1174 | $\gamma(CH_2)$ |
| | 47 | 1197 | $\gamma(CH_2)$ |

| 1253 | 48 | 1253 | $\gamma(CH_2)$ |
|---|---|---|---|
| 1336 | 49 | 1313 | $\nu_{sym}(COO^-)$, $\delta(CH_2)$ |
| | 50 | 1332 | $\nu_{sym}(COO^-)$, $\delta(CH_2)$ |
| | 51 | 1336 | $\nu_{sym}(COO^-)$, $\delta(CH_2)$ |
| | 52 | 1353 | $\nu_{sym}(COO^-)$, $\delta(CH_2)$ |
| 1379 | 53 | 1381 | $\nu_{sym}(COO^-)$, $\delta(CH_2)$ |
| 1478 | 54 | 1477 | $\nu_{asym}(COO^-)$, $\gamma(CH_2)$ |
| | 55 | 1497 | $\nu_{asym}(COO^-)$, $\gamma(CH_2)$ |
| | 56 | 1524 | $\nu_{asym}(COO^-)$, $\gamma(CH_2)$ |

**Table S4**. The vibrational modes contained in each simulated SERS peak of Er-citrate

| Simulated SERS peak (cm$^{-1}$) | Vibrational mode (number) | Frequency (cm$^{-1}$) | Assignment |
|---|---|---|---|
| 636 | 32 | 636 | $\delta(COO^-)$ |
| 681 | 33 | 681 | $\delta(COO^-)$ |
| 715 | 34 | 715 | $\delta(COO^-)+$ $\gamma(COO^-)$ |
| 749 | 35 | 749 | $\delta(COO^-)+$ $\gamma(COO^-)$ |
| 847 | 36 | 839 | $\nu$(CCCC-O) |
| | 37 | 848 | $\nu$(CCCC-O) |
| 923, 939 | 38 | 923 | $\nu$(C-COO$^-$) |
| | 39 | 931 | $\delta$(CH$_2$) |
| | 40 | 939 | $\delta$(CH$_2$) |
| 974 | 41 | 974 | $\nu$(C-COO$^-$) |
| 1001 | 42 | 1001 | $\gamma$(CH$_2$) |
| | 43 | 1012 | $\gamma$(CH$_2$) |
| 1113 | 44 | 1102 | $\gamma$(CH$_2$) |
| | 45 | 1116 | $\nu$(C-O$\cdots$Er) |
| 1174, 1197 | 46 | 1174 | $\gamma$(CH$_2$) |
| | 47 | 1197 | $\gamma$(CH$_2$) |
| 1253 | 48 | 1253 | $\gamma$(CH$_2$) |
| 1336 | 49 | 1313 | $\nu_{sym}$(COO$^-$), $\delta$(CH$_2$) |
| | 50 | 1332 | $\nu_{sym}$(COO$^-$), $\delta$(CH$_2$) |
| | 51 | 1336 | $\nu_{sym}$(COO$^-$), $\delta$(CH$_2$) |
| | 52 | 1353 | $\nu_{sym}$(COO$^-$), $\delta$(CH$_2$) |
| 1379 | 53 | 1381 | $\nu_{sym}$(COO$^-$), $\delta$(CH$_2$) |
| 1478 | 54 | 1477 | $\nu_{asym}$(COO$^-$), $\gamma$(CH$_2$) |
| | 55 | 1499 | $\nu_{asym}$(COO$^-$), $\gamma$(CH$_2$) |

| | 56 | 1525 | $\nu_{asym}(COO^-)$, $\gamma(CH_2)$ |
|---|---|---|---|

Table S5. The vibrational modes contained in each simulated SERS peak of Tm-citrate

| Simulated SERS peak ($cm^{-1}$) | Vibrational mode (number) | Frequency ($cm^{-1}$) | Assignment |
|---|---|---|---|
| 636 | 32 | 636 | $\delta(COO^-)$ |
| 681 | 33 | 681 | $\delta(COO^-)$ |
| 716 | 34 | 716 | $\delta(COO^-)+$ $\gamma(COO^-)$ |
| 749 | 35 | 749 | $\delta(COO^-)+$ $\gamma(COO^-)$ |
| 846 | 36 | 839 | $\nu(CCCC-O)$ |
| | 37 | 848 | $\nu(CCCC-O)$ |
| 923, 939 | 38 | 923 | $\nu(C-COO^-)$ |
| | 39 | 931 | $\delta(CH_2)$ |
| | 40 | 939 | $\delta(CH_2)$ |
| 974 | 41 | 974 | $\nu(C-COO^-)$ |
| 1001 | 42 | 1001 | $\gamma(CH_2)$ |
| | 43 | 1012 | $\gamma(CH_2)$ |
| 1113 | 44 | 1102 | $\gamma(CH_2)$ |
| | 45 | 1116 | $\nu(C-O\cdots Tm)$ |
| 1174, 1197 | 46 | 1174 | $\gamma(CH_2)$ |
| | 47 | 1197 | $\gamma(CH_2)$ |
| 1253 | 48 | 1253 | $\gamma(CH_2)$ |
| 1334, 1379 | 49 | 1313 | $\nu_{sym}(COO^-)$, |

| | | | $\delta(CH_2)$ |
|---|---|---|---|
| | 50 | 1331 | $\nu_{sym}(COO^-)$, $\delta(CH_2)$ |
| | 51 | 1336 | $\nu_{sym}(COO^-)$, $\delta(CH_2)$ |
| | 52 | 1352 | $\nu_{sym}(COO^-)$, $\delta(CH_2)$ |
| | 53 | 1381 | $\nu_{sym}(COO^-)$, $\delta(CH_2)$ |
| 1478, 1500 | 54 | 1477 | $\nu_{asym}(COO^-)$, $\gamma(CH_2)$ |
| | 55 | 1500 | $\nu_{asym}(COO^-)$, $\gamma(CH_2)$ |
| | 56 | 1527 | $\nu_{asym}(COO^-)$, $\gamma(CH_2)$ |

**Table S6**. The vibrational modes contained in each simulated SERS peak of Yb-citrate

| Simulated SERS peak ($cm^{-1}$) | Vibrational mode (number) | Frequency ($cm^{-1}$) | Assignment |
|---|---|---|---|
| 636 | 32 | 636 | $\delta(COO^-)$ |
| 681 | 33 | 681 | $\delta(COO^-)$ |
| 716 | 34 | 716 | $\delta(COO^-)+\gamma(COO^-)$ |
| 749 | 35 | 749 | $\delta(COO^-)+\gamma(COO^-)$ |
| 846 | 36 | 838 | $\nu(CCCC-O)$ |

| | 37 | 847 | $\nu(CCCC-O)$ |
|---|---|---|---|
| 926 | 38 | 923 | $\nu(C-COO^-)$ |
| | 39 | 931 | $\delta(CH_2)$ |
| | 40 | 939 | $\delta(CH_2)$ |
| 973 | 41 | 973 | $\nu(C-COO^-)$ |
| 1001 | 42 | 1001 | $\gamma(CH_2)$ |
| | 43 | 1012 | $\gamma(CH_2)$ |
| 1113 | 44 | 1102 | $\gamma(CH_2)$ |
| | 45 | 1117 | $\nu(C-O\cdots Yb)$ |
| 1174, 1197 | 46 | 1174 | $\gamma(CH_2)$ |
| | 47 | 1197 | $\gamma(CH_2)$ |
| 1252 | 48 | 1252 | $\gamma(CH_2)$ |
| 1334, 1379 | 49 | 1314 | $\nu_{sym}(COO^-)$, $\delta(CH_2)$ |
| | 50 | 1331 | $\nu_{sym}(COO^-)$, $\delta(CH_2)$ |
| | 51 | 1335 | $\nu_{sym}(COO^-)$, $\delta(CH_2)$ |
| | 52 | 1351 | $\nu_{sym}(COO^-)$, $\delta(CH_2)$ |
| | 53 | 1381 | $\nu_{sym}(COO^-)$, $\delta(CH_2)$ |
| 1478, 1501 | 54 | 1477 | $\nu_{asym}(COO^-)$, $\gamma(CH_2)$ |
| | 55 | 1501 | $\nu_{asym}(COO^-)$, $\gamma(CH_2)$ |
| | 56 | 1527 | $\nu_{asym}(COO^-)$, |

| | | | $\gamma(CH_2)$ |
|---|---|---|---|

**Table S7**. The vibrational modes contained in each simulated SERS peak of Lu-citrate

| Simulated SERS peak (cm$^{-1}$) | Vibrational mode (number) | Frequency (cm$^{-1}$) | Assignment |
|---|---|---|---|
| 637 | 32 | 637 | $\delta(COO^-)$ |
| 679 | 33 | 679 | $\delta(COO^-)$ |
| 715 | 34 | 715 | $\delta(COO^-)+$ $\gamma(COO^-)$ |
| 748 | 35 | 748 | $\delta(COO^-)+$ $\gamma(COO^-)$ |
| 843 | 36 | 837 | $\nu(CCCC-O)$ |
| | 37 | 844 | $\nu(CCCC-O)$ |
| 925 | 38 | 922 | $\nu(C-COO^-)$ |
| | 39 | 931 | $\delta(CH_2)$ |
| | 40 | 941 | $\delta(CH_2)$ |
| 972 | 41 | 972 | $\nu(C-COO^-)$ |
| 1001 | 42 | 1000 | $\gamma(CH_2)$ |
| | 43 | 1015 | $\gamma(CH_2)$ |
| 1113 | 44 | 1103 | $\gamma(CH_2)$ |
| | 45 | 1117 | $\nu(C-O\cdots Lu)$ |
| 1172, 1198 | 46 | 1172 | $\gamma(CH_2)$ |
| | 47 | 1198 | $\gamma(CH_2)$ |
| 1254 | 48 | 1254 | $\gamma(CH_2)$ |
| 1318, 1379 | 49 | 1304 | $\nu_{sym}(COO^-),$ $\delta(CH_2)$ |
| | 50 | 1317 | $\nu_{sym}(COO^-)$, $\delta(CH_2)$ |
| | 51 | 1331 | $\nu_{sym}(COO^-)$, $\delta(CH_2)$ |
| | 52 | 1344 | $\nu_{sym}(COO^-)$, $\delta(CH_2)$ |
| | 53 | 1382 | $\nu_{sym}(COO^-)$, $\delta(CH_2)$ |
| 1482, 1524, 1552 | 54 | 1482 | $\nu_{asym}(COO^-)$, $\gamma(CH_2)$ |
| | 55 | 1524 | $\nu_{asym}(COO^-)$, $\gamma(CH_2)$ |
| | 56 | 1552 | $\nu_{asym}(COO^-)$, $\gamma(CH_2)$ |

References

(1) Jin, H.; Itoh, T.; Yamamoto, Y. S. Classification of La3+ and Gd3+ Rare-Earth Ions Using Surface-Enhanced Raman Scattering. The Journal of Physical Chemistry C 2024, 128 (13), 5611–5620.

(2) Hao Jin, Tamitake Itoh, Yuko S Yamamoto; Surface-enhanced Raman scattering and density functional theory study of selected-lanthanide-citrate complexes (lanthanide: La, Ce, Pr, Nd, Sm, Eu, and Gd), submitted to The Journal of Physical Chemistry C

(3)Itoh, T.; Procházka, M.; Dong, Z.-C.; Ji, W.; Yamamoto, Y. S.; Zhang, Y.; Ozaki, Y.

Toward a New Era of SERS and TERS at the Nanometer Scale: From Fundamentals to Innovative Applications. Chem. Rev. 2023, 123 (4), 1552– 1634

(4)Yamamoto, Y. S.; Itoh, T. Why and How Do the Shapes of Surface-Enhanced Raman Scattering Spectra Change? Recent Progress from Mechanistic Studies. Journal of Raman Spectroscopy 2016, 47 (1), 78–88.

(5)Polavarapu, P. L. Ab Initio Vibrational Raman and Raman Optical Activity Spectra. The Journal of Physical Chemistry 1990, 94 (21), 8106–8112.

## Modelling data

Here we add modelling data with structures as follows:

**SERS simulations:**

**Tb-citrate-Ag$_{11}$ (mode 1)**

| Ag | 0  | 1.94984500  | -0.85360300 | -0.79718900 |
| Ag | -1 | 0.42812600  | 1.32210200  | -0.55406200 |
| Ag | -1 | -1.29467200 | 3.63813400  | -0.43315100 |
| Ag | -1 | -0.09507900 | -2.62348700 | -1.47345200 |
| Ag | -1 | -1.81856000 | -0.30814100 | -1.35215000 |
| Ag | -1 | -0.23621400 | -1.05548100 | 0.94825700  |

| Ag | -1 | -1.96169400 | 1.26212600 | 1.06856500 |
| Ag | -1 | -3.54159100 | 2.00713200 | -1.23195300 |
| Ag | -1 | -2.48487500 | -2.68459200 | 0.14925100 |
| Ag | -1 | -4.20802400 | -0.36933900 | 0.26961300 |
| Ag | 0 | -2.49240400 | -1.07152000 | 2.50341200 |
| C | 0 | 3.93423200 | -3.00667300 | -0.31734500 |
| C | 0 | 5.26216900 | -2.41427700 | 0.11589000 |
| C | 0 | 5.25802600 | -0.88004600 | 0.21067400 |
| C | 0 | 6.61772700 | -0.45774400 | 0.79884400 |
| C | 0 | 6.87747800 | 1.02966300 | 0.95617800 |
| C | 0 | 5.15375400 | -0.24142900 | -1.19931100 |
| H | 0 | 5.53466900 | -2.82260900 | 1.09665000 |
| H | 0 | 6.02565800 | -2.73233700 | -0.60826000 |
| H | 0 | 6.69748200 | -0.89878900 | 1.80400300 |
| H | 0 | 7.44048700 | -0.87659400 | 0.20574600 |
| O | 0 | 3.40673200 | -2.53827400 | -1.38613600 |
| O | 0 | 3.43160400 | -3.93392100 | 0.36374300 |
| O | 0 | 8.04639200 | 1.39100300 | 1.21505500 |

| O  | 0  | 5.90650500  | 1.85920500   | 0.83904700   |
|----|----|-------------|--------------|--------------|
| O  | 0  | 5.92046200  | -0.60407900  | -2.11119800  |
| O  | 0  | 4.29167800  | 0.70626200   | -1.32214800  |
| O  | 0  | 4.20999300  | -0.43654700  | 1.02642900   |
| Tb | 0  | 3.60717200  | 1.63427100   | 0.73151200   |

**Tb-citrate-Ag₁₁ (mode 2)**

| Ag | 0  | 1.83420600  | 0.08022900   | -2.29549200  |
|----|----|-------------|--------------|--------------|
| Ag | -1 | -0.13938000 | 1.65252700   | -1.37361400  |
| Ag | -1 | -2.34116100 | 3.28672400   | -0.46362800  |
| Ag | -1 | 0.36843300  | -2.32064900  | -2.18038800  |
| Ag | -1 | -1.83276000 | -0.68615100  | -1.26972900  |
| Ag | -1 | 0.67846400  | -0.65630200  | 0.16013800   |
| Ag | -1 | -1.52552200 | 0.97877200   | 1.07080400   |
| Ag | -1 | -4.03417100 | 0.94811100   | -0.36036400  |
| Ag | -1 | -1.01632500 | -2.99394300  | 0.26421500   |
| Ag | -1 | -3.21780100 | -1.35976600  | 1.17385500   |
| Ag | 0  | -0.63374600 | -1.22145100  | 2.50005200   |
| C  | 0  | 4.73275900  | -0.83795600  | -1.75786600  |

| C | 0 | 4.48461500 | -1.54260000 | -0.43245000 |
| C | 0 | 3.98998400 | -0.54396900 | 0.63081100 |
| C | 0 | 3.56978500 | -1.30105800 | 1.89577100 |
| C | 0 | 2.97050900 | -0.46780300 | 3.02016700 |
| C | 0 | 5.10640900 | 0.46248400 | 1.01393900 |
| H | 0 | 3.70440500 | -2.30421400 | -0.55738100 |
| H | 0 | 5.40947800 | -2.02429000 | -0.09302300 |
| H | 0 | 2.79500800 | -2.02897300 | 1.60413400 |
| H | 0 | 4.40767900 | -1.87439000 | 2.31245000 |
| O | 0 | 5.77011000 | -0.13861000 | -1.87517200 |
| O | 0 | 3.87282300 | -0.98657700 | -2.69458700 |
| O | 0 | 2.70885100 | -1.05807600 | 4.09171900 |
| O | 0 | 2.74808600 | 0.78293700 | 2.84624500 |
| O | 0 | 6.21205900 | 0.04198900 | 1.40569900 |
| O | 0 | 4.78620100 | 1.70342200 | 0.94402100 |
| O | 0 | 2.89935400 | 0.15437500 | 0.11859600 |
| Tb | 0 | 2.46135500 | 2.11293600 | 0.95699300 |

**Tb-citrate-Ag$_{II}$ (mode 3)**

| Ag | 0 | -1.09587400 | -2.71475500 | -0.88778400 |
|----|----|----|----|----|
| Ag | -1 | -1.02777800 | 0.06531200 | -0.80650200 |
| Ag | -1 | -0.89994600 | 2.94538200 | -0.90263700 |
| Ag | -1 | 0.86439700 | -2.84022100 | 1.36143500 |
| Ag | -1 | 0.99046400 | 0.04469400 | 1.26252200 |
| Ag | -1 | 1.37482900 | -1.50285300 | -1.14618000 |
| Ag | -1 | 1.50611200 | 1.38149700 | -1.25043300 |
| Ag | -1 | 1.11965400 | 2.92938300 | 1.16128800 |
| Ag | -1 | 3.39564200 | -1.52022600 | 0.91764400 |
| Ag | -1 | 3.52292500 | 1.36368400 | 0.81831000 |
| Ag | 0 | 3.80001300 | -0.14863900 | -1.56730500 |
| C | 0 | -3.80292000 | 3.27956000 | 0.08757600 |
| C | 0 | -3.25565700 | 2.46745000 | 1.24894300 |
| C | 0 | -3.54906700 | 0.96666400 | 1.08909000 |
| C | 0 | -2.73805800 | 0.20878600 | 2.14959000 |
| C | 0 | -2.86379100 | -1.30348400 | 2.16565800 |
| C | 0 | -5.05720500 | 0.68770700 | 1.32850300 |
| H | 0 | -2.16827300 | 2.60251500 | 1.32170300 |

| H | 0 | -3.71221600 | 2.82718500 | 2.17982200 |
|---|---|---|---|---|
| H | 0 | -1.67205100 | 0.43438200 | 1.97606900 |
| H | 0 | -2.97614300 | 0.57420600 | 3.15696400 |
| O | 0 | -5.04730600 | 3.32367600 | -0.07640800 |
| O | 0 | -2.97568500 | 3.89511200 | -0.67032500 |
| O | 0 | -2.24495700 | -1.92672800 | 3.05586500 |
| O | 0 | -3.58836000 | -1.88581100 | 1.28373600 |
| O | 0 | -5.60466300 | 1.09809500 | 2.36971700 |
| O | 0 | -5.64654500 | -0.01034800 | 0.42800300 |
| O | 0 | -3.20736100 | 0.52908300 | -0.19738500 |
| Tb | 0 | -4.18804200 | -1.30178700 | -0.88820300 |

**Tb-citrate-Ag$_{11}$ (mode 4)**

| Ag | 0 | -0.76151300 | 2.85263700 | -0.13856600 |
|---|---|---|---|---|
| Ag | -1 | -1.04671700 | 0.16490700 | -0.01928000 |
| Ag | -1 | -1.26599300 | -2.71529800 | 0.01701800 |
| Ag | -1 | 1.71254400 | 2.83449200 | -1.41821800 |
| Ag | -1 | 1.49306000 | -0.04585800 | -1.38080200 |
| Ag | -1 | 1.29491800 | 1.45165900 | 1.08275800 |

| Ag | -1 | 1.07535300 | -1.43153600 | 1.11933500 |
| Ag | -1 | 1.27331400 | -2.92589600 | -1.34442700 |
| Ag | -1 | 3.83411400 | 1.23845500 | -0.27836800 |
| Ag | -1 | 3.61452400 | -1.64145000 | -0.24188500 |
| Ag | 0 | 3.29029200 | -0.08009700 | 2.17642000 |
| C | 0 | -3.79539800 | 3.35563500 | -0.07128600 |
| C | 0 | -3.67043400 | 2.53457500 | 1.20050100 |
| C | 0 | -4.00341700 | 1.05201900 | 0.97001200 |
| C | 0 | -3.62407800 | 0.28092700 | 2.24303100 |
| C | 0 | -3.88732300 | -1.21444500 | 2.26302100 |
| C | 0 | -5.52629500 | 0.88646700 | 0.72684100 |
| H | 0 | -2.64746800 | 2.60134700 | 1.59433800 |
| H | 0 | -4.35638700 | 2.94442900 | 1.95307600 |
| H | 0 | -2.54008800 | 0.40975700 | 2.39378700 |
| H | 0 | -4.11898600 | 0.71727700 | 3.11978600 |
| O | 0 | -4.92601800 | 3.45926800 | -0.60669700 |
| O | 0 | -2.74518500 | 3.91831200 | -0.53810700 |
| O | 0 | -3.72149200 | -1.81499800 | 3.34672600 |

| O  | 0  | -4.25374100 | -1.81189700 | 1.18917000  |
|----|----|-------------|-------------|-------------|
| O  | 0  | -6.33890800 | 1.35262400  | 1.54692000  |
| O  | 0  | -5.85555600 | 0.21724900  | -0.31799900 |
| O  | 0  | -3.30824200 | 0.55010900  | -0.13839000 |
| Tb | 0  | -4.18854400 | -1.24621500 | -1.05975300 |

**Tb-citrate-Ag$_{\text{II}}$ (mode 5)**

| Ag | 0  | 0.76826300  | -2.84429100 | -1.20464000 |
|----|----|-------------|-------------|-------------|
| Ag | -1 | 1.00948100  | -0.12310100 | -1.07254100 |
| Ag | -1 | 1.30838400  | 2.74473300  | -0.87983200 |
| Ag | -1 | -2.14466900 | -2.67065400 | -1.57508300 |
| Ag | -1 | -1.84532100 | 0.19672100  | -1.38227200 |
| Ag | -1 | -0.79979200 | -1.50817700 | 0.70200300  |
| Ag | -1 | -0.50063100 | 1.35825700  | 0.89547800  |
| Ag | -1 | -1.54612000 | 3.06333500  | -1.18921500 |
| Ag | -1 | -3.65446200 | -1.19007400 | 0.39363300  |
| Ag | -1 | -3.35518700 | 1.67697300  | 0.58622700  |
| Ag | 0  | -2.16158800 | 0.00159100  | 2.58252200  |
| C  | 0  | 4.31891700  | 2.83106900  | -0.50327500 |
| C | 0 | 4.66829000 | 1.55603100 | -1.25416600 |
| C | 0 | 4.38573300 | 0.29810500 | -0.41821400 |
| C | 0 | 4.59782700 | -0.93908300 | -1.30284800 |
| C | 0 | 4.39707400 | -2.29344100 | -0.64515700 |
| C | 0 | 5.36032700 | 0.20518900 | 0.78561400 |
| H | 0 | 4.07381300 | 1.49514700 | -2.17394500 |
| H | 0 | 5.73159700 | 1.58960300 | -1.52675300 |
| H | 0 | 3.86941100 | -0.88203900 | -2.12778700 |
| H | 0 | 5.59658000 | -0.93119200 | -1.75684400 |
| O | 0 | 4.92226700 | 3.08194500 | 0.56748800 |
| O | 0 | 3.43282200 | 3.60457100 | -1.01308000 |
| O | 0 | 4.75651500 | -3.30771600 | -1.28274600 |
| O | 0 | 3.86172100 | -2.35945900 | 0.51730100 |
| O | 0 | 6.59097700 | 0.26847700 | 0.59198300 |
| O | 0 | 4.81589900 | 0.00992500 | 1.92981300 |
| O | 0 | 3.07431800 | 0.33688100 | 0.05629200 |
| Tb | 0 | 2.61601000 | -0.88377700 | 1.80664700 |

**Dy-citrate-Ag$_{II}$ (mode 1)**

| Ag | 0 | 1.94569100 | -0.86387800 | -0.76748100 |
|----|----|----|----|----|
| Ag | -1 | 0.43084600 | 1.31649600 | -0.53607000 |
| Ag | -1 | -1.28826000 | 3.63648000 | -0.44115500 |
| Ag | -1 | -0.09095700 | -2.63112800 | -1.44748800 |
| Ag | -1 | -1.81075100 | -0.31182700 | -1.35218600 |
| Ag | -1 | -0.25423300 | -1.05455500 | 0.96725700 |
| Ag | -1 | -1.97601000 | 1.26700700 | 1.06153700 |
| Ag | -1 | -3.53008300 | 2.00739600 | -1.25798300 |
| Ag | -1 | -2.49779200 | -2.68174500 | 0.15019900 |
| Ag | -1 | -4.21724500 | -0.36254200 | 0.24456500 |
| Ag | 0 | -2.52640700 | -1.05878000 | 2.49911500 |
| C | 0 | 3.93593000 | -3.01714900 | -0.36427100 |
| C | 0 | 5.26063000 | -2.41883000 | 0.07041100 |
| C | 0 | 5.24892800 | -0.88614300 | 0.18355700 |
| C | 0 | 6.60972300 | -0.46671700 | 0.77247700 |
| C | 0 | 6.86698800 | 1.01892600 | 0.94787800 |
| C | 0 | 5.13872300 | -0.22801200 | -1.21703400 |
| H | 0 | 5.54006400 | -2.83708200 | 1.04504200 |

| H | 0 | 6.02308200 | -2.72434300 | -0.66025800 |
|---|---|---|---|---|
| H | 0 | 6.69339000 | -0.91974400 | 1.77202600 |
| H | 0 | 7.43146900 | -0.87660900 | 0.17180400 |
| O | 0 | 3.39321800 | -2.53251500 | -1.41816400 |
| O | 0 | 3.45113500 | -3.96578000 | 0.30010600 |
| O | 0 | 8.03750200 | 1.38055400 | 1.19808500 |
| O | 0 | 5.89204400 | 1.84711400 | 0.85504400 |
| O | 0 | 5.89476300 | -0.58350600 | -2.14032900 |
| O | 0 | 4.28392900 | 0.72909200 | -1.31897200 |
| O | 0 | 4.20153500 | -0.45729200 | 1.00826700 |
| Dy | 0 | 3.60648500 | 1.61585600 | 0.73918800 |

**Dy-citrate-Ag$_{11}$ (mode 2)**

| Ag | 0 | 1.82965200 | 0.07184200 | -2.29841100 |
|---|---|---|---|---|
| Ag | -1 | -0.14399300 | 1.64731800 | -1.38107700 |
| Ag | -1 | -2.34334300 | 3.28579400 | -0.47291000 |
| Ag | -1 | 0.35987700 | -2.32861100 | -2.17668500 |
| Ag | -1 | -1.83888300 | -0.68983200 | -1.26784600 |
| Ag | -1 | 0.67435400 | -0.65750300 | 0.15842300 |

| Ag | -1 | -1.52719900 | 0.98185500 | 1.06727100 |
| Ag | -1 | -4.03786400 | 0.94870700 | -0.36030100 |
| Ag | -1 | -1.02194400 | -2.99361400 | 0.27184600 |
| Ag | -1 | -3.22098900 | -1.35515900 | 1.17966700 |
| Ag | 0 | -0.63572600 | -1.21581900 | 2.50197000 |
| C | 0 | 4.73412100 | -0.84603000 | -1.76363700 |
| C | 0 | 4.47672600 | -1.54934100 | -0.43931300 |
| C | 0 | 3.98399700 | -0.55025700 | 0.62439200 |
| C | 0 | 3.56527400 | -1.30729800 | 1.88983200 |
| C | 0 | 2.96804100 | -0.47208000 | 3.01363000 |
| C | 0 | 5.10043800 | 0.45622000 | 1.00672300 |
| H | 0 | 3.69227900 | -2.30612700 | -0.56741500 |
| H | 0 | 5.39751800 | -2.03693000 | -0.09727800 |
| H | 0 | 2.78969300 | -2.03482500 | 1.59941300 |
| H | 0 | 4.40335400 | -1.88075600 | 2.30598400 |
| O | 0 | 5.78165700 | -0.16177300 | -1.88042700 |
| O | 0 | 3.87229400 | -0.98092900 | -2.70058100 |
| O | 0 | 2.70624500 | -1.06014300 | 4.08624200 |

| O | 0 | 2.74787300 | 0.77887400 | 2.83797100 |
|---|---|---|---|---|
| O | 0 | 6.20685400 | 0.03638400 | 1.39702300 |
| O | 0 | 4.77920700 | 1.69689800 | 0.93781200 |
| O | 0 | 2.89314400 | 0.14904300 | 0.11439400 |
| Dy | 0 | 2.46109200 | 2.09910000 | 0.95452400 |

**Dy-citrate-Ag$_{11}$ (mode 3)**

| Ag | 0 | -1.03679500 | -2.72625000 | -1.12996700 |
|---|---|---|---|---|
| Ag | -1 | -0.98978600 | 0.04378300 | -1.00787800 |
| Ag | -1 | -0.89411000 | 2.92593200 | -1.06814800 |
| Ag | -1 | 0.76754300 | -2.85837500 | 1.27536200 |
| Ag | -1 | 0.86194400 | 0.02868700 | 1.21170800 |
| Ag | -1 | 1.45273300 | -1.49018100 | -1.17317900 |
| Ag | -1 | 1.55263400 | 1.39655600 | -1.24180000 |
| Ag | -1 | 0.95936600 | 2.91567900 | 1.14611700 |
| Ag | -1 | 3.30751800 | -1.50179000 | 1.04104600 |
| Ag | -1 | 3.40301900 | 1.38442000 | 0.97719800 |
| Ag | 0 | 3.87888100 | -0.09278300 | -1.40300500 |
| C | 0 | -3.77273400 | 3.26598100 | -0.03765700 |
| C | 0 | -3.17170700 | 2.53771100 | 1.15172500 |
| C | 0 | -3.43407100 | 1.02398000 | 1.08768800 |
| C | 0 | -2.55257400 | 0.34955200 | 2.14974300 |
| C | 0 | -2.61339300 | -1.16248000 | 2.26381700 |
| C | 0 | -4.91696300 | 0.72600200 | 1.44052800 |
| H | 0 | -2.08631000 | 2.70396300 | 1.18442300 |
| H | 0 | -3.60993400 | 2.93985600 | 2.07393300 |
| H | 0 | -1.50288600 | 0.60429800 | 1.91936700 |
| H | 0 | -2.75847800 | 0.76683500 | 3.14376500 |
| O | 0 | -5.02195400 | 3.26850000 | -0.16846200 |
| O | 0 | -2.98288100 | 3.85587200 | -0.85363900 |
| O | 0 | -1.95388900 | -1.69906200 | 3.18025300 |
| O | 0 | -3.32153200 | -1.83595500 | 1.43354200 |
| O | 0 | -5.41860200 | 1.21008000 | 2.47261000 |
| O | 0 | -5.53208400 | -0.07011800 | 0.64400400 |
| O | 0 | -3.15527900 | 0.52145200 | -0.19120500 |
| Dy | 0 | -4.11427900 | -1.37382600 | -0.67096500 |

**Dy-citrate-Ag$_{11}$ (mode 4)**

| | | | | |
|---|---|---|---|---|
| Ag | 0 | -0.74459900 | 2.86002200 | -0.13168900 |
| Ag | -1 | -1.03978500 | 0.17455500 | -0.01111000 |
| Ag | -1 | -1.27149500 | -2.70470400 | 0.02297200 |
| Ag | -1 | 1.72716700 | 2.83349900 | -1.41511100 |
| Ag | -1 | 1.49525200 | -0.04590400 | -1.37991200 |
| Ag | -1 | 1.31048600 | 1.45001200 | 1.08566000 |
| Ag | -1 | 1.07847600 | -1.43223500 | 1.12001900 |
| Ag | -1 | 1.26307300 | -2.92499300 | -1.34575200 |
| Ag | -1 | 3.84493300 | 1.22711600 | -0.28274700 |
| Ag | -1 | 3.61291200 | -1.65184000 | -0.24847800 |
| Ag | 0 | 3.30298900 | -0.09169100 | 2.17113700 |
| C | 0 | -3.78118800 | 3.36225400 | -0.08071500 |
| C | 0 | -3.66359200 | 2.54559000 | 1.19454300 |
| C | 0 | -3.99846800 | 1.06297000 | 0.96811200 |
| C | 0 | -3.62483900 | 0.29582800 | 2.24529000 |
| C | 0 | -3.88868100 | -1.19925400 | 2.26787000 |
| C | 0 | -5.52048300 | 0.89776500 | 0.71995900 |
| H | 0 | -2.64218100 | 2.61168900 | 1.59249200 |
| H | 0 | -4.35179500 | 2.95982300 | 1.94268800 |
| H | 0 | -2.54126100 | 0.42429100 | 2.39941200 |
| H | 0 | -4.12271500 | 0.73509400 | 3.11888700 |
| O | 0 | -4.90892900 | 3.46442800 | -0.62252300 |
| O | 0 | -2.72830700 | 3.92317500 | -0.54352300 |
| O | 0 | -3.73209400 | -1.79673700 | 3.35457200 |
| O | 0 | -4.24542000 | -1.80008300 | 1.19262500 |
| O | 0 | -6.33641400 | 1.37223900 | 1.53180000 |
| O | 0 | -5.84603200 | 0.21912500 | -0.32019900 |
| O | 0 | -3.30036600 | 0.55585700 | -0.13582700 |
| Dy | 0 | -4.18076200 | -1.23691200 | -1.04687500 |

**Dy-citrate-Ag$_{11}$ (mode 5)**

| | | | | |
|---|---|---|---|---|
| Ag | 0 | 1.13983000 | -2.68400900 | -0.28191800 |
| Ag | -1 | 1.04485000 | 0.08026400 | -0.25708700 |
| Ag | -1 | 0.95388300 | 2.96498900 | -0.11357300 |
| Ag | -1 | -1.51031300 | -2.82888500 | -1.55990100 |
| Ag | -1 | -1.60142000 | 0.05521900 | -1.41607400 |
| Ag | -1 | -1.04935600 | -1.49267700 | 0.95958500 |

| Ag | -1 | -1.14051400 | 1.39073400 | 1.10384700 |
| Ag | -1 | -1.69258200 | 2.93908400 | -1.27185300 |
| Ag | -1 | -3.69606500 | -1.51915700 | -0.19800800 |
| Ag | -1 | -3.78702400 | 1.36502600 | -0.05447100 |
| Ag | 0 | -3.09584100 | -0.14714200 | 2.27004800 |
| C | 0 | 4.00610800 | 3.27394000 | 0.05592500 |
| C | 0 | 4.10169100 | 2.33031500 | -1.13140400 |
| C | 0 | 4.25702900 | 0.86191400 | -0.71086600 |
| C | 0 | 4.14440900 | -0.00865100 | -1.97222900 |
| C | 0 | 4.35648900 | -1.50328400 | -1.80847700 |
| C | 0 | 5.64820200 | 0.61581800 | -0.06812400 |
| H | 0 | 3.19653800 | 2.41772200 | -1.74574400 |
| H | 0 | 4.96354800 | 2.62956900 | -1.74294700 |
| H | 0 | 3.12184000 | 0.12132300 | -2.36217700 |
| H | 0 | 4.83120500 | 0.34376200 | -2.75138900 |
| O | 0 | 4.95181800 | 3.30204300 | 0.88039200 |
| O | 0 | 2.97021400 | 4.02084800 | 0.15062400 |
| O | 0 | 4.51160200 | -2.18417700 | -2.84549400 |

| O  | 0  | 4.35361500  | -2.01935400 | -0.63484000 |
| O  | 0  | 6.68104000  | 0.95151700  | -0.67814600 |
| O  | 0  | 5.64031900  | 0.01226900  | 1.06572300  |
| O  | 0  | 3.26803400  | 0.51579900  | 0.21473000  |
| Dy | 0  | 3.70021400  | -1.23524600 | 1.45002100  |

**Ho-citrate-Ag$_{11}$ (mode 1)**

| Ag | 0  | 1.93967900  | -0.87749700 | -0.75705100 |
| Ag | -1 | 0.43194600  | 1.30867200  | -0.53264800 |
| Ag | -1 | -1.27964400 | 3.63465700  | -0.44948700 |
| Ag | -1 | -0.10019600 | -2.63961300 | -1.43517900 |
| Ag | -1 | -1.81247700 | -0.31430900 | -1.35163100 |
| Ag | -1 | -0.26593600 | -1.05602000 | 0.97480000  |
| Ag | -1 | -1.98019000 | 1.27155000  | 1.05731300  |
| Ag | -1 | -3.52429400 | 2.01091400  | -1.26918100 |
| Ag | -1 | -2.51231500 | -2.67786400 | 0.15486500  |
| Ag | -1 | -4.22425200 | -0.35266100 | 0.23747800  |
| Ag | 0  | -2.54333900 | -1.04780200 | 2.49965400  |
| C  | 0  | 3.93939200  | -3.02451800 | -0.38970000 |
| C | 0 | 5.26142700 | -2.42091400 | 0.04599900 |
| C | 0 | 5.24324500 | -0.88918800 | 0.16864100 |
| C | 0 | 6.60265000 | -0.46804700 | 0.75959700 |
| C | 0 | 6.84968300 | 1.01658800 | 0.95632500 |
| C | 0 | 5.12939900 | -0.22183400 | -1.22727300 |
| H | 0 | 5.54478600 | -2.84386300 | 1.01746100 |
| H | 0 | 6.02403400 | -2.71865600 | -0.68774000 |
| H | 0 | 6.69350300 | -0.93501800 | 1.75204200 |
| H | 0 | 7.42569500 | -0.86269700 | 0.15039900 |
| O | 0 | 3.38776900 | -2.53272500 | -1.43562600 |
| O | 0 | 3.46591000 | -3.98496700 | 0.26576100 |
| O | 0 | 8.01547000 | 1.38177500 | 1.22156000 |
| O | 0 | 5.87100200 | 1.84102600 | 0.86477500 |
| O | 0 | 5.88041800 | -0.57255900 | -2.15614400 |
| O | 0 | 4.27648200 | 0.73842300 | -1.31914800 |
| O | 0 | 4.19460000 | -0.46992800 | 0.99729800 |
| Ho | 0 | 3.60241500 | 1.59668800 | 0.73609900 |

**Ho-citrate-Ag$_{II}$ (mode 2)**

| Ag | 0  | 1.79583600  | 0.06062000   | -2.34304800 |
|----|----|-------------|--------------|-------------|
| Ag | -1 | -0.18174000 | 1.63442700   | -1.42431900 |
| Ag | -1 | -2.36813800 | 3.27616000   | -0.49110400 |
| Ag | -1 | 0.32073700  | -2.34918700  | -2.18143100 |
| Ag | -1 | -1.86505800 | -0.70714100  | -1.24756100 |
| Ag | -1 | 0.67685800  | -0.64834800  | 0.12606700  |
| Ag | -1 | -1.51173100 | 0.99426700   | 1.05999800  |
| Ag | -1 | -4.05109900 | 0.93464900   | -0.31497600 |
| Ag | -1 | -1.00786700 | -2.98888000  | 0.30303300  |
| Ag | -1 | -3.19396700 | -1.34717100  | 1.23589600  |
| Ag | 0  | -0.58814100 | -1.18744000  | 2.50438400  |
| C  | 0  | 4.68693300  | -0.83336900  | -1.79150500 |
| C  | 0  | 4.46833800  | -1.54712600  | -0.46586400 |
| C  | 0  | 3.98562900  | -0.55386700  | 0.60691200  |
| C  | 0  | 3.57091100  | -1.31542000  | 1.87085300  |
| C  | 0  | 2.98681600  | -0.48295300  | 3.00342200  |
| C  | 0  | 5.10798200  | 0.44591600   | 0.98869900  |
| H  | 0  | 3.69291400  | -2.31500100  | -0.58109900 |

| H | 0 | 5.40296200 | -2.02289000 | -0.14482800 |
|---|---|---|---|---|
| H | 0 | 2.78811000 | -2.03563800 | 1.58150900 |
| H | 0 | 4.40782300 | -1.89727100 | 2.27764000 |
| O | 0 | 5.69328300 | -0.09064300 | -1.90894800 |
| O | 0 | 3.83246700 | -1.01789300 | -2.72722000 |
| O | 0 | 2.73434600 | -1.07377700 | 4.07663200 |
| O | 0 | 2.76707700 | 0.76919700 | 2.83429400 |
| O | 0 | 6.21783200 | 0.02060400 | 1.36261400 |
| O | 0 | 4.78722500 | 1.68797900 | 0.94007400 |
| O | 0 | 2.89467700 | 0.15283400 | 0.10574300 |
| Ho | 0 | 2.48128300 | 2.08691600 | 0.96209000 |

**Ho-citrate-Ag$_{11}$ (mode 3)**

| Ag | 0 | -1.04189800 | -2.71958200 | -1.12943200 |
|---|---|---|---|---|
| Ag | -1 | -0.98508700 | 0.05036300 | -1.00496900 |
| Ag | -1 | -0.87997200 | 2.93234100 | -1.06446900 |
| Ag | -1 | 0.76521600 | -2.85840200 | 1.27492700 |
| Ag | -1 | 0.86873300 | 0.02843000 | 1.21256500 |
| Ag | -1 | 1.45253400 | -1.49125400 | -1.17369000 |

| Ag | -1 | 1.56166700 | 1.39511700 | -1.24098100 |
| Ag | -1 | 0.97554400 | 2.91509200 | 1.14812900 |
| Ag | -1 | 3.30928200 | -1.50986500 | 1.03888700 |
| Ag | -1 | 3.41405400 | 1.37597900 | 0.97619000 |
| Ag | 0 | 3.88315600 | -0.10109000 | -1.40505500 |
| C | 0 | -3.76127500 | 3.27456200 | -0.04040800 |
| C | 0 | -3.16362800 | 2.54591300 | 1.15043600 |
| C | 0 | -3.42811000 | 1.03263300 | 1.08732200 |
| C | 0 | -2.54845300 | 0.35775500 | 2.15059200 |
| C | 0 | -2.61176300 | -1.15402700 | 2.26438600 |
| C | 0 | -4.91145600 | 0.73588900 | 1.43803700 |
| H | 0 | -2.07809200 | 2.71065000 | 1.18500600 |
| H | 0 | -3.60281700 | 2.94939200 | 2.07160500 |
| H | 0 | -1.49813300 | 0.61077500 | 1.92114800 |
| H | 0 | -2.75485100 | 0.77548100 | 3.14434400 |
| O | 0 | -5.01020000 | 3.27847800 | -0.17392800 |
| O | 0 | -2.96890400 | 3.86347800 | -0.85464500 |
| O | 0 | -1.95497000 | -1.69218500 | 3.18159900 |
| O | 0 | -3.31988100 | -1.82618500 | 1.43270100 |
| O | 0 | -5.41626100 | 1.22383500 | 2.46663700 |
| O | 0 | -5.52392600 | -0.06420500 | 0.64324500 |
| O | 0 | -3.14884100 | 0.52840100 | -0.19103900 |
| Ho | 0 | -4.10396800 | -1.35971100 | -0.66223300 |

**Ho-citrate-Ag$_{11}$ (mode 4)**

| | | | | |
|---|---|---|---|---|
| Ag | 0 | -0.72674700 | 2.86756000 | -0.12166600 |
| Ag | -1 | -1.03231100 | 0.18382500 | -0.00091700 |
| Ag | -1 | -1.27612300 | -2.69446500 | 0.03046500 |
| Ag | -1 | 1.74095500 | 2.83274200 | -1.41139900 |
| Ag | -1 | 1.49694200 | -0.04569300 | -1.37889800 |
| Ag | -1 | 1.32719400 | 1.44794600 | 1.08913300 |
| Ag | -1 | 1.08306900 | -1.43333200 | 1.12078800 |
| Ag | -1 | 1.25266300 | -2.92381200 | -1.34743600 |
| Ag | -1 | 3.85584800 | 1.21599300 | -0.28845500 |
| Ag | -1 | 3.61172600 | -1.66199400 | -0.25688500 |
| Ag | 0 | 3.31762900 | -0.10368200 | 2.16463500 |
| C | 0 | -3.76645900 | 3.36813200 | -0.09360100 |

| C | 0 | -3.65892000 | 2.55661300 | 1.18580300 |
| C | 0 | -3.99633100 | 1.07401700 | 0.96413700 |
| C | 0 | -3.62965300 | 0.31160100 | 2.24619800 |
| C | 0 | -3.89361800 | -1.18321200 | 2.27233300 |
| C | 0 | -5.51727600 | 0.90948200 | 0.71001700 |
| H | 0 | -2.63981400 | 2.62185100 | 1.58972700 |
| H | 0 | -4.35041300 | 2.97613700 | 1.92795900 |
| H | 0 | -2.54676000 | 0.44022900 | 2.40504100 |
| H | 0 | -4.13170000 | 0.75418900 | 3.11572500 |
| O | 0 | -4.89004300 | 3.46795100 | -0.64448000 |
| O | 0 | -2.71012000 | 3.92765800 | -0.55002900 |
| O | 0 | -3.74600600 | -1.77674000 | 3.36232800 |
| O | 0 | -4.24082500 | -1.78848400 | 1.19633600 |
| O | 0 | -6.33707500 | 1.39505800 | 1.51097600 |
| O | 0 | -5.83855800 | 0.21801200 | -0.32352200 |
| O | 0 | -3.29434900 | 0.56051200 | -0.13448300 |
| Ho | 0 | -4.17450200 | -1.22750800 | -1.03365600 |

**Ho-citrate-Ag$_{II}$ (mode 5)**

| | | | | |
|---|---|---|---|---|
| Ag | 0 | 0.75456600 | -2.84192600 | -1.22446600 |
| Ag | -1 | 0.99741400 | -0.12038000 | -1.09037600 |
| Ag | -1 | 1.29996000 | 2.74530600 | -0.89841800 |
| Ag | -1 | -2.16160200 | -2.66716100 | -1.57244200 |
| Ag | -1 | -1.85923200 | 0.19891700 | -1.37974300 |
| Ag | -1 | -0.79925900 | -1.50804000 | 0.69513800 |
| Ag | -1 | -0.49685900 | 1.35889900 | 0.88925700 |
| Ag | -1 | -1.55684900 | 3.06620400 | -1.18630600 |
| Ag | -1 | -3.65594900 | -1.18714100 | 0.40868800 |
| Ag | -1 | -3.35352900 | 1.67952100 | 0.60120200 |
| Ag | 0 | -2.14724100 | 0.00118800 | 2.58619100 |
| C | 0 | 4.30653100 | 2.83809200 | -0.50031500 |
| C | 0 | 4.66873900 | 1.56453800 | -1.24762000 |
| C | 0 | 4.37759600 | 0.30450700 | -0.41770200 |
| C | 0 | 4.60754500 | -0.93024600 | -1.30104200 |
| C | 0 | 4.38674800 | -2.28559600 | -0.65230000 |
| C | 0 | 5.33324100 | 0.21454800 | 0.80103200 |
| H | 0 | 4.08703100 | 1.50356300 | -2.17551000 |

| H | 0 | 5.73565600 | 1.60059300 | -1.50544700 |
|---|---|---|---|---|
| H | 0 | 3.90035100 | -0.86877200 | -2.14371700 |
| H | 0 | 5.61721500 | -0.92427300 | -1.73050600 |
| O | 0 | 4.89635200 | 3.09024800 | 0.57767700 |
| O | 0 | 3.42414000 | 3.60891800 | -1.02048200 |
| O | 0 | 4.74561800 | -3.29989900 | -1.28959100 |
| O | 0 | 3.83701500 | -2.35283100 | 0.50367800 |
| O | 0 | 6.56631000 | 0.28373700 | 0.62840400 |
| O | 0 | 4.77061600 | 0.01576900 | 1.93648300 |
| O | 0 | 3.05913600 | 0.33734200 | 0.03744200 |
| Ho | 0 | 2.59965800 | -0.88217400 | 1.76901400 |

**Er-citrate-Ag₁₁ (mode 1)**

| Ag | 0 | 1.93045900 | -0.90295400 | -0.76130000 |
|---|---|---|---|---|
| Ag | -1 | 0.43400000 | 1.29448300 | -0.54382800 |
| Ag | -1 | -1.26080800 | 3.63261900 | -0.45695400 |
| Ag | -1 | -0.12957400 | -2.65113200 | -1.43872900 |
| Ag | -1 | -1.82509900 | -0.31370600 | -1.35147900 |
| Ag | -1 | -0.27442700 | -1.06314000 | 0.96971000 |

| Ag | -1 | -1.97188200 | 1.27655200 | 1.05593300 |
|---|---|---|---|---|
| Ag | -1 | -3.52014100 | 2.02363600 | -1.26533200 |
| Ag | -1 | -2.53546600 | -2.67023200 | 0.16110900 |
| Ag | -1 | -4.23063300 | -0.33290300 | 0.24740100 |
| Ag | 0 | -2.54690300 | -1.03746800 | 2.50323400 |
| C | 0 | 3.95880700 | -3.03071000 | -0.38296300 |
| C | 0 | 5.27831600 | -2.41807800 | 0.04851400 |
| C | 0 | 5.24761600 | -0.88675600 | 0.16871700 |
| C | 0 | 6.60136000 | -0.45013800 | 0.76092900 |
| C | 0 | 6.82547200 | 1.03641300 | 0.97236200 |
| C | 0 | 5.12851800 | -0.22340500 | -1.22855500 |
| H | 0 | 5.56658400 | -2.83764500 | 1.01995200 |
| H | 0 | 6.04142900 | -2.71131200 | -0.68647600 |
| H | 0 | 6.70203500 | -0.92474100 | 1.74874900 |
| H | 0 | 7.42964600 | -0.82557900 | 0.14658000 |
| O | 0 | 3.39985000 | -2.54236700 | -1.42654400 |
| O | 0 | 3.49483100 | -3.99486000 | 0.27379200 |
| O | 0 | 7.98251600 | 1.41539800 | 1.25378400 |

| O | 0 | 5.83663400 | 1.84930400 | 0.87642500 |
|---|---|---|---|---|
| O | 0 | 5.87946400 | -0.57157900 | -2.15800100 |
| O | 0 | 4.27031000 | 0.73257300 | -1.32069500 |
| O | 0 | 4.19442700 | -0.47591400 | 0.99607100 |
| Er | 0 | 3.59246700 | 1.57080100 | 0.72278800 |

**Er-citrate-Ag$_{11}$ (mode 2)**

| Ag | 0 | 1.81766600 | 0.03938400 | -2.30513200 |
|---|---|---|---|---|
| Ag | -1 | -0.14195900 | 1.63199900 | -1.39368900 |
| Ag | -1 | -2.33214800 | 3.28978200 | -0.49762300 |
| Ag | -1 | 0.33878600 | -2.35229700 | -2.16023100 |
| Ag | -1 | -1.85055200 | -0.69432000 | -1.26344800 |
| Ag | -1 | 0.66243200 | -0.66584800 | 0.16264200 |
| Ag | -1 | -1.52923600 | 0.99248200 | 1.05940000 |
| Ag | -1 | -4.04005500 | 0.96333500 | -0.36805000 |
| Ag | -1 | -1.04653900 | -2.99168300 | 0.29307800 |
| Ag | -1 | -3.23646400 | -1.33387800 | 1.18870500 |
| Ag | 0 | -0.65140800 | -1.20201400 | 2.51020600 |
| C | 0 | 4.72551700 | -0.86679400 | -1.77908400 |

| C | 0 | 4.46626200 | -1.56389900 | -0.45181300 |
| C | 0 | 3.97789600 | -0.56236200 | 0.61127500 |
| C | 0 | 3.56740600 | -1.31632300 | 1.88094400 |
| C | 0 | 2.97695100 | -0.47444000 | 3.00261900 |
| C | 0 | 5.09248600 | 0.44855600 | 0.98473400 |
| H | 0 | 3.67956900 | -2.31887900 | -0.57681000 |
| H | 0 | 5.38584200 | -2.05334600 | -0.10924100 |
| H | 0 | 2.79063300 | -2.04532600 | 1.59788100 |
| H | 0 | 4.40830300 | -1.88705500 | 2.29512000 |
| O | 0 | 5.77713600 | -0.18981400 | -1.90069200 |
| O | 0 | 3.86134300 | -1.00035400 | -2.71410800 |
| O | 0 | 2.71563900 | -1.05470400 | 4.07906600 |
| O | 0 | 2.76332100 | 0.77717200 | 2.82069000 |
| O | 0 | 6.20355000 | 0.03570200 | 1.36795800 |
| O | 0 | 4.76376800 | 1.68767200 | 0.91601000 |
| O | 0 | 2.88243400 | 0.13478500 | 0.10706800 |
| Er | 0 | 2.46477100 | 2.06435700 | 0.94526000 |

**Er-citrate-Ag$_{11}$ (mode 3)**

| | | | | |
|---|---|---|---|---|
| Ag | 0 | -1.04521400 | -2.71280000 | -1.12771200 |
| Ag | -1 | -0.97951200 | 0.05666900 | -1.00514700 |
| Ag | -1 | -0.86680900 | 2.93843400 | -1.06095000 |
| Ag | -1 | 0.76304300 | -2.85961200 | 1.27107900 |
| Ag | -1 | 0.87416100 | 0.02701500 | 1.21242000 |
| Ag | -1 | 1.45404900 | -1.49114200 | -1.17576100 |
| Ag | -1 | 1.57078100 | 1.39501600 | -1.23934900 |
| Ag | -1 | 0.98857300 | 2.91346600 | 1.15168600 |
| Ag | -1 | 3.31065900 | -1.51747600 | 1.03685300 |
| Ag | -1 | 3.42302900 | 1.36816100 | 0.97785800 |
| Ag | 0 | 3.88828100 | -0.10663000 | -1.40531800 |
| C | 0 | -3.75091300 | 3.28067600 | -0.04443600 |
| C | 0 | -3.15577300 | 2.55304300 | 1.14833100 |
| C | 0 | -3.42128700 | 1.04011600 | 1.08695900 |
| C | 0 | -2.54303900 | 0.36539400 | 2.15135700 |
| C | 0 | -2.60859100 | -1.14616300 | 2.26417600 |
| C | 0 | -4.90476500 | 0.74378600 | 1.43587500 |
| H | 0 | -2.07022300 | 2.71716100 | 1.18470200 |

| H | 0 | -3.59626300 | 2.95795800 | 2.06822300 |
|---|---|---|---|---|
| H | 0 | -1.49227000 | 0.61708800 | 1.92242400 |
| H | 0 | -2.74975800 | 0.78336200 | 3.14488200 |
| O | 0 | -4.99951000 | 3.28403700 | -0.18078300 |
| O | 0 | -2.95677900 | 3.86927400 | -0.85712000 |
| O | 0 | -1.95495600 | -1.68661800 | 3.18190300 |
| O | 0 | -3.31611300 | -1.81637600 | 1.42992500 |
| O | 0 | -5.41292800 | 1.23468800 | 2.46111200 |
| O | 0 | -5.51393900 | -0.06037900 | 0.64232900 |
| O | 0 | -3.14145300 | 0.53335200 | -0.19042400 |
| Er | 0 | -4.09307000 | -1.34532800 | -0.65213600 |

**Er-citrate-Ag₁₁ (mode 4)**

| Ag | 0 | -0.71016600 | 2.87405900 | -0.11220200 |
|---|---|---|---|---|
| Ag | -1 | -1.02489400 | 0.19239200 | 0.00837100 |
| Ag | -1 | -1.27979400 | -2.68493000 | 0.03568700 |
| Ag | -1 | 1.75420200 | 2.83285700 | -1.40622300 |
| Ag | -1 | 1.49903300 | -0.04466600 | -1.37804600 |
| Ag | -1 | 1.34308300 | 1.44552700 | 1.09332000 |

| Ag | -1 | 1.08776600 | -1.43490700 | 1.12065200 |
|----|----|------------|-------------|------------|
| Ag | -1 | 1.24363500 | -2.92178500 | -1.35070200 |
| Ag | -1 | 3.86635400 | 1.20594700 | -0.29285200 |
| Ag | -1 | 3.61108100 | -1.67108300 | -0.26546800 |
| Ag | 0 | 3.33110700 | -0.11582700 | 2.15907200 |
| C | 0 | -3.75317500 | 3.37179900 | -0.11009400 |
| C | 0 | -3.65565100 | 2.56782300 | 1.17490200 |
| C | 0 | -3.99434900 | 1.08484800 | 0.95975500 |
| C | 0 | -3.63274800 | 0.32809600 | 2.24659200 |
| C | 0 | -3.89631100 | -1.16651400 | 2.27668000 |
| C | 0 | -5.51417100 | 0.91877500 | 0.70111700 |
| H | 0 | -2.63916600 | 2.63362500 | 1.58525000 |
| H | 0 | -4.35110500 | 2.99309900 | 1.91000500 |
| H | 0 | -2.55037100 | 0.45731500 | 2.40850300 |
| H | 0 | -4.13792900 | 0.77410800 | 3.11251000 |
| O | 0 | -4.87253900 | 3.46774600 | -0.67025700 |
| O | 0 | -2.69356500 | 3.92927400 | -0.56113200 |
| O | 0 | -3.75475700 | -1.75674400 | 3.36901000 |

| O | 0 | -4.23672500 | -1.77552800 | 1.20036800 |
|---|---|---|---|---|
| O | 0 | -6.33789500 | 1.41461200 | 1.49108500 |
| O | 0 | -5.83068400 | 0.21298600 | -0.32483200 |
| O | 0 | -3.28907300 | 0.56452700 | -0.13363200 |
| Er | 0 | -4.16804700 | -1.21747800 | -1.01794900 |

**Er-citrate-Ag$_{11}$ (mode 5)**

| Ag | 0 | 0.66509400 | -2.84841100 | -1.36352100 |
|---|---|---|---|---|
| Ag | -1 | 1.00185400 | -0.16103600 | -1.14546600 |
| Ag | -1 | 1.36520600 | 2.69987800 | -0.96921000 |
| Ag | -1 | -2.21718700 | -2.64457700 | -1.54947400 |
| Ag | -1 | -1.85358900 | 0.21610600 | -1.37278300 |
| Ag | -1 | -0.78258700 | -1.50594300 | 0.68417000 |
| Ag | -1 | -0.41915500 | 1.35437300 | 0.86182200 |
| Ag | -1 | -1.49010300 | 3.07664000 | -1.19559900 |
| Ag | -1 | -3.63796500 | -1.12936900 | 0.45889400 |
| Ag | -1 | -3.27440600 | 1.73122700 | 0.63536800 |
| Ag | 0 | -2.05767300 | 0.03959200 | 2.59781900 |
| C | 0 | 4.43204600 | 2.59907100 | -0.77252600 |

| C  | 0 | 4.72512000 | 1.23488800  | -1.37740700 |
|----|---|------------|-------------|-------------|
| C  | 0 | 4.33009000 | 0.10201500  | -0.41656000 |
| C  | 0 | 4.44729700 | -1.24209800 | -1.14889100 |
| C  | 0 | 4.11420200 | -2.48943400 | -0.34780500 |
| C  | 0 | 5.27474300 | 0.06113500  | 0.81265100  |
| H  | 0 | 4.15312000 | 1.11014300  | -2.30492800 |
| H  | 0 | 5.79556300 | 1.16952600  | -1.61166200 |
| H  | 0 | 3.73694100 | -1.21917000 | -1.99182300 |
| H  | 0 | 5.44933700 | -1.37260900 | -1.57571700 |
| O  | 0 | 5.06437700 | 2.94451400  | 0.25558200  |
| O  | 0 | 3.55854700 | 3.33783000  | -1.34882700 |
| O  | 0 | 4.41460300 | -3.59614400 | -0.84544500 |
| O  | 0 | 3.52904800 | -2.37830000 | 0.78843900  |
| O  | 0 | 6.50903800 | 0.00744200  | 0.64568600  |
| O  | 0 | 4.69647800 | 0.02543400  | 1.95744400  |
| O  | 0 | 3.02060400 | 0.30395400  | 0.02008800  |
| Er | 0 | 2.46333100 | -0.66732000 | 1.85336300  |

**Tm-citrate-Ag$_{11}$ (mode 1)**

| Ag | 0 | 1.92219000 | -0.92450900 | -0.76146100 |
| Ag | -1 | 0.43588500 | 1.28257700 | -0.55042100 |
| Ag | -1 | -1.24460400 | 3.63101700 | -0.46182600 |
| Ag | -1 | -0.15347800 | -2.66020800 | -1.44084900 |
| Ag | -1 | -1.83464700 | -0.31255500 | -1.35185500 |
| Ag | -1 | -0.28353100 | -1.06943600 | 0.96661600 |
| Ag | -1 | -1.96664200 | 1.28052000 | 1.05452500 |
| Ag | -1 | -3.51535100 | 2.03505600 | -1.26400500 |
| Ag | -1 | -2.55598600 | -2.66349600 | 0.16419100 |
| Ag | -1 | -4.23679800 | -0.31590800 | 0.25220200 |
| Ag | 0 | -2.55360600 | -1.02932700 | 2.50450400 |
| C | 0 | 3.97185900 | -3.03870200 | -0.37472100 |
| C | 0 | 5.28968400 | -2.41797500 | 0.05094500 |
| C | 0 | 5.24833100 | -0.88707100 | 0.16824300 |
| C | 0 | 6.59821600 | -0.43663900 | 0.75877900 |
| C | 0 | 6.80563000 | 1.05208700 | 0.97363100 |
| C | 0 | 5.12273400 | -0.22728400 | -1.22997900 |
| H | 0 | 5.58355800 | -2.83406300 | 1.02219500 |

| H | 0 | 6.05212700 | -2.70788100 | -0.68599800 |
|---|---|---|---|---|
| H | 0 | 6.70609700 | -0.91200700 | 1.74546200 |
| H | 0 | 7.42983300 | -0.80104800 | 0.14225700 |
| O | 0 | 3.40517600 | -2.55398000 | -1.41576800 |
| O | 0 | 3.51707400 | -4.00569300 | 0.28421300 |
| O | 0 | 7.95760200 | 1.44316100 | 1.25775300 |
| O | 0 | 5.80815000 | 1.85520000 | 0.87790200 |
| O | 0 | 5.87108100 | -0.57436300 | -2.16161100 |
| O | 0 | 4.26145500 | 0.72637200 | -1.32016200 |
| O | 0 | 4.19256600 | -0.48334500 | 0.99601900 |
| Tm | 0 | 3.58367500 | 1.54789000 | 0.71451800 |

**Tm-citrate-Ag₁₁ (mode 2)**

| Ag | 0 | 1.73473700 | 0.04743200 | -2.41871200 |
|---|---|---|---|---|
| Ag | -1 | -0.24387100 | 1.61450300 | -1.48716000 |
| Ag | -1 | -2.41358300 | 3.25764900 | -0.51737500 |
| Ag | -1 | 0.26568100 | -2.37861100 | -2.18628600 |
| Ag | -1 | -1.90314900 | -0.73522400 | -1.21584700 |
| Ag | -1 | 0.67871100 | -0.63264000 | 0.07770800 |

| Ag | -1 | -1.49247500 | 1.01108100 | 1.04820600 |
| Ag | -1 | -4.07221100 | 0.90779900 | -0.24676900 |
| Ag | -1 | -0.98097700 | -2.98185200 | 0.34917600 |
| Ag | -1 | -3.15044000 | -1.33868400 | 1.31850600 |
| Ag | 0 | -0.51683300 | -1.14299100 | 2.50518400 |
| C | 0 | 4.63157600 | -0.80202700 | -1.83762500 |
| C | 0 | 4.44799500 | -1.53728200 | -0.51831600 |
| C | 0 | 3.98360400 | -0.55767700 | 0.57328300 |
| C | 0 | 3.57973800 | -1.33071100 | 1.83367800 |
| C | 0 | 3.01442500 | -0.50690100 | 2.98194100 |
| C | 0 | 5.11164500 | 0.43484000 | 0.95468200 |
| H | 0 | 3.67626500 | -2.30966800 | -0.62687500 |
| H | 0 | 5.39297300 | -2.01075300 | -0.22488100 |
| H | 0 | 2.78837700 | -2.04217800 | 1.54595600 |
| H | 0 | 4.41763900 | -1.92232500 | 2.22381200 |
| O | 0 | 5.60021600 | -0.00980100 | -1.94593500 |
| O | 0 | 3.78605800 | -1.01849700 | -2.77459800 |
| O | 0 | 2.77856400 | -1.10464100 | 4.05459300 |

| O | 0 | 2.79194000 | 0.74710600 | 2.82582200 |
|---|---|---|---|---|
| O | 0 | 6.22952900 | 0.00620300 | 1.29808200 |
| O | 0 | 4.78455500 | 1.67682600 | 0.94322100 |
| O | 0 | 2.89039600 | 0.16033300 | 0.09220800 |
| Tm | 0 | 2.50229100 | 2.05631200 | 0.97909100 |

**Tm-citrate-Ag$_{11}$ (mode 3)**

| Ag | 0 | -1.04722600 | -2.70703500 | -1.12529900 |
|---|---|---|---|---|
| Ag | -1 | -0.97403600 | 0.06195400 | -1.00581200 |
| Ag | -1 | -0.85522000 | 2.94354100 | -1.05799600 |
| Ag | -1 | 0.76190200 | -2.86085000 | 1.26710100 |
| Ag | -1 | 0.87914500 | 0.02560700 | 1.21206600 |
| Ag | -1 | 1.45627200 | -1.49079400 | -1.17789900 |
| Ag | -1 | 1.57913000 | 1.39518700 | -1.23786200 |
| Ag | -1 | 0.99968200 | 2.91188100 | 1.15495600 |
| Ag | -1 | 3.31240000 | -1.52382200 | 1.03503100 |
| Ag | -1 | 3.43089400 | 1.36164100 | 0.97965800 |
| Ag | 0 | 3.89340200 | -0.11074400 | -1.40527400 |
| C | 0 | -3.74148600 | 3.28674100 | -0.04872500 |

| C  | 0 | -3.14844100 | 2.56055800  | 1.14597700  |
| C  | 0 | -3.41408000 | 1.04780700  | 1.08621500  |
| C  | 0 | -2.53647200 | 0.37369400  | 2.15146900  |
| C  | 0 | -2.60368700 | -1.13761400 | 2.26446400  |
| C  | 0 | -4.89742900 | 0.75119100  | 1.43413900  |
| H  | 0 | -2.06298500 | 2.72475200  | 1.18411300  |
| H  | 0 | -3.59046700 | 2.96658000  | 2.06463600  |
| H  | 0 | -1.48543300 | 0.62425600  | 1.92235200  |
| H  | 0 | -2.74306100 | 0.79241900  | 3.14466800  |
| O  | 0 | -4.98978200 | 3.28940000  | -0.18761200 |
| O  | 0 | -2.94581500 | 3.87487800  | -0.86020100 |
| O  | 0 | -1.95383900 | -1.67917200 | 3.18389300  |
| O  | 0 | -3.30933900 | -1.80708000 | 1.42763600  |
| O  | 0 | -5.40862300 | 1.24515000  | 2.45616800  |
| O  | 0 | -5.50338800 | -0.05743900 | 0.64249100  |
| O  | 0 | -3.13405100 | 0.53904700  | -0.19043600 |
| Tm | 0 | -4.08142000 | -1.33259600 | -0.64217400 |

**Tm-citrate-Ag$_{11}$ (mode 4)**

| | | | | |
|---|---|---|---|---|
| Ag | 0 | -0.69444700 | 2.87992600 | -0.10534200 |
| Ag | -1 | -1.01773400 | 0.19998300 | 0.01472300 |
| Ag | -1 | -1.28259700 | -2.67648400 | 0.03789300 |
| Ag | -1 | 1.76765600 | 2.83295700 | -1.40129200 |
| Ag | -1 | 1.50250300 | -0.04359000 | -1.37734500 |
| Ag | -1 | 1.35692900 | 1.44310000 | 1.09681100 |
| Ag | -1 | 1.09157100 | -1.43654700 | 1.11992800 |
| Ag | -1 | 1.23708600 | -2.91991500 | -1.35420100 |
| Ag | -1 | 3.87641000 | 1.19690900 | -0.29505600 |
| Ag | -1 | 3.61116800 | -1.67924200 | -0.27180600 |
| Ag | 0 | 3.34187600 | -0.12687700 | 2.15546800 |
| C | 0 | -3.73990000 | 3.37534800 | -0.12739300 |
| C | 0 | -3.65143100 | 2.57969500 | 1.16345900 |
| C | 0 | -3.99103500 | 1.09617100 | 0.95562000 |
| C | 0 | -3.63255900 | 0.34543600 | 2.24691600 |
| C | 0 | -3.89669300 | -1.14882400 | 2.28244600 |
| C | 0 | -5.51017800 | 0.92812400 | 0.69504600 |
| H | 0 | -2.63746000 | 2.64683000 | 1.57970200 |

| H | 0 | -4.35066600 | 3.01075000 | 1.89157100 |
|---|---|---|---|---|
| H | 0 | -2.55036800 | 0.47479700 | 2.41005600 |
| H | 0 | -4.13919800 | 0.79563700 | 3.10977000 |
| O | 0 | -4.85530000 | 3.46698400 | -0.69620500 |
| O | 0 | -2.67739700 | 3.93071700 | -0.57395800 |
| O | 0 | -3.76138400 | -1.73466100 | 3.37770100 |
| O | 0 | -4.23097600 | -1.76255200 | 1.20667500 |
| O | 0 | -6.33644000 | 1.43422300 | 1.47528500 |
| O | 0 | -5.82356200 | 0.20791400 | -0.32227200 |
| O | 0 | -3.28390400 | 0.56931600 | -0.13358800 |
| Tm | 0 | -4.16216800 | -1.20839900 | -1.00125300 |

**Tm-citrate-Ag$_{11}$ (mode 5)**

| Ag | 0 | 0.66348400 | -2.83595900 | -1.39358500 |
|---|---|---|---|---|
| Ag | -1 | 0.99861800 | -0.15256700 | -1.15601000 |
| Ag | -1 | 1.35405200 | 2.70818200 | -0.97922700 |
| Ag | -1 | -2.21423200 | -2.64679900 | -1.54791500 |
| Ag | -1 | -1.85893200 | 0.21424800 | -1.37044600 |
| Ag | -1 | -0.77306600 | -1.50604200 | 0.67994200 |

| Ag | -1 | -0.41770900 | 1.35580800 | 0.85877300 |
| Ag | -1 | -1.50360400 | 3.07637000 | -1.19227200 |
| Ag | -1 | -3.63058500 | -1.13782400 | 0.46834800 |
| Ag | -1 | -3.27524600 | 1.72373400 | 0.64563000 |
| Ag | 0 | -2.04373200 | 0.03481400 | 2.60062800 |
| C | 0 | 4.42025600 | 2.59551600 | -0.78128500 |
| C | 0 | 4.72219800 | 1.22877800 | -1.37619000 |
| C | 0 | 4.31974600 | 0.10120100 | -0.41254100 |
| C | 0 | 4.43612300 | -1.24637500 | -1.13920300 |
| C | 0 | 4.09978500 | -2.49048900 | -0.33468600 |
| C | 0 | 5.25701100 | 0.06092200 | 0.82190200 |
| H | 0 | 4.15922900 | 1.09817200 | -2.30839000 |
| H | 0 | 5.79485700 | 1.16394900 | -1.60013800 |
| H | 0 | 3.72586200 | -1.22510000 | -1.98242100 |
| H | 0 | 5.43815000 | -1.37962400 | -1.56494200 |
| O | 0 | 5.03880000 | 2.94606000 | 0.25356100 |
| O | 0 | 3.55331800 | 3.32991900 | -1.37281400 |
| O | 0 | 4.41290000 | -3.59871700 | -0.82051200 |

| O | 0 | 3.49755000 | -2.37575600 | 0.79255100 |
| O | 0 | 6.49240100 | 0.01258600 | 0.66411200 |
| O | 0 | 4.67020700 | 0.01885500 | 1.96259900 |
| O | 0 | 3.00905200 | 0.30939300 | 0.01795500 |
| Tm | 0 | 2.44785700 | -0.65927100 | 1.83995100 |

**Yb-citrate-Ag$_{11}$ (mode 1)**

| Ag | 0 | 1.91301600 | -0.95848400 | -0.74934100 |
| Ag | -1 | 0.44165300 | 1.26427000 | -0.55209100 |
| Ag | -1 | -1.21558900 | 3.62928800 | -0.46317900 |
| Ag | -1 | -0.18667600 | -2.67219700 | -1.44307400 |
| Ag | -1 | -1.84456800 | -0.30814100 | -1.35374400 |
| Ag | -1 | -0.30110200 | -1.08059500 | 0.96475400 |
| Ag | -1 | -1.96095000 | 1.28582400 | 1.05276500 |
| Ag | -1 | -3.50197300 | 2.05596300 | -1.26569400 |
| Ag | -1 | -2.58914500 | -2.65207800 | 0.16192400 |
| Ag | -1 | -4.24668400 | -0.28798200 | 0.25017900 |
| Ag | 0 | -2.57151700 | -1.01873800 | 2.50217100 |
| C | 0 | 4.00879300 | -3.04928900 | -0.35186500 |

| C | 0 | 5.32599300 | -2.41200000 | 0.05252100 |
|---|---|---|---|---|
| C | 0 | 5.26383400 | -0.88195400 | 0.16371900 |
| C | 0 | 6.61032200 | -0.40312900 | 0.73977300 |
| C | 0 | 6.79093700 | 1.09016000 | 0.95389600 |
| C | 0 | 5.11415500 | -0.23136700 | -1.23578500 |
| H | 0 | 5.63750200 | -2.82030000 | 1.02157200 |
| H | 0 | 6.08194700 | -2.69586100 | -0.69336800 |
| H | 0 | 6.73877500 | -0.87595600 | 1.72525600 |
| H | 0 | 7.44277900 | -0.75029600 | 0.11440500 |
| O | 0 | 3.42023900 | -2.57426100 | -1.38499100 |
| O | 0 | 3.57649900 | -4.02019200 | 0.31639300 |
| O | 0 | 7.93725200 | 1.50211300 | 1.22952800 |
| O | 0 | 5.77818500 | 1.87626200 | 0.86674800 |
| O | 0 | 5.85515600 | -0.57337000 | -2.17478900 |
| O | 0 | 4.24042300 | 0.71154600 | -1.31952800 |
| O | 0 | 4.20873800 | -0.49313800 | 0.99985300 |
| Yb | 0 | 3.57527100 | 1.51358700 | 0.71019100 |

**Yb-citrate-Ag$_{II}$ (mode 2)**

| | | | | |
|---|---|---|---|---|
| Ag | 0 | 1.68497400 | 0.08075500 | -2.47640400 |
| Ag | -1 | -0.31390000 | 1.61602500 | -1.52564500 |
| Ag | -1 | -2.48831800 | 3.22972100 | -0.51274100 |
| Ag | -1 | 0.24584500 | -2.37201600 | -2.20701400 |
| Ag | -1 | -1.92593200 | -0.75870400 | -1.19373400 |
| Ag | -1 | 0.68355700 | -0.60017600 | 0.03318300 |
| Ag | -1 | -1.48824800 | 1.01195200 | 1.04588100 |
| Ag | -1 | -4.09798100 | 0.85355300 | -0.18204800 |
| Ag | -1 | -0.92281600 | -2.97794600 | 0.36432700 |
| Ag | -1 | -3.09754900 | -1.36396400 | 1.37581100 |
| Ag | 0 | -0.44734200 | -1.11568600 | 2.49479200 |
| C | 0 | 4.59525400 | -0.75266800 | -1.88857100 |
| C | 0 | 4.42866900 | -1.51263100 | -0.58095500 |
| C | 0 | 3.98252900 | -0.55217400 | 0.53417100 |
| C | 0 | 3.58191300 | -1.34842200 | 1.78176700 |
| C | 0 | 3.03385500 | -0.54860900 | 2.95528800 |
| C | 0 | 5.11872000 | 0.42446700 | 0.92894500 |
| H | 0 | 3.65424000 | -2.28180000 | -0.69314900 |

| H  | 0  | 5.37687100  | -1.99343600 | -0.31044400 |
|----|----|-------------|-------------|-------------|
| H  | 0  | 2.78038200  | -2.04477100 | 1.48464800  |
| H  | 0  | 4.41700400  | -1.95796300 | 2.14957900  |
| O  | 0  | 5.55309400  | 0.05397700  | -1.98705800 |
| O  | 0  | 3.74846100  | -0.96472200 | -2.82515400 |
| O  | 0  | 2.81921700  | -1.16653200 | 4.02033800  |
| O  | 0  | 2.80044300  | 0.70747500  | 2.82677900  |
| O  | 0  | 6.24142400  | -0.01524500 | 1.24024200  |
| O  | 0  | 4.79112600  | 1.66611400  | 0.96606400  |
| O  | 0  | 2.89323900  | 0.18662800  | 0.07558400  |
| Yb | 0  | 2.52046100  | 2.04159300  | 1.01828700  |

**Yb-citrate-Ag$_{11}$ (mode 3)**

| Ag | 0  | -1.04755900 | -2.70195600 | -1.12254700 |
|----|----|-------------|-------------|-------------|
| Ag | -1 | -0.96830400 | 0.06649700  | -1.00837400 |
| Ag | -1 | -0.84515300 | 2.94798700  | -1.05563900 |
| Ag | -1 | 0.76127400  | -2.86268000 | 1.26118300  |
| Ag | -1 | 0.88286000  | 0.02368700  | 1.21107400  |
| Ag | -1 | 1.45984400  | -1.48957900 | -1.18091100 |

| Ag | -1 | 1.58704800 | 1.39631200 | -1.23594500 |
| Ag | -1 | 1.00774200 | 2.90987100 | 1.15889300 |
| Ag | -1 | 3.31396200 | -1.52906600 | 1.03359600 |
| Ag | -1 | 3.43679800 | 1.35630600 | 0.98314900 |
| Ag | 0 | 3.89907000 | -0.11243400 | -1.40383200 |
| C | 0 | -3.73370400 | 3.29096500 | -0.05417300 |
| C | 0 | -3.14227200 | 2.56724900 | 1.14281200 |
| C | 0 | -3.40687200 | 1.05450200 | 1.08551200 |
| C | 0 | -2.52981800 | 0.38238300 | 2.15248800 |
| C | 0 | -2.59779900 | -1.12865100 | 2.26673900 |
| C | 0 | -4.88983500 | 0.75654900 | 1.43182700 |
| H | 0 | -2.05705000 | 2.73244400 | 1.18276500 |
| H | 0 | -3.58624500 | 2.97454700 | 2.05997700 |
| H | 0 | -1.47855700 | 0.63241600 | 1.92359100 |
| H | 0 | -2.73724100 | 0.80262100 | 3.14487700 |
| O | 0 | -4.98168900 | 3.29191800 | -0.19557800 |
| O | 0 | -2.93689200 | 3.87883300 | -0.86467700 |
| O | 0 | -1.95324800 | -1.67055200 | 3.18925700 |
| O | 0 | -3.29950700 | -1.79791500 | 1.42583900 |
| O | 0 | -5.40490000 | 1.25312400 | 2.45039200 |
| O | 0 | -5.49152400 | -0.05707500 | 0.64184800 |
| O | 0 | -3.12634600 | 0.54322700 | -0.19024600 |
| Yb | 0 | -4.06718000 | -1.32064200 | -0.63182200 |

**Yb-citrate-Ag$_{11}$ (mode 4)**

| | | | | |
|---|---|---|---|---|
| Ag | 0 | -0.68134100 | 2.88425600 | -0.10075100 |
| Ag | -1 | -1.01092100 | 0.20548700 | 0.01743700 |
| Ag | -1 | -1.28319800 | -2.67033600 | 0.03490000 |
| Ag | -1 | 1.78018500 | 2.83427700 | -1.39522500 |
| Ag | -1 | 1.50767400 | -0.04169700 | -1.37688500 |
| Ag | -1 | 1.36759500 | 1.44026800 | 1.10033900 |
| Ag | -1 | 1.09489000 | -1.43868500 | 1.11779200 |
| Ag | -1 | 1.23490500 | -2.91733600 | -1.35939500 |
| Ag | -1 | 3.88552100 | 1.19053300 | -0.29374600 |
| Ag | -1 | 3.61288700 | -1.68499100 | -0.27621200 |
| Ag | 0 | 3.34939900 | -0.13671000 | 2.15440600 |
| C | 0 | -3.72888400 | 3.37722000 | -0.14955500 |
| C | 0 | -3.64901800 | 2.59225400 | 1.14839000 |
| C | 0 | -3.98760600 | 1.10745700 | 0.95046000 |
| C | 0 | -3.62974300 | 0.36546900 | 2.24709000 |
| C | 0 | -3.89420600 | -1.12834200 | 2.29130900 |
| C | 0 | -5.50594700 | 0.93463700 | 0.69006300 |
| H | 0 | -2.63791900 | 2.66291100 | 1.57090800 |
| H | 0 | -4.35290500 | 3.02928900 | 1.86840700 |
| H | 0 | -2.54754400 | 0.49572600 | 2.40956900 |
| H | 0 | -4.13680800 | 0.82148900 | 3.10663800 |
| O | 0 | -4.84056100 | 3.46376700 | -0.72650000 |
| O | 0 | -2.66353800 | 3.92925400 | -0.59321600 |
| O | 0 | -3.76410200 | -1.70807100 | 3.39011000 |
| O | 0 | -4.22344200 | -1.74814700 | 1.21712300 |
| O | 0 | -6.33457700 | 1.45024500 | 1.46091900 |
| O | 0 | -5.81628000 | 0.19857400 | -0.31731300 |
| O | 0 | -3.27953600 | 0.57334700 | -0.13500500 |
| Yb | 0 | -4.15547300 | -1.20044800 | -0.98080000 |

**Yb-citrate-Ag₁₁ (mode 5)**

| | | | | |
|----|----|----|----|----|
| Ag | 0 | 1.09399000 | -2.67626500 | -0.89383000 |
| Ag | -1 | 0.99064000 | 0.09160900 | -0.86039200 |
| Ag | -1 | 0.93490400 | 2.97524400 | -0.69229400 |
| Ag | -1 | -1.79162300 | -2.81498600 | -1.57110000 |
| Ag | -1 | -1.84717400 | 0.06855900 | -1.40245500 |
| Ag | -1 | -0.78137400 | -1.48617000 | 0.78620500 |
| Ag | -1 | -0.83711000 | 1.39712400 | 0.95594700 |
| Ag | -1 | -1.90287300 | 2.95217200 | -1.23315000 |
| Ag | -1 | -3.61927700 | -1.50944200 | 0.24641200 |
| Ag | -1 | -3.67478000 | 1.37409000 | 0.41492500 |
| Ag | 0 | -2.46065300 | -0.14969400 | 2.52969800 |
| C | 0 | 3.82597400 | 3.27090100 | 0.19860100 |
| C | 0 | 4.31924800 | 2.30267400 | -0.86422800 |
| C | 0 | 4.34043900 | 0.84458200 | -0.39134900 |
| C | 0 | 4.70379700 | -0.04327100 | -1.59245200 |
| C | 0 | 4.82203400 | -1.53507700 | -1.33739200 |
| C | 0 | 5.41224600 | 0.62973200 | 0.70944600 |
| H | 0 | 3.67012700 | 2.36878100 | -1.74645300 |

| H | 0 | 5.33168300 | 2.60964600 | -1.16119300 |
| H | 0 | 3.91233100 | 0.08529400 | -2.34723300 |
| H | 0 | 5.64078000 | 0.29111100 | -2.05488900 |
| O | 0 | 4.37689500 | 3.26536200 | 1.32546400 |
| O | 0 | 2.88305200 | 4.07653300 | -0.12312700 |
| O | 0 | 5.31545300 | -2.24507700 | -2.23846100 |
| O | 0 | 4.40681300 | -2.02135700 | -0.22404700 |
| O | 0 | 6.58613700 | 0.99832200 | 0.51568400 |
| O | 0 | 5.01349400 | 0.00664400 | 1.76020500 |
| O | 0 | 3.09043400 | 0.48628400 | 0.12224500 |
| Yb | 0 | 3.09435700 | -1.21644400 | 1.41016900 |

**Lu-citrate-Ag$_{\text{II}}$ (mode 1)**

| Ag | 0 | 1.78210200 | -0.03303000 | -2.14830200 |
| Ag | -1 | -0.19135200 | 1.55498600 | -1.41633800 |
| Ag | -1 | -2.39318700 | 3.28780700 | -0.71171100 |
| Ag | -1 | 0.28780200 | -2.47866400 | -1.84694800 |
| Ag | -1 | -1.91425200 | -0.74677500 | -1.14162800 |
| Ag | -1 | 0.56383400 | -0.61193900 | 0.33994200 |

| Ag | -1 | -1.64099400 | 1.12134100 | 1.04503600 |
| Ag | -1 | -4.11624700 | 0.98501100 | -0.43754400 |
| Ag | -1 | -1.16136000 | -2.91355100 | 0.61451900 |
| Ag | -1 | -3.36342700 | -1.18176100 | 1.31884900 |
| Ag | 0 | -0.81970300 | -0.94704800 | 2.69193500 |
| C | 0 | 3.46475900 | -2.30583900 | -1.17191100 |
| C | 0 | 4.37845500 | -2.05076200 | 0.01719100 |
| C | 0 | 4.31502900 | -0.59818800 | 0.50817400 |
| C | 0 | 5.19158200 | -0.44132200 | 1.76077700 |
| C | 0 | 5.18719500 | 0.92489300 | 2.43323400 |
| C | 0 | 4.88527200 | 0.36119000 | -0.57033500 |
| H | 0 | 4.09648300 | -2.71190700 | 0.84455400 |
| H | 0 | 5.40916400 | -2.28816800 | -0.27992100 |
| H | 0 | 4.83742200 | -1.16064800 | 2.51386900 |
| H | 0 | 6.23563200 | -0.69804700 | 1.54235600 |
| O | 0 | 3.58361700 | -1.53165600 | -2.17621400 |
| O | 0 | 2.64614900 | -3.26241800 | -1.10787200 |
| O | 0 | 6.00908800 | 1.12135500 | 3.34764400 |

| O   | 0  | 4.34445100  | 1.82862100  | 2.06206300  |
|-----|----|-------------|-------------|-------------|
| O   | 0  | 6.00097400  | 0.15643200  | -1.06750100 |
| O   | 0  | 4.16192600  | 1.40155000  | -0.83597300 |
| O   | 0  | 2.99607300  | -0.24666600 | 0.84152800  |
| Lu  | 0  | 2.62518500  | 1.80382300  | 0.68366000  |

**Lu-citrate-Ag$_{11}$ (mode 2)**

| Ag  | 0  | 1.79209200  | -0.19853400 | -2.36882700 |
|-----|----|-------------|-------------|-------------|
| Ag  | -1 | -0.10548500 | 1.51042600  | -1.52453400 |
| Ag  | -1 | -2.19839600 | 3.31005400  | -0.67528200 |
| Ag  | -1 | 0.18973300  | -2.52958600 | -2.06408500 |
| Ag  | -1 | -1.90358000 | -0.72930100 | -1.21459000 |
| Ag  | -1 | 0.64781500  | -0.72686600 | 0.14614300  |
| Ag  | -1 | -1.45043000 | 1.07509700  | 0.99543100  |
| Ag  | -1 | -3.99760500 | 1.07098400  | -0.36602100 |
| Ag  | -1 | -1.15589300 | -2.96337200 | 0.45556400  |
| Ag  | -1 | -3.24867700 | -1.16392200 | 1.30465400  |
| Ag  | 0  | -0.63842600 | -1.08465900 | 2.55415800  |
| C   | 0  | 4.69376200  | -0.97800200 | -1.68683900 |

| C | 0 | 4.49707700 | -1.60170800 | -0.31192700 |
| C | 0 | 3.97379400 | -0.54461700 | 0.67329600 |
| C | 0 | 3.53121400 | -1.21194400 | 1.98105100 |
| C | 0 | 2.85188400 | -0.32485500 | 3.01384800 |
| C | 0 | 5.06180800 | 0.50888200 | 1.00106200 |
| H | 0 | 3.75283400 | -2.40531200 | -0.36973700 |
| H | 0 | 5.44684700 | -2.01383600 | 0.04978600 |
| H | 0 | 2.79437100 | -1.98934300 | 1.72125600 |
| H | 0 | 4.37197400 | -1.71730900 | 2.47257200 |
| O | 0 | 5.65909100 | -0.19055100 | -1.84790600 |
| O | 0 | 3.86058600 | -1.27745300 | -2.61038700 |
| O | 0 | 2.57686100 | -0.82892700 | 4.11752200 |
| O | 0 | 2.56608400 | 0.90124000 | 2.72861800 |
| O | 0 | 6.18487500 | 0.15790800 | 1.38939700 |
| O | 0 | 4.68332300 | 1.73994600 | 0.89180000 |
| O | 0 | 2.88197600 | 0.10475400 | 0.08866100 |
| Lu | 0 | 2.47761200 | 2.00533600 | 0.83998400 |

**Lu-citrate-Ag$_{II}$ (mode 3)**

| Ag | 0 | -1.05850100 | -2.69700700 | -0.99737200 |
| Ag | -1 | -0.96948300 | 0.06856800 | -0.89487200 |
| Ag | -1 | -0.85237400 | 2.95042900 | -0.93422900 |
| Ag | -1 | 0.86127000 | -2.86818800 | 1.28370700 |
| Ag | -1 | 0.97668600 | 0.01856000 | 1.24160000 |
| Ag | -1 | 1.45092500 | -1.48297000 | -1.18017300 |
| Ag | -1 | 1.57174000 | 1.40333300 | -1.22744700 |
| Ag | -1 | 1.09530400 | 2.90513900 | 1.19727800 |
| Ag | -1 | 3.39982600 | -1.52962800 | 0.95125000 |
| Ag | -1 | 3.51647700 | 1.35612900 | 0.90874600 |
| Ag | 0 | 3.87813300 | -0.10517400 | -1.50248700 |
| C | 0 | -3.78635800 | 3.28257500 | -0.06823300 |
| C | 0 | -3.25109000 | 2.54064000 | 1.14466500 |
| C | 0 | -3.51864100 | 1.03262200 | 1.05139400 |
| C | 0 | -2.68317800 | 0.33192500 | 2.13320700 |
| C | 0 | -2.73740800 | -1.18244600 | 2.19264000 |
| C | 0 | -5.01949100 | 0.73476700 | 1.31619200 |
| H | 0 | -2.16867000 | 2.69997800 | 1.23600800 |
| H | 0 | -3.73449900 | 2.93499200 | 2.04728000 |
| H | 0 | -1.62698700 | 0.59757300 | 1.95871200 |
| H | 0 | -2.94242600 | 0.71370300 | 3.12902700 |
| O | 0 | -5.02609700 | 3.28593400 | -0.26521700 |
| O | 0 | -2.95084600 | 3.87931600 | -0.83089800 |
| O | 0 | -2.10196600 | -1.75725700 | 3.09357100 |
| O | 0 | -3.42991000 | -1.82893600 | 1.31600900 |
| O | 0 | -5.59043700 | 1.21225200 | 2.30622300 |
| O | 0 | -5.58085500 | -0.07107700 | 0.47609400 |
| O | 0 | -3.18124400 | 0.54374100 | -0.22491900 |
| Lu | 0 | -4.12041900 | -1.27324700 | -0.67210100 |

**Lu-citrate-Ag$_{11}$ (mode 4)**

| | | | | |
|---|---|---|---|---|
| Ag | 0 | -0.76975300 | 2.81685600 | -0.04548400 |
| Ag | -1 | -1.00802800 | 0.11619200 | 0.04204100 |
| Ag | -1 | -1.16214500 | -2.76864400 | 0.03181000 |
| Ag | -1 | 1.66229200 | 2.87080100 | -1.36421400 |
| Ag | -1 | 1.50773600 | -0.01415500 | -1.37341100 |
| Ag | -1 | 1.32653000 | 1.43627600 | 1.11981300 |

| Ag | -1 | 1.17178800 | -1.45134400 | 1.10993900 |
|----|----|-----------|------------|------------|
| Ag | -1 | 1.35285700 | -2.89839400 | -1.38339100 |
| Ag | -1 | 3.84156200 | 1.30380900 | -0.29516200 |
| Ag | -1 | 3.68689300 | -1.58055400 | -0.30524500 |
| Ag | 0 | 3.37498600 | -0.06970900 | 2.14903800 |
| C | 0 | -3.78909200 | 3.34362400 | -0.28662100 |
| C | 0 | -3.77576200 | 2.61212700 | 1.04495100 |
| C | 0 | -4.06885000 | 1.11604900 | 0.88008400 |
| C | 0 | -3.74836400 | 0.41829300 | 2.21053300 |
| C | 0 | -3.99291100 | -1.07610300 | 2.29048400 |
| C | 0 | -5.56927900 | 0.89094700 | 0.55353300 |
| H | 0 | -2.79732400 | 2.72328400 | 1.53031300 |
| H | 0 | -4.53550200 | 3.05436400 | 1.70171100 |
| H | 0 | -2.67480700 | 0.57028400 | 2.40562500 |
| H | 0 | -4.29399300 | 0.89136500 | 3.03627500 |
| O | 0 | -4.87187700 | 3.41068100 | -0.91773600 |
| O | 0 | -2.69866900 | 3.86809200 | -0.70030800 |
| O | 0 | -3.90283500 | -1.63083600 | 3.39904000 |

| O  | 0  | -4.27325600 | -1.72959900 | 1.21304200  |
|----|----|-------------|-------------|-------------|
| O  | 0  | -6.44986000 | 1.42386600  | 1.24270800  |
| O  | 0  | -5.80565800 | 0.08122100  | -0.42490700 |
| O  | 0  | -3.29756800 | 0.56578500  | -0.15968400 |
| Lu | 0  | -4.08333500 | -1.23599800 | -0.89124600 |

**Lu-citrate-Ag$_{11}$ (mode 5)**

| Ag | 0  | 0.98201400  | -2.77002900 | -1.25615700 |
|----|----|-------------|-------------|-------------|
| Ag | -1 | 0.83897500  | -0.02780500 | -1.38499200 |
| Ag | -1 | 1.12731900  | 2.82581200  | -1.03782900 |
| Ag | -1 | -2.32407500 | -2.59848800 | -1.66887800 |
| Ag | -1 | -2.03555800 | 0.25497900  | -1.32134000 |
| Ag | -1 | -0.72164800 | -1.56213800 | 0.49946000  |
| Ag | -1 | -0.43320500 | 1.29103000  | 0.84802900  |
| Ag | -1 | -1.74706300 | 3.10850800  | -0.97326500 |
| Ag | -1 | -3.59599500 | -1.27963900 | 0.56518800  |
| Ag | -1 | -3.30750600 | 1.57386300  | 0.91249600  |
| Ag | 0  | -1.82079900 | -0.24135200 | 2.64847600  |
| C  | 0  | 4.06162000  | 2.94138200  | -0.39065200 |

| C | 0 | 4.85580200 | 1.76830400 | -0.94433600 |
| C | 0 | 4.51244000 | 0.46752400 | -0.20647700 |
| C | 0 | 5.17982400 | -0.71523000 | -0.92615300 |
| C | 0 | 4.93074500 | -2.10168500 | -0.35596200 |
| C | 0 | 5.04997100 | 0.50310600 | 1.24924900 |
| H | 0 | 4.62582900 | 1.63436600 | -2.00778000 |
| H | 0 | 5.92811200 | 1.98204500 | -0.84259000 |
| H | 0 | 4.79889500 | -0.73255800 | -1.95862600 |
| H | 0 | 6.26543200 | -0.57158600 | -0.98995700 |
| O | 0 | 4.18568400 | 3.22892700 | 0.82379000 |
| O | 0 | 3.30641000 | 3.58012600 | -1.20574700 |
| O | 0 | 5.62540800 | -3.04299100 | -0.78200500 |
| O | 0 | 4.00745300 | -2.27605300 | 0.52774600 |
| O | 0 | 6.23845500 | 0.78236200 | 1.47426700 |
| O | 0 | 4.20396200 | 0.15827600 | 2.16089300 |
| O | 0 | 3.12361500 | 0.29253400 | -0.18419700 |
| Lu | 0 | 2.45177300 | -0.96184400 | 1.33325600 |